\documentclass[twocolumn, times]{aastex631}

\usepackage{amsmath}
\usepackage{xspace}
\usepackage{comment}
\shorttitle{eDisk First-Look Results of IRAS 16544-1604}
\shortauthors{Kido et al.}
\begin{document}
\title{Early Planet Formation in Embedded Disks (eDisk) VII: \\ Keplerian Disk, Disk Substructure, and Accretion Streamers in the Class 0 Protostar IRAS 16544-1604 in CB 68}

\correspondingauthor{Miyu Kido}
\email{k3394334@kadai.jp}

\author[0000-0002-2902-4239]{Miyu Kido}
\affiliation{Department of Physics and Astronomy, Graduate School of Science and Engineering, Kagoshima University, 1-21-35 Korimoto, Kagoshima,Kagoshima 890-0065, Japan}

\author[0000-0003-0845-128X]{Shigehisa Takakuwa}
\affiliation{Department of Physics and Astronomy, Graduate School of Science and Engineering, Kagoshima University, 1-21-35 Korimoto, Kagoshima,Kagoshima 890-0065, Japan}
\affiliation{Academia Sinica Institute of Astronomy \& Astrophysics,
11F of Astronomy-Mathematics Building, AS/NTU, No.1, Sec. 4, Roosevelt Rd, Taipei 10617, Taiwan, R.O.C.}

\author[0000-0003-1549-6435]{Kazuya Saigo}
\affiliation{Department of Physics and Astronomy, Graduate School of Science and Engineering, Kagoshima University, 1-21-35 Korimoto, Kagoshima,Kagoshima 890-0065, Japan}

\author[0000-0003-0998-5064]{Nagayoshi Ohashi}
\affiliation{Academia Sinica Institute of Astronomy \& Astrophysics,
11F of Astronomy-Mathematics Building, AS/NTU, No.1, Sec. 4, Roosevelt Rd, Taipei 10617, Taiwan, R.O.C.}

\author[0000-0002-6195-0152]{John J. Tobin}
\affil{National Radio Astronomy Observatory, 520 Edgemont Rd., Charlottesville, VA 22903 USA} 

\author[0000-0001-9133-8047]{Jes K. J{\o}rgensen}
\affil{Niels Bohr Institute, University of Copenhagen, {\O}ster Voldgade 5--7, DK~1350 Copenhagen K., Denmark}

\author[0000-0003-3283-6884]{Yuri Aikawa}
\affiliation{Department of Astronomy, Graduate School of Science, The University of Tokyo, 7-3-1 Hongo, Bunkyo-ku, Tokyo 113-0033, Japan}

\author[0000-0002-8238-7709]{Yusuke Aso}
\affiliation{Korea Astronomy and Space Science Institute, 776 Daedeok-daero, Yuseong-gu, Daejeon 34055, Republic of Korea}

\author[0000-0002-3566-6270]{Frankie J. Encalada}
\affiliation{Department of Astronomy, University of Illinois, 1002 West Green St, Urbana, IL 61801, USA}

\author[0000-0002-8591-472X]{Christian Flores}
\affiliation{Academia Sinica Institute of Astronomy \& Astrophysics,
11F of Astronomy-Mathematics Building, AS/NTU, No.1, Sec. 4, Roosevelt Rd, Taipei 10617, Taiwan, R.O.C.}

\author[0000-0001-5782-915X]{Sacha Gavino}
\affiliation{Niels Bohr Institute, University of Copenhagen, {\O}ster Voldgade 5--7, DK~1350 Copenhagen K., Denmark}

\author[0000-0003-4518-407X]{Itziar de Gregorio-Monsalvo}
\affiliation{European Southern Observatory, Alonso de Cordova 3107, Casilla 19, Vitacura, Santiago, Chile}

\author[0000-0002-9143-1433]{Ilseung Han}
\affiliation{Division of Astronomy and Space Science, University of Science and Technology, 217 Gajeong-ro, Yuseong-gu, Daejeon 34113, Republic of Korea}
\affiliation{Korea Astronomy and Space Science Institute, 776 Daedeok-daero, Yuseong-gu, Daejeon 34055, Republic of Korea}

\author[0000-0002-4317-767X]{Shingo Hirano}
\affiliation{Department of Astronomy, Graduate School of Science, The University of Tokyo, 7-3-1 Hongo, Bunkyo-ku, Tokyo 113-0033, Japan}

%\author{Chul-Hwan Kim}
%\affiliation{Department of Physics and Astronomy, Seoul National University, Seoul National University, 1 Gwanak-ro, Gwanak-gu, Seoul 08826, Korea}

\author[0000-0003-2777-5861]{Patrick M. Koch}
\affil{Academia Sinica Institute of Astronomy \& Astrophysics,
11F of Astronomy-Mathematics Building, AS/NTU, No.1, Sec. 4, Roosevelt Rd, Taipei 10617, Taiwan, R.O.C.}

\author[0000-0003-4022-4132]{Woojin Kwon}
\affiliation{Department of Earth Science Education, Seoul National University, 1 Gwanak-ro, Gwanak-gu, Seoul 08826, Republic of Korea}
\affiliation{SNU Astronomy Research Center, Seoul National University, 1 Gwanak-ro, Gwanak-gu, Seoul 08826, Republic of Korea}

\author[0000-0001-5522-486X]{Shih-Ping Lai}
\affiliation{Institute of Astronomy, National Tsing Hua University, No. 101, Section 2, Kuang-Fu Road, Hsinchu 30013, Taiwan}
\affiliation{Center for Informatics and Computation in Astronomy, National Tsing Hua University, No. 101, Section 2, Kuang-Fu Road, Hsinchu 30013, Taiwan}
\affiliation{Department of Physics, National Tsing Hua University, No. 101, Section 2, Kuang-Fu Road, Hsinchu 30013, Taiwan}
\affiliation{Academia Sinica Institute of Astronomy and Astrophysics, P.O. Box 23-141, 10617 Taipei, Taiwan}

\author[0000-0002-3179-6334]{Chang Won Lee}
\affiliation{Division of Astronomy and Space Science, University of Science and Technology, 217 Gajeong-ro, Yuseong-gu, Daejeon 34113, Republic of Korea}
\affiliation{Korea Astronomy and Space Science Institute, 776 Daedeok-daero, Yuseong-gu, Daejeon 34055, Republic of Korea}

\author[0000-0003-3119-2087]{Jeong-Eun Lee}
\affiliation{Department of Physics and Astronomy, Seoul National University, 1 Gwanak-ro, Gwanak-gu, Seoul 08826, Korea}

\author[0000-0002-7402-6487]{Zhi-Yun Li}
\affiliation{University of Virginia, 530 McCormick Rd., Charlottesville, Virginia 22904, USA}

\author[0000-0001-7233-4171]{Zhe-Yu Daniel Lin}
\affiliation{University of Virginia, 530 McCormick Rd., Charlottesville, Virginia 22904, USA}

\author[0000-0002-4540-6587]{Leslie W.  Looney}
\affiliation{Department of Astronomy, University of Illinois, 1002 West Green St, Urbana, IL 61801, USA}

\author[0000-0002-7002-939X]{Shoji Mori}
\affiliation{Astronomical Institute, Graduate School of Science, Tohoku University, Sendai 980-8578, Japan}

%\author[0000-0002-0554-1151]{Mayank Narang}
%\affiliation{Academia Sinica Institute of Astronomy \& Astrophysics,
%11F of Astronomy-Mathematics Building, AS/NTU, No.1, Sec. 4, Roosevelt Rd, Taipei 10617, Taiwan, R.O.C.}

\author[0000-0002-0244-6650]{Suchitra Narayanan}
\affiliation{Institute for Astronomy, University of Hawai‘i at Mānoa, 2680 Woodlawn Dr., Honolulu, HI 96822, USA}

\author[0000-0002-9912-5705]{Adele L. Plunkett}
\affiliation{National Radio Astronomy Observatory, 520 Edgemont Rd., Charlottesville, VA 22903 USA}

\author[0000-0002-4372-5509]{Nguyen Thi Phuong}
\affiliation{Korea Astronomy and Space Science Institute, 776 Daedeok-daero, Yuseong-gu, Daejeon 34055, Republic of Korea}
\affiliation{Department of Astrophysics, Vietnam National Space Center, Vietnam Academy of Science and Techonology, 18 Hoang Quoc Viet, Cau Giay, Hanoi, Vietnam}

\author[0000-0003-4361-5577]{Jinshi Sai (Insa Choi)}
\affiliation{Academia Sinica Institute of Astronomy \& Astrophysics,
11F of Astronomy-Mathematics Building, AS/NTU, No.1, Sec. 4, Roosevelt Rd, Taipei 10617, Taiwan, R.O.C.}

\author[0000-0001-6267-2820]{Alejandro Santamaría-Miranda}
\affiliation{European Southern Observatory, Alonso de Cordova 3107, Casilla 19, Vitacura, Santiago, Chile}

\author[0000-0002-0549-544X]{Rajeeb Sharma}
\affiliation{Niels Bohr Institute, University of Copenhagen, {\O}ster Voldgade 5--7, DK~1350 Copenhagen K., Denmark}

\author[0000-0002-9209-8708]{Patrick Sheehan}
\affiliation{National Radio Astronomy Observatory, 520 Edgemont Rd., Charlottesville, VA 22903 USA}

\author[0000-0003-0334-1583]{Travis J. Thieme}
\affiliation{Institute of Astronomy, National Tsing Hua University, No. 101, Section 2, Kuang-Fu Road, Hsinchu 30013, Taiwan}
\affiliation{Center for Informatics and Computation in Astronomy, National Tsing Hua University, No. 101, Section 2, Kuang-Fu Road, Hsinchu 30013, Taiwan}
\affiliation{Department of Physics, National Tsing Hua University, No. 101, Section 2, Kuang-Fu Road, Hsinchu 30013, Taiwan}

\author[0000-0001-8105-8113]{Kengo Tomida}
\affiliation{Astronomical Institute, Graduate School of Science, Tohoku University, Sendai 980-8578, Japan}

\author[0000-0002-2555-9869]{Merel L.R. van 't Hoff}
\affiliation{Department of Astronomy, University of Michigan, 1085 S. University Ave., Ann Arbor, MI 48109-1107, USA}

\author[0000-0001-5058-695X]{Jonathan P. Williams}
\affiliation{Institute for Astronomy, University of Hawai‘i at Mānoa, 2680 Woodlawn Dr., Honolulu, HI 96822, USA}

\author[0000-0003-4099-6941]{Yoshihide Yamato}
\affiliation{Department of Astronomy, Graduate School of Science, The University of Tokyo, 7-3-1 Hongo, Bunkyo-ku, Tokyo 113-0033, Japan}

\author[0000-0003-1412-893X]{Hsi-Wei Yen}
\affiliation{Academia Sinica Institute of Astronomy \& Astrophysics,
11F of Astronomy-Mathematics Building, AS/NTU, No.1, Sec. 4, Roosevelt Rd, Taipei 10617, Taiwan, R.O.C.}
%%%%%%%%%%%%%%%%%%%%%%%%%%%%%%%%%%%%%%%%%%%%%%%%%%%%%%%%%%
%\textcolor{red,blue}-->the sentence changed in the first revision

%\textcolor{teal}-->the sentence changed in the second revision
%%%%%%%%%%%%%%%%%%%%%%%%%%%%%%%%%%%%%%%%%%%%%%%%%%%%%%%%%%

%%%%%%%%%%%%%%%%%%%%%%%%%%%%%%%%%%%%%%%%%%%%%%



\begin{abstract}
We present observations of the Class 0 protostar IRAS 16544-1604 in CB 68 from the ``Early Planet Formation in Embedded Disks (eDisk)'' ALMA Large program. 
%The 1.3-mm dust-continuum image with a high spatial resolution offered by eDisk ($\sim$10 au) 　reveals a dusty protostellar disk of radius $\sim$30 au　close to edge-on for the first time.
The ALMA observations target continuum and lines at 1.3-mm with an angular resolution of $\sim$5 au.
The continuum image reveals a dusty protostellar disk with a radius of $\sim$30 au seen close to edge-on, and asymmetric structures both along the major and minor axes.
%While the asymmetry along the minor axis can be　interpreted as the effect of the dust flaring, the asymmetry along the major axis reveals a real non-axisymmetric structure.
While the asymmetry along the minor axis can be interpreted as the effect of the dust flaring, the asymmetry along the major axis comes from a real non-axisymmetric structure.
%The CO (2--1) isotopologue lines trace different components associated with the disk, outflow, and infalling streamers.
%The CO image cubes clearly show the gas in the disk that follows a Keplerian rotation pattern,　and the mass of the central protostar is estimated to be $\sim$0.12 $M_{\odot}$ from the fitting to the emission in the Position-Velocity diagrams.
The C$^{18}$O image cubes clearly show the gas in the disk that follows a Keplerian rotation pattern around a $\sim$0.14 $M_{\odot}$ central protostar.
Furthermore, there are $\sim$1500 au-scale streamer-like features of gas connecting from North-East, North-North-West, and North-West to the disk, as well as the bending outflow as seen in the $^{12}$CO (2--1) emission. At the apparent landing point of NE streamer, there are SO (6$_5$--5$_4$) and SiO (5--4) emission detected.
The spatial and velocity structure of NE streamer can be interpreted as a free-falling gas with a conserved specific angular momentum, and the detection of the SO and SiO emission at the tip of the streamer implies presence of accretion shocks.
Our eDisk observations have unveiled that the Class 0 protostar in CB 68 has a Keplerian rotating disk with flaring and non-axisymmetric structure associated with accretion streamers and outflows.
%The SO, SiO, DCN, and
%H$_2$CO lines also trace such streamers and the landing point of the streamer onto the disk,
%while CH$_{3}$OH lines the disk rotation.
%Our eDisk observations of IRAS 16544 have revealed that the Class 0 protostar with active　molecular outflow and accretion possesses a well-developed Keplerian disk with flaring and　internal substructure.
%These results suggest that planet formation should be initiated at the early protostellar stage associated with mass accretion from the surrounding material.
\end{abstract}


\section{Introduction} \label{sec:intro}
%There are now growing pieces of observational evidence that planet formation should be initiated at the protostellar stage.
There is now growing observational evidence that planet formation is initiated in circumstellar disks around protostars in their deeply embedded (Class 0/I) stages.
%Concentric ring and gap features are ubiquitously seen in protoplanetary disks around Class II sources \citep{2013Fukagawa,2015ALMA,2016Pinte,2018Andrews,2018Long2}.
%Concentric ring and gap features \textcolor{red}{in dust continuum emission are thought to be related to the formation of planetesimals ubiquitously seen in protoplanetary disks around Class II young stellar objects (YSOs) in dusy continuum emission} 
%In the gap of the protoplanetary disk in PDS 70, the 1.3-mm continuum emission associated with a protoplanet has been found \citep{2019Isella,2021Benisty}.
%This is clear evidence for presence of a planet in the gap of the disk.
%although the time-variable nature of the 1.3-mm continuum emission suggests that
%the 1.3-mm continuum emission does not trace a circumplanetary disk but a free-free emission associated with the mass accretion \citep{2022Casassus}.
%Furthermore, the mass of the dust disks around Class II YSOs is low in comparison with solid mass of exoplanets systems, which suggests that planet formation should happen earlier than the Class II stage \citep{2018tychoniec,2018wardduong,2020Tychoniec}.
For example, the mass of the disks around Class II young stellar objects (YSOs), so-called protoplanetary disks, is low compared to the total mass of solids in exoplanet systems, which suggests that planet formation should happen earlier than the Class II stage \citep{2018Tychoniec,2018Wardduong,2020Tychoniec}.
The concentric ring and gap features found in dust disks around Class II YSOs are thought be related to planet formation processes; $e.g.,$ the gaps reflect the orbits of protoplanets and rings are possible growth sites of planetesimals \citep{2015ALMA,2016Pinte,2018Andrews,2018Long2}.
Consequently, it is of great interest to characterize the structures of disks around deeply embedded protostars, and to investigate whether they also show ring-like structures as those seen around Class II YSOs - or perhaps precursors to these.
%and all the inconsistencies to explain different architectures of planetary systems,

%These results imply that planet formation should be initiated early than the Class II stage,
%$i.e.,$ Class 0 and I protostellar stages.

%Our ALMA Large Program, Early Planet Formation in Embedded Disks (eDisk), is designed to perform a systematic study for signs for signs of planet formation in Class 0 and I protostars in the 1.3-mm dust-continuum emission and three CO (2--1) isotopologue lines, and the other Band 6 lines at a spatial resolution of $\sim$5 au.
The purpose of the Atacama Large Millimeter/submillimeter Array (ALMA) Large Program ``Early Planet Formation in Embedded Disks (eDisk)'' is to systematically characterize the structures of disks around a sample of 19 protostars in Class 0/I stages through high spatial resolution images of 1.3 mm continuum and line emission  \citep{2023Ohashi}.
Specific interests of the target sources of eDisk include: (1) the presence or absence of a Keplerian rotating disk, as well as its size and internal structure \citep{2012Tobin,2015Tobin,2020Tobin,2022Sheehan}; (2) spatial and velocity structure of the protostellar envelope, including the accretion streamers onto the disk, which are found in recent ALMA studies \citep{2014Yen,2019Yen,2020Pineda,2022Thieme}; (3) evolutionary stage of the sources in the context of planet formation.
%in comparison with the other eDisk targets \citep{2023Ohashi} and Class II sources \citep{2016Andrews,2018Andrews,2020Andrews,2021Andrews,2018HuangII,2018HuangIII,2018HuangTW,2020Huang,2017Long,2018Long2,2018Long1,2019Long,2022Long}.
%The details of the program can be found in \citet{2023Ohashi}.
%As one of the series of the eDisk first-look papers, we report the results of the Class 0 protostar CB 68.
In this paper we present the eDisk results for the Class 0 protostar IRAS 16544-1604 embedded in the Bok globule CB 68, located in the southwestern outskirt of the Ophiuchus North region \citep{1997Launhardt,1998Launhardt,2010Launhardt}.
In our eDisk papers, we simply refer to the Class 0 protostar IRAS 16544-1604 as IRAS 16544.
%IRAS 16544 is \textcolor{red}{one of the Class 0 sources} among the eDisk sample with a bolometric temperature $T_{bol}$ = 50 K and bolometric luminosity $L_{bol}$ = 0.89 $L_{\odot}$, which are re-estimated using the archival photometric data by the eDisk team.
IRAS 16544, one of the Class 0 sources among the eDisk sample, has a bolometric temperature $T_{bol} =$ 50 K and bolometric luminosity $L_{bol} =$ 0.89 $L_{\odot}$, which are re-estimated using the archival photometric data by the eDisk team \citep[see][for details]{2023Ohashi}.
Distance estimates to the $\rho$~Ophiuchus region range from 160 pc \citep{1981Chini} to 119$\pm$6 pc \citep{2008Lombardi}; the latter based on Hipparcos and Tycho parallax measurements.
A recent VLBA parallax measurement yielded a distance of 137.3$\pm$1.2 pc \citep{2017Ortiz}. The distance to the cloud region closest to IRAS 16544 in Ophiuchus North was reported to be $151^{+3}_{-5}$ pc by \citet{2020Zucker}, who adopted an advanced method that combines stellar photometric data with parallax measurements. In this paper we adopt this distance of 151 pc determined within a $\sim$5 \% accuracy.
IRAS 16544 drives
%collimated powerful molecular
an expansive bipolar outflow along the northwest to southeast direction at a position angle of $\sim$140$\degr$, where the northwestern outflow lobe is redshifted and southeastern lobe blueshifted \citep{1996Wu,2000Vallee,2007Vallee}.
%\textcolor{red}{This source} drives collimated powerful molecular outflows along the northwest to southeast direction at a position angle of $\sim$140$\degr$, where the northwestern outflow lobe is redshifted and southeastern lobe blueshifted \citep{1996Wu,2000Vallee,2007Vallee}.
%Observational signatures of a rotating flattened protostellar envelope associated with IRAS 16544 have been found by \citet{2007Vallee}.
Recently, \citet{2022Imai} reported results for IRAS 16544 at a spatial resolution of $\sim$70 au from the ALMA Large Program ``Fifty AU Study of the chemistry in the disk/envelope system of Solar-like protostars (FAUST)".
%They have revealed compact ($\lesssim$60 au), almost unresolved 1.3-mm dust-continuum emission, and a 1000-au scale infalling and rotating protostellar envelope in the C$^{18}$O (2--1), $c$-C$_3$H$_2$ (6$_{0,6}$--5$_{1,5}$; 6$_{1,6}$--5$_{0,5}$), and CCH ($N$=3--2, $J$=7/2--5/2) emission. 
%Toward the continuum emission they have identified a gas velocity gradient along the northeast (blueshifted) to southwest (redshifted) direction in the C$^{18}$O (2--1), OCS (19--18), and CH$_3$OH (4$_{2}$--3$_{1}$ $E$) lines.
%The central continuum source is also associated with complex organic molecular lines such as higher-excitation CH$_3$OH (16$_{2}$--15$_{3}$ $E$, etc) lines and HCOOCH$_3$ (20$_{11,9}$--19$_{11,8}$ $E$) and CH$_3$OCH$_3$ (14$_{1,14}$--13$_{0,13}$) lines, suggesting CB 68 harbours a hot corino \citep{2022Imai}.
They revealed the presence of 
%a flattened envelope of $\sim$1000 au scale 
a $\sim$1000-au scale infalling and rotating protostellar envelope by C$^{18}$O and the existence of a velocity gradient on a smaller scale ($\sim$100 au), potentially related to disk rotation, by C$^{18}$O, CH$_3$OH, and OCS.
Compact emission of various complex organic molecules are also found, suggesting that IRAS 16544 harbors a hot corino.

We here report the $\sim$5 au resolution observations for IRAS 16544 from eDisk revealing the presence of a protostellar disk in Keplerian rotation as well as accretion streamers from the larger scale cloud to this disk for the first time. 
The paper is organized as follows:
Section \ref{sec:data} describes eDisk observations of IRAS 16544, and parameters of the imaging.
Section \ref{subsec:continuum} shows the results of the 1.3-mm dust-continuum emission.
Images and velocity structures of the $^{12}$CO (2--1) emission are shown in Section \ref{subsec:CO-outflow}, while those of the $^{13}$CO (2--1), C$^{18}$O (2--1)
%and SO (6$_5$--5$_4$)
emission are in Section \ref{subsec:c18o}.
%These results are discussed with our analytical tools in Sections \ref{subsec:discdisc} and \ref{subsec:streamer}, and then we discuss the evolutionary stage of CB 68 in the context of planet formation in Section \ref{subsec:planet}.
In section \ref{subsec:discdisc} and \ref{subsec:streamer}, we analyze the Keplerian disk and streamers in the envelope, respectively. We further discuss the nature of the streamers in
section \ref{subsec:discussion_streamer}, and
an overall physical picture of the Class 0 protostellar system IRAS 16544 in section \ref{subsec:planet}.
Section \ref{sec:summary} summarizes our main results.
Appendix \ref{sec:appendixlines} show the moment maps of the SO, $c$-C$_3$H$_2$, and H$_2$CO lines and the channel maps of the SiO, CH$_3$OH, and DCN lines.
The full set of the velocity channel maps of the CO isotopologue lines
is presented in Appendix \ref{sec:appendixchannelmap}.

\begin{deluxetable*}{lcccc}
\tablewidth{\textwidth}
\caption{Observational Parameters}
\label{table:obs}
\tablehead{
\colhead{Project code} & \multicolumn{3}{c}{2019.1.00261.L} & \colhead{2019.A.00034.S}
}
\startdata
Observing date & 24 Aug. 2021 & 04 Oct. 2021 & 05 Oct. 2021 & 14 Jun. 2022\\
Number of antennas & 50 & 45 & 46 & 42\\
Configuration & \multicolumn{3}{c}{C-8} & C-4\\
Phase center & \multicolumn{4}{c}{( 16$^h$57$^m$19$\fs$64, $-$16$^d$09$^m$23$\fs$9) (ICRS)}\\ 
Synthesized beam (continuum, robust$=$0.0) & \multicolumn{4}{c}{0$\farcs$036 $\times$ 0$\farcs$027 (P.A.$=$69$\degr$)}\\%restoring beam
Conversion factor (continuum, robust$=$0.0) & \multicolumn{4}{c}{1 Jy beam$^{-1}$$=$24836 $\rm{K}$}\\
Noise level (continuum, robust$=$0.0) & \multicolumn{4}{c}{21 $\mu$Jy beam$^{-1}$$=$0.5 $\rm{K}$}
\enddata
%\tablenotetext{a}{Using the Rayleigh-Jeans approximately.}
\end{deluxetable*}

\section{ALMA Observations and Data Reduction} \label{sec:data}

The IRAS 16544 data presented in this paper were taken from two observing projects;
one from the ALMA large program eDisk (project code: 2019.1.00261.L, PI: N. Ohashi) and the other from the Director's Discretionary Time (DDT) program (project code: 2019.A.00034.S, PI: J. J. Tobin). 
The eDisk observations were conducted to achieve high-resolution ($<$0$\farcs$1) imaging.
The DDT observation was made to supplement the eDisk observations with  a more compact antenna configuration C-4, which is sensitive to extended structures of $\sim$5$\arcsec$.
% sensitivity to more extended ({more like $\sim$5$\arcsec$}) structures.
Details of these observations are summarized in Table \ref{table:obs}.
Hereafter, we call the eDisk observations/data LB (long baseline) observations/data, and the DDT observation/data SB (short baseline) observation/data.
%\textcolor{red}{Since the minimum projected baseline length of the SB data is 11.5$k\lambda$, our observations of CB 68 are insensitive to structures larger than 4$\farcs$9 \citep{1994Wilne}.%15m
The correlator setup included seven spectral windows (spws).
Two spws have a bandwidth of 1.875 GHz to increase the continuum sensitivity, and four spws have a bandwidth of 58.594 MHz to ensure a high enough velocity resolution of the main target lines, and the other one has a 937.5 MHz bandwidth for the $^{12}$CO (2--1) line.
%The correlator settings in both the LB and SB observations include 1.3-mm continuum,
%CO (2--1) isotopologue lines, SO (6$_5$--5$_4$), SiO (5--4), and other Band 6 lines,
Between the LB and SB observations a slight change of the correlator setting was made for the $^{13}$CO (2--1) emission.
%, which results in a slight change of the frequency coverage of the continuum data.}
Table \ref{table:CB68_molecular_line} lists the detected molecular lines, along with angular and velocity resolutions and rms noise levels.
%Other details of the observations, such as calibrators, weather conditions, as well as the spectroscopic data, are summarized in \citet{2023Ohashi}.
% and their observational parameters.

% SB-only and SB+LB reduction
%The delivered visibility data after the standard pipeline calibrations were further edited and self-calibrated with a unique version of CASA tailored to the eDisk project, which is based on the CASA version 6.2.1-7.
The delivered visibility data after the standard pipeline calibrations were further edited and self-calibrated with the Common Astronomy Software Applications (CASA) version 6.2.1.
The continuum-only visibility data were extracted from the channels without spectral emission lines detected.
The SB continuum visibility data was Fourier-transformed and CLEANed with the CASA task $tclean$, and the SB-only continuum image was made.
Then, self-calibrations of the SB continuum data only were conducted.
The phase-only self-calibration was first conducted with progressively shorter solution intervals, which improved the S/N until the solution interval of ``int". The total number of the iteration was six.
Then a single amplitude and phase self-calibration was conducted.
%The phase-only self-calibration was followed by a single amplitude and phase self-calibration.}
%The iterations of the phase-only self-calibration were performed with 
%solint=['inf', 'inf', '90.72s', '42.34s', '18.14s', 'int'] and combine=['spw,scan', 'spw', 'spw', 'spw', 'spw', 'spw'], respectively, followed by the amplitude and phase self-calibration with solint=['inf'] and selfcalmode=['ap'].
Next, from the self-calibrated SB visibility data and the three LB visibility data individual 1.3-mm continuum images were made.
Through the 2-dimensional Gaussian fitting to the images, the centroid positions of the continuum images were derived, and the alignment of the image center was applied to the visibilities.
%to correct for 
% the relative amplitude scales.
%the image center was aligned to investigate the relative intensity scales of the visibilities
%and to scale the visibility amplitudes.
%with a common central coordinates of ($\alpha_{ICRS}$, $\delta_{ICRS}$) =(16$^{h}$57$^{m}$19.642$^{s}$, -16${\degr}$09$^{m}$24.012$^{s}$) and applied to all the visibility data \textcolor{red}{for checking the relative flux scale}.
%The visibility amplitudes as a function of the $uv$ distance were also checked and the amplitudes are consistent within $\sim$10$\%$ among the different execution blocks (NEED TO CHECK), which is consistent with the uncertainty of the absolute flux calibrations of ALMA observations at Band 6.
Then, phase-only self-calibrations of the combined SB$+$LB data were conducted.
We found that while the 1st and 2nd round phase-only self-calibration improved the S/N, the third round of the phase-only calibration at the intermediate solution interval worsened the S/N. 
%Amplitude and phase self-calibration was not performed for the SB + LB data, because the improvement of the signal-to-noise ratio was already saturated in the phase-only self-calibration.
Amplitude and phase self-calibration was thus not performed for the SB$+$LB data.
%because \textbf{the first and second phase-only self-calibrations improved S/N, but the third lowered S/N.
%Therefore, we skipped phase and amplitude calibration.}
%followed by a single amplitude and phase self-calibration.
%with solint=['inf', 'inf', '18.14s'] and combine=['spw,scan', 'spw', 'spw'].
% followed by the amplitude and phase self-calibration with solint=['18.14s'] and combine=['spw'].
The calibration tables and the image centering were applied to the line visibility data.
A more comprehensive description of the data reduction process can be found in \citet{2023Ohashi}, where the standard eDisk data reduction script for each source is also linked.

With the self-calibrated SB+LB datasets, the final 1.3-mm continuum image was made with the Briggs robust parameter of 0.0, which gives the best compromise between the spatial resolution and signal-to-noise ratio (Figure \ref{fig_cont}).
For the line imaging robust$=$2.0, $i.e.$, Natural weighting, and $uv$-tapering at 2000 $k\lambda$  are adopted, except for the LB-only P-V diagram shown in Figure \ref{fig_C18O_pv}$b$ which adopts robust$=$1.5 and $uv$-tapering of 2000 $k\lambda$. We also made line images with robust$=$0.5 and the same $uv$-tapering.
We found, however, that those higher-resolution line images show patchy features presumably due to the more severe effect of the missing flux. 
We thus adopted robust$=$2.0 and $uv$-tapering at 2000 $k\lambda$ for the line imaging.
%various Briggs parameters were adopted to find the best compromise between the spatial resolution and sensitivity, depending on the intensities and structures of interest in that line.

%CB 68 was observed by the ALMA large project eDisk
%(project code:2019.1.00261.L, PI:N.Ohashi) and the dedicated Director's Discretionary Time (DDT) program (project code:2019.A.00034.S, PI:J. Tobin).
%The former was observed using the long baseline and the latter using the short baseline, three times and once with the Band 6, respectively. 
%These maximum projected baseline lengths are equivalent to an FWHM of $\sim$ ?$\falcs$? and $\sim$ ?$\falcs$?.
%For more information about the observation setting, observing dates, configuration, baselines, number of antennas, elevation, precipitable water varpor (PWV), see \cite{Ohashi22}  

%The results of the adopted the receiver, 1.3-mm dust continuum and 13 molecular lines listed in Table \ref{table:CB68_molecular_line} were detected.
%These data was performed by calibrating raw visibility data through pipeline, then self-calibration to improve signal-to-noise-ratio , and Fourier-transforming by Common Astronomy Software Applications (CASA, version 5.?.?). 
%All subsequent continuum and molecular line images were produced using CLEAN algorithm, adopting different robust parameters for each emission.
%For the continuum imaging, as a result of changing  various robust parameters to explore substructures in the protostar disks, which is one of our team science goals, we confirmed that in this paper, the robust=? image for the discussion. 

\begin{deluxetable*}{ccccccccc}
%\centering
\caption{Observed Molecular Lines}
\label{table:CB68_molecular_line}
\tablehead{
    \colhead{Molecule} & \colhead{Transition} & \colhead{Frequency} & \colhead{$E_u$$^{a,b}$} & \colhead{$A^{b,c}$} & \colhead{$n_{crit}$$^{b,d}$} & \colhead{Beam Size} & \colhead{$\Delta$$v^e$} & \colhead{rms}\\
    \colhead{} &\colhead{} &\colhead{(GHz)} &\colhead{(K)} &\colhead{(s$^{-1}$)} &\colhead{(cm$^{-3}$)} &\colhead{} & \colhead{(km s$^{-1}$)} & \colhead{(mJy beam$^{-1}$)}
}
\startdata
    $^{12}$CO & 2--1 & 230.538 & 16.60 & 6.910 $\times$ 10$^{-7}$ & 1.1 $\times$ 10$^4$ & 0$\farcs$24 $\times$ 0$\farcs$18 (P.A.$=$79$\degr$) & 0.635 & 1.2\\
    $^{13}$CO & 2--1 & 220.399 & 15.87 & 6.038 $\times$ 10$^{-7}$ & 9.4 $\times$ 10$^3$ & 0$\farcs$24 $\times$ 0$\farcs$18 (P.A.$=$81$\degr$) & 0.167 & 2.5\\
    C$^{18}$O & 2--1 & 219.560 & 15.81 & 6.011 $\times$ 10$^{-7}$ & 9.3 $\times$ 10$^3$ & 0$\farcs$27 $\times$ 0$\farcs$20 (P.A.$=$77$\degr$)& 0.167 & 1.9\\
    SO & 6$_5$--5$_4$ & 219.995 & 35.0 & 1.335 $\times$ 10$^{-4}$ & 2.3 $\times$ 10$^6$ & 0$\farcs$27 $\times$ 0$\farcs$20 (P.A.$=$77$\degr$) & 0.167 & 2.4\\
    %not (high T) ver.
    SiO & 5--4 & 219.949 & 31.26 & 5.197 $\times$ 10$^{-4}$ & 2.4 $\times$ 10$^6$ & 0$\farcs$25 $\times$ 0$\farcs$19 (P.A.$=$79$\degr$) & 1.34 & 0.7\\
    %hcn.dat (265.88643390 GHz)
    DCN$^f$ & 3--2 & 217.239 & 20.85 & 4.57 $\times$ 10$^{-4}$ & 3.7 $\times$ 10$^7$ & 0$\farcs$25  $\times$ 0$\farcs$19 (P.A.$=$79$\degr$) & 1.34 & 0.7\\
    %e-CH3OH
    CH$_3$OH & 4$_2$--3$_1$ & 218.440 & 45.5 & 4.686 $\times$ 10$^{-5}$ & 2.0 $\times$ 10$^7$ & 0$\farcs$25  $\times$ 0$\farcs$19 (P.A.$=$79$\degr$) & 1.34 & 0.6\\
    %p-C3H2
    $c$-C$_3$H$_2$ & 6$_{0,6}$--5$_{1,5}$ & 217.822 & 38.61 & 5.393 $\times$ 10$^{-4}$ & 4.5 $\times$ 10$^7$ & 0$\farcs$25 $\times$ 0$\farcs$19 (P.A.$=$79$\degr$) & 1.34 & 0.7\\
            %o-C3H2
    & 5$_{1,4}$--4$_{2,3}$ & 217.940 & 35.42 & 4.023 $\times$ 10$^{-4}$ & 4.8 $\times$ 10$^7$ & 0$\farcs$25 $\times$ 0$\farcs$19 (P.A.$=$79$\degr$) & 1.34 & 0.6\\
             %p-C3H2
    & 5$_{2,4}$--4$_{1,3}$ & 218.160 & 35.42 & 4.039 $\times$ 10$^{-4}$ & 4.8 $\times$ 10$^7$ & 0$\farcs$25  $\times$ 0$\farcs$19 (P.A.$=$79$\degr$) & 1.34 & 0.6\\
    %ph2co-h2.dat         
    H$_2$CO & 3$_{0,3}$--2$_{0,2}$ & 218.222 & 21.0 & 2.818 $\times$ 10$^{-4}$ & 2.6 $\times$ 10$^6$ & 0$\farcs$25  $\times$ 0$\farcs$19 (P.A.$=$79$\degr$) & 1.34 & 0.6\\
          %ph2co-h2.dat  
    & 3$_{2,2}$--2$_{2,1}$ & 218.476 & 68.1 & 1.571 $\times$ 10$^{-4}$ & 3.0 $\times$ 10$^6$ & 0$\farcs$35  $\times$ 0$\farcs$27 (P.A.$=$78$\degr$) & 1.34 & 0.6\\
          %ph2co-h2.dat  
    & 3$_{2,1}$--2$_{2,0}$ & 218.760 & 68.1 & 1.577 $\times$ 10$^{-4}$ & 3.4 $\times$ 10$^6$& 0$\farcs$26  $\times$ 0$\farcs$20 (P.A.$=$77$\degr$) & 0.167 & 1.8\\
\enddata 
\tablecomments{$^a$Upper state energy of the line.\\
$^b$From the LAMDA database \citep{Schoier2005}.\\
$^c$Einstein $A$ coefficient.\\
$^d$Critical density at $T_K =$ 20 K, except for that of the SO and $c$-C$_3$H$_2$ lines which adopt 60 K and 30 K, respectively.\\ 
\hspace{4pt}Calculated from the Einstein $A$- and $C$-coefficients from the LAMDA database.\\
$^e$Velocity resolution.\\
$^f$$E_u$ and $A$ values are from the CDMS database \citep{2001Muller,2005Muller,2016Endres}.\\
\hspace{4pt}The $C$-coefficient of the HCN line is adopted.
}
\end{deluxetable*}

\section{RESULTS} \label{sec:results}
\subsection{1.3-mm Dust-Continuum Emission} \label{subsec:continuum}
%%%%%%%%%%%%
%final ver.%
%%%%%%%%%%%%
\begin{figure*}[t]
\centering
\includegraphics[width=180mm, angle=0]{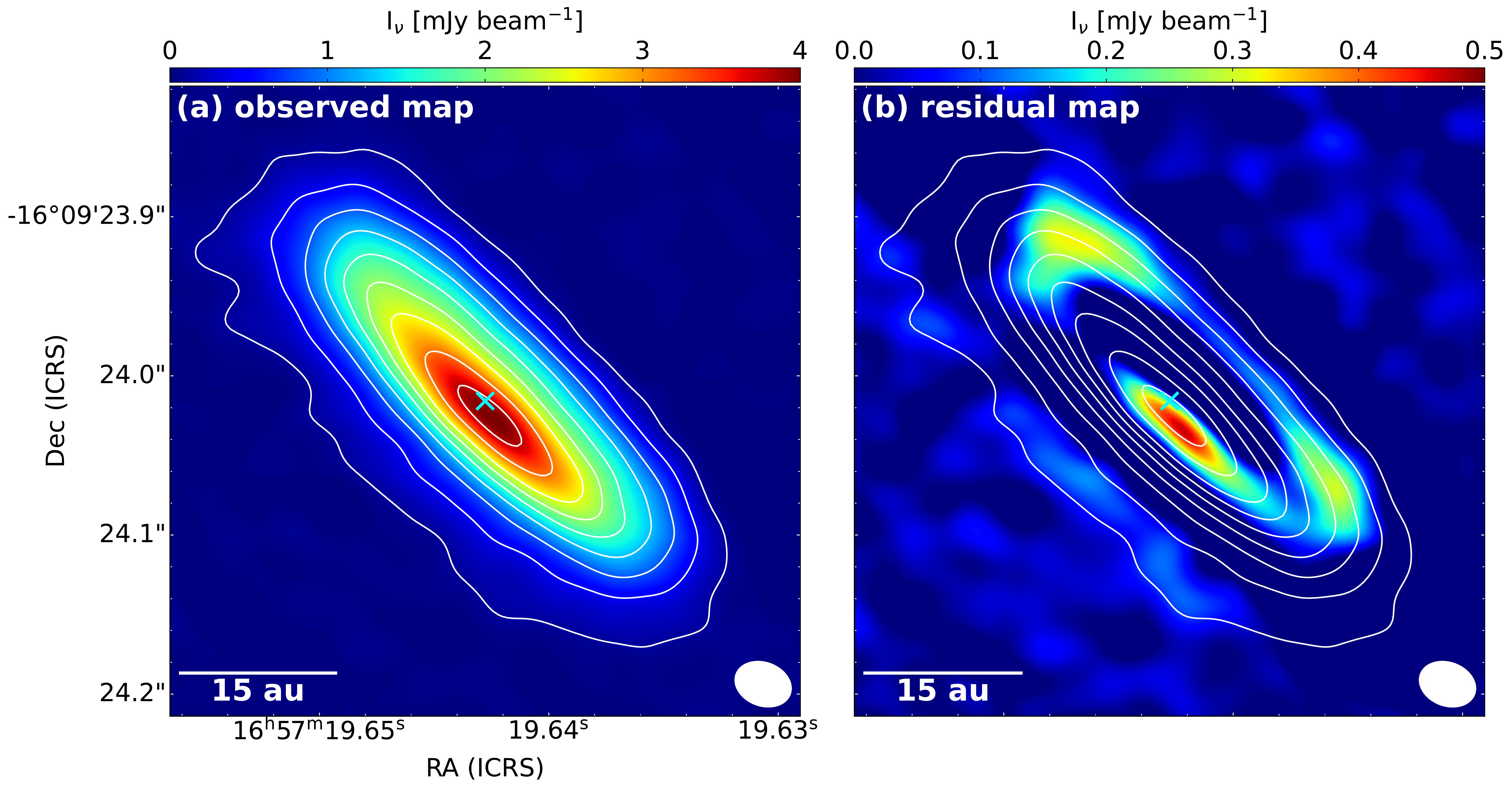}\\
\caption{a) 1.3-mm dust-continuum image of IRAS 16544 with a robust parameter 0.0.
Contour levels are 5$\sigma$, 20$\sigma$, 40$\sigma$, 60$\sigma$, 80$\sigma$, 100$\sigma$, 120$\sigma$, 150$\sigma$, and 180$\sigma$ (1$\sigma$ $=$ 21 $\mu$Jy beam$^{-1}$). A cross at the center denotes the centroid position of the 1.3-mm dust-continuum emission as derived from the two-dimensional
Gaussian fitting, which is regarded as the protostellar position.
A filled ellipse at the bottom-right corner shows the synthesized beam (0$\farcs$036 $\times$ 0$\farcs$027; P.A. $=$ 69$\degr$).
b) Residual 1.3-mm dust-continuum image after subtracting
the fitted two-dimensional Gaussian (color),
overlaid with the observed image in white contours. Contour levels are the same as those in a).}
\label{fig_cont}
\end{figure*}
%The size of the disk is $\sim$ 20 au at 120 pc and P.A. is 47$\degr$.
%The deconvolved major axes and minor as D$_{maj}$ and D$_{min}$, D$_{maj}$ $\times$ D$_{min}$ = 0$\farcs$28 $\times$ 0$\farcs$19. 
%The inclination angle calculated using these values is 69$\degr$.
%Considering that the synthesized beam size is 0$\farcs$20 $\times$ 0$\farcs$17, it is difficult to make clear the detailed structure in the disk from this image because the both sizes are almost the same.
Figure \ref{fig_cont}$a$ shows the 1.3-mm dust-continuum image of IRAS 16544 at an angular resolution of 0$\farcs$036 $\times$ 0$\farcs$027 (5.4 au $\times$ 4.1 au) (P.A. $=$ 69$\degr$).
The 1.3-mm dust emission is elongated along northeast to southwest.
The dust emission likely traces a protostellar disk close to edge-on.
Previous ALMA observations of IRAS 16544 in the dust-continuum emission could not resolve such an elongated structure in the vicinity of the protostar \citep{2022Imai}.
Our high-resolution eDisk observations have resolved the elongated disk feature around the protostar IRAS 16544 for the first time.
From a two-dimensional Gaussian fitting with the CASA task $imfit$, the deconvolved full width at half maximum (FWHM) of the 1.3-mm continuum emission along the major $\times$ minor axes is derived to be 0$\farcs$207 $\times$ 0$\farcs$060 (31.3 au $\times$ 9.1 au) at a position angle of 45$\degr$,
where the fitting region is set to include the entire emission extent down to 3$\sigma$.
%Using the derived deconvolved size, the inclination angle of the dust disk is
% estimated to be $\gtrsim$73$\degr$.
Assuming that the dust grains are settled to the midplane, the aspect ratio equals $\cos i$, where $i$ is the inclination angle ($i=0\degr$ means face-on), and $i$ is estimated to be $\gtrsim$73$\degr$.
Since the grains are not completely settled, this serves as a lower limit.
%The radius and the inclination angle of the dusty disk are thus $\sim$20 au and $\sim$73.24$\degr$, respectively.
The location of the emission centroid derived from the two-dimensional Gaussian fitting is ($\alpha_{ICRS}$, $\delta_{ICRS}$)$=$(16$^{h}$57$^{m}$19.6428$^{s}$, $-$16$^{d}$09$^{m}$24.016$^{s}$), which we regard as the position of the central protostar (crosses in Figure \ref{fig_cont}).
The centroid position of the Gaussian is slightly ($\sim$0$\farcs$01) offset from the position where the peak intensity of the map is seen. %$\sim$0$\farcs$075
This reflects that the 2D Gaussian structure does not completely represent the actual intensity distribution as also illustrated in the residual map (Figure \ref{fig_cont}$b$) after subtracting the fitted Gaussian.

From the 1.3-mm flux density of 50.63 mJy ($\equiv S_\nu$) as derived from the integration over the emission area above 3$\sigma$,
the dust mass of the disk ($\equiv M_{\rm dust}$) is calculated by the following formula,
%51.62 mJy(from Gaussian fitting)
\begin{equation}
M_{\rm dust}=\frac{S_{\nu}d^2}{\kappa_{\nu}B_{\nu}(T_{\rm dust})},
\label{eq2}
\end{equation}

\noindent where $\nu$ is the frequency ($=$ 225 GHz), $d$ is the distance ($=$ 151 pc), $\kappa_\nu$ is the dust mass opacity, and $B_\nu(T_{\rm dust})$ is the Planck function at dust temperature $T_{\rm dust}$.
The dust opacity of $\kappa_{\rm 225 GHz} =$ 2.3 cm$^2$ g$^{-1}$ is adopted \citep{Beckwith1990}.
Adopting a typical value of the dust temperature in Class II disks, $i.e.,$ $T_{\rm dust}$ $=$ 20 K \citep{2016Pascucci},
$M_{\rm dust}$ is estimated to be $\sim$ 34.1 $M_{\Earth}$.
%Since the peak brightness temperature of the 1.3-mm dust-continuum emission is as high as $\sim$100 K (see Figure \ref{fig_radial_profile}), we adopt $T_{\rm d}$ = 20 K, which yields $M_{\rm d}$ = 34.8 $M_{\Earth}$.
%We adopt two values of $T_{\rm d}$ = 20 K and $T_{\rm d}$ = 43 K
%$\times$ $(\frac {L_{bol}}{L_{sun}})^{0.25}$ \citep{2020Tobin},  which yields $M_{\rm d}$ = 1.0 $\times$ $10^{-4}$ $M_{\odot}$ and 1.0 $\times$ $10^{-4}$ $M_{\odot}$, respectively.
%Assuming the gas-to-dust mass ratio of 100, this dust mass yields the total gas + dust mass of 1.04 $\times$ 10$^{-2}$ $M_{\odot}$.
%$T_{\rm d}$ = 20 K, and T = 43 K $\times$ $(\frac {L_{bol}}{L_{sun}})^{0.25}$.
%On the assumption of the gas-to-dust volume ratio of 100, from $\kappa_{225 GHz}$ of gas is 0.023 cm$^2$ g$^{-1}$ by \cite{Beckwith1990}, a $\kappa_{225 GHz}$ of dust is 2.3 cm$^2$ g$^{-1}$.
%As for T$_d$, to compare the masses with Class $\rm I \hspace{-.01em}I $ disks, we adopted the temperature of the former, which one is typical for them.
%For reference, the latter temperature that luminosity depends on the protostar, was also used.
%Substituting the above values into (\ref{eq1}), the mass of the dust disk is estimated to be approximately 1.0 $\times$ $10^{-4}$ $M_{\odot}$.
\citet{2020Tobin} derived the scaling relation between the bolometric luminosity ($\equiv~L_{bol}$) and the dust temperature as;
\begin{equation}
T_{\rm dust} = 43 (L_{\rm bol}/L_{\odot})^{1/4} ~ {\rm K}.
\label{eq1}
\end{equation}
For IRAS 16544, $T_{\rm dust}$ is estimated to be 42 K with $L_{bol}$ $=$ 0.89 $L_{\odot}$.
This dust temperature gives $M_{\rm dust}\sim$ 14.1 $M_{\Earth}$.
On the other hand, the peak brightness temperature of the dust emission is as high as 90 K (see Figure \ref{fig_radial_profile}), suggesting that the dust temperature in the central region should be higher than 90 K.
$T_{\rm dust}$ $=$ 100 K gives the dust mass of $M_{\rm dust} \sim$5.4 $M_{\Earth}$.
Thus, with a fixed value of $\kappa_{\rm 225 GHz}$ $=$ 2.3 cm$^2$ g$^{-1}$, the range of the dust mass is $M_{\rm dust}\sim$5.4--34.1 $M_{\Earth}$.
Assuming the gas-to-dust mass ratio of 100, this dust mass yields the total gas + dust mass of 1.63$\times$10$^{-3}$--1.02$\times$10$^{-2}$ $M_{\odot}$.
Note that all of these estimates assume optically-thin 1.3-mm dust emission.
Since the high brightness temperature implies the optically thick 1.3-mm dust-continuum emission at least partially, these mass estimates should be regarded as lower limits.

%\textcolor{red}{Additionally, using a temperature that estimated based on the typically higher temperature one arrives at by scaling with luminosity \citep{2020Tobin} of 38 K derived from 
%\begin{equation}
%T=43(L_{bol}/L_{\odot})^{1/4},
%\label{eq1}
%\end{equation}
%where L$_{bol}$ is 0.89 L$_{\odot}$ \citep{2023Ohashi}, and the dust mass is estimated M$_{\rm %dust}$ $\sim$ 15.8 $M_{\Earth}$.
%Note that the estimated value is a lower limit due to the asymmetry along the minor axis is clear, suggesting that the dust continuum emission is optically thick and the disk has high inclination angle.}

\subsection{$^{12}$CO ($J$=2--1) Emission} \label{subsec:CO-outflow}
%%%%%%%%%%%%
%final ver.%
%%%%%%%%%%%%
\begin{figure*}
\centering
\includegraphics[width=180mm, angle=0]{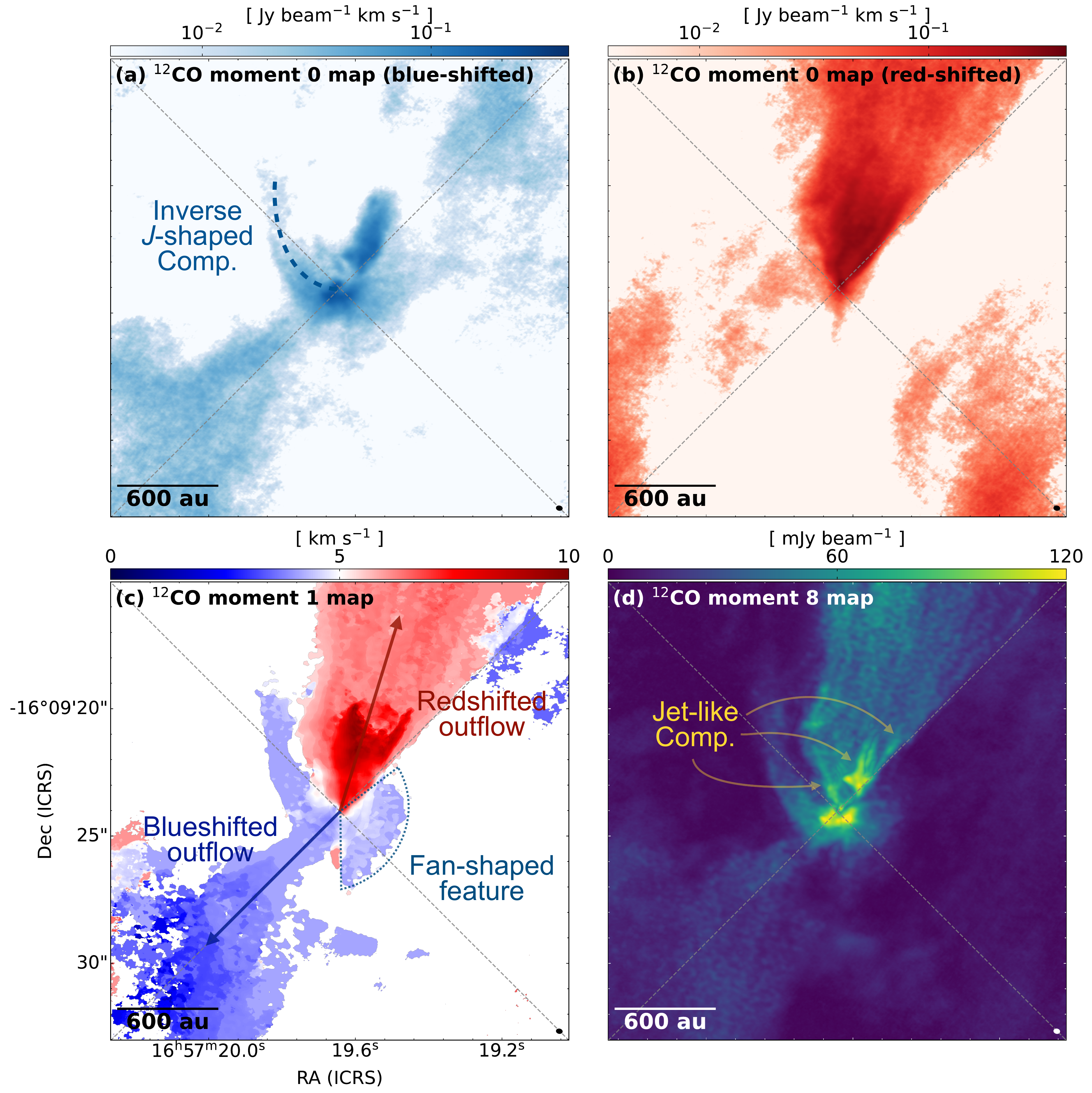}\\
\caption{Various maps of the $^{12}$CO (2--1) emission in IRAS 16544.
a), b) Moment 0 maps in the blueshifted ($V_{\rm LSR}=-$2.21--4.78 km s$^{-1}$) and redshifted ($V_{\rm LSR}=$ 5.41--23.19 km s$^{-1}$) velocity ranges. 
The color ranges from 4.0$\times$$10^{-3}$ to 4.0$\times$$10^{-1}$ Jy beam$^{-1}$ km s$^{-1}$.
The apparent redshifted components to the southeastern and southwestern ends are interferometric artifacts.
c) Intensity-weighted mean velocity map (Moment 1 map).
10$\sigma$ clipping is adopted to make this map (1$\sigma =$ 1.2 mJy beam$^{-1}$).
%\textcolor{red}{As in the case of panel b), the apparent components
%located to the image edges are interferometric artifacts.}
d) Map of the peak intensity in the spectra (Moment 8 map).
In all the panels, 
%contours denote the distribution of the 1.3-mm
%continuum emission at 10$\sigma$ (1$\sigma$ = 21 $\mu$Jy beam$^{-1}$).
the gray dashed lines show the major and minor axes of the dust-continuum image, and the filled ellipse at the bottom-right corner the synthesized beam. A blue dashed curve in panel a) delineates the inverse $J$-shaped feature.}
\label{fig_12CO}
\end{figure*}

%\begin{figure*}
%\centering
%\includegraphics[width=180mm, angle=0]{./12CO_pvdiagram.png}
%\caption{
%Position-Velocity diagrams of $^{12}$CO (2--1).
%}
%\label{fig_12CO_pv}
%\end{figure*}

Figure \ref{fig_12CO} shows various maps of the $^{12}$CO (2--1) emission in IRAS 16544.
From the analyses of the Position-Velocity diagrams of the C$^{18}$O emission with SLAM
(see Section \ref{subsec:discdisc}), the systemic velocity is derived to be 4.96 km s$^{-1}$, which is consistent with that in previous studies \citep{1997Codella,2022Imai}.
%5.2 km s-1: Wang1995, Codella&Muders 1997
Hereafter in this paper we use $V_{LSR}$ $=$ 5.0 km s$^{-1}$ as a systemic velocity of IRAS 16544.
%\textcolor{red}{From the symmetric center of the velocity features of the CO isotopologue emission, 
%The line profiles are made by spatial integration for each CO isotope, 
%$V_{LSR}$ = 5.0 km s$^{-1}$ is adopted as the systemic velocity.}
%Figure \ref{fig_12CO}a and \ref{fig_12CO}b compares integrated-intensity maps of the blueshifted and redshifted $^{12}$CO emission. 
%Overall, the blueshifted and redshifted $^{12}$CO emission are located toward the southeast and northwest of the protostar, respectively, and there is a clear bipolarity of the $^{12}$CO emission centered on the protostar.
%The moment 1 map of the $^{12}$CO emission (Figure \ref{fig_12CO}c) shows that the northwestern component is redshifted and the southeastern component blueshifted.
%These results indicate that the $^{12}$CO emission in CB 68 primarily traces the molecular outflow associated with the Class 0 protostar.
Figure \ref{fig_12CO}$a$ and \ref{fig_12CO}$b$ compare the integrated-intensity maps of the blueshifted and redshifted $^{12}$CO emission, while Figure \ref{fig_12CO}$c$ shows the moment 1 map. 
%The $^{12}$CO emission is predominantly distributed in a bipolar structure with the blueshifted emission is in the southeast, normal to the extended continuum structure, and the red-shifted emission to the northwest.
The $^{12}$CO emission is predominantly distributed in a bipolar structure, normal to the elongated continuum structure, with the blueshifted and the redshifted emissions located to the southeast and northwest, respectively.
This distribution suggests that the $^{12}$CO emission toward IRAS 16544 primarily traces the molecular outflow associated with the Class 0 protostar.
%There is also blueshifted and redshifted $^{12}$CO emission to the northwest and southeast of the protostar, respectively.
%These results imply that the outflow axis is close to the plane of the sky, consistent with the geometry of the dusty disk close to edge-on ($i \sim$73$\degr$).
There are also blueshifted emission toward the northwest and redshifted emission toward the southeast. 
It indicates that the outflow axis is close to the plane of the sky, which is consistent with the geometry of the dust disk close to edge-on ($i \gtrsim$73$\degr$).
%the moment 0 (Figure \ref{fig_12CO}a), 8 (Figure \ref{fig_12CO}b),
%and 1 maps (Figure \ref{fig_12CO}c,d) of the $^{12}$CO (2--1) emission
%in CB 68. These maps reveal spectacular outflow features associated with the Class 0 protostar
%along the northwest to southeast direction, over the entire extent of $\sim$? au.
%approximately perpendicular to the major axis of the dust disk (black contours in Figure \ref{fig_12CO}).
%Overall, the $^{12}$CO
%emission from the northwestern, redshifted outflow lobe is more intense than that from
%the southeastern, blueshifted outflow lobe.

%The redshifted outflow lobe to the northwest appears to consist of a number of needle-like emission components, which are most clearly visible in the peak intensity map (moment 8 map; Figure \ref{fig_12CO}d).
The redshifted outflow lobe to the northwest appears to consist of a number of jet-like emission components, which are most clearly visible in the peak intensity map (moment 8 map; Figure \ref{fig_12CO}$d$).
%These emission components point toward a variety of directions over the entire opening angle of \textcolor{red}{$\sim$62 $\degr$}, suggesting change of the outflow directions. %(red:~62, blue:~54)
%These emission components point toward a variety of directions over the entire opening angle of \textcolor{red}{$\sim$62$\degr$}, suggesting that the outflow is not orthogonal to the disk. %(red:~62, blue:~54)
These emission components point toward a variety of directions, suggesting that the outflow direction may have changed during the protostar's lifetime.
From the visual inspection of the moment 0 map (Figure \ref{fig_12CO}$b$), the mean position angle of the redshifted outflow lobe is derived to be $\sim$$-$17$\degr$.
On the other hand, from the investigation of the $^{12}$CO velocity channel maps we found that the $^{12}$CO map at $V_{LSR}$ $=$ 6.68 km s$^{-1}$ shows the $V$-shaped feature of the outflow cavity most unambiguously (Figure \ref{co_ch}).
%\footnote{ The full $^{12}$CO velocity channel maps for the entire and zoom-in regions are available in electronic journal.}).
From this map the opening angle of the $V$-shape is inferred to be $\sim$60$\degr$ for a spatial extent of $\sim$300 au at the outer intensity threshold of $\sim$50 $\sigma$, which corresponds the lowest contour level of the $V$-shaped cavity feature.
%From the investigation of the $^{12}$CO velocity channel maps, the opening angle of the redshifted outflow lobe is inferred to be $\sim$60$\degr$ at the intensity threshold of $\sim$50 $\sigma$ and over the spatial extent of \textbf{$\sim$300 au}, and the position angle is $\sim$$-$17$\degr$.
%\textbf{The opening angle is measured from the $^{12}$CO channel map at V$_{LSR}$ = 6.68 km s$^{-1}$, which shows the outflow cavity is most clearly visible. We then set the edge of the outflow at the $\sim$50 sigma counter.}
%Adopting the redshifted outflow lobe as shown in Figure \ref{fig_12CO}b and the intensity threshold of $\sim$50 $\sigma$, the opening angle is derived to be $\sim$63$\degr$.}
The mean central axis of the redshifted outflow lobe appears to be tilted with respect to the minor axis of the dust disk (dashed line from northwest to southeast in Figure \ref{fig_12CO}).
%The blueshifted outflow lobe in contrast does not clearly exhibit such jet-like components toward different directions.

The blueshifted outflow lobe in contrast does not clearly exhibit such jet-like components toward different directions. The mean position angle of the blueshifted outflow is $\sim$135$\degr$ from the visual inspection of Figure \ref{fig_12CO}$a$, which is not that of the redshifted lobe plus 180$\degr$.
% \textbf{The position angles of the blue- and redshifted outflow lobe are estimated from the $^{12}$CO moment 0 maps in Figure \ref{fig_12CO}$a$ and $b$.}
% This is partially due to the weaker blueshifted emission.
%and the redshifted outflow axis is more east toward the positive position angle than the disk minor axis.
This suggests the presence of the outflow bending, which is also seen in other protostellar sources \cite[$e.g.,$][]{2018Aso, 2019Aso}. 
If the magnetic-field structure in a protostellar system is not orthogonal to the disk rotational axis, misalignments between the outflow and disk rotational axes could be produced, and the observed bending outflow structure could reflect such a misalignment \cite[$e.g.,$][]{2004Matsumoto,2019Hirano,2020Hirano}.
% \textcolor{red}{One of the possible mechanisms to
% produce outflow bending is the binary orbital motion, although binarity of CB 68 has not been % identified to date.
%The detected non-axissymmetric structure of the dusty disk could be related to the binarity.}

% \textcolor{red}{adding the possible physical origin of the outflow bending}

In addition to these outflow components, the $^{12}$CO (2--1) emission also traces distinct components. There is an inverse $J$-shaped blueshifted component (dashed curve in Figure \ref{fig_12CO}$a$) starting from northeast from the protostar toward the southeastern end of the protostar.
There is another, fan-shaped blueshifted component to the west of the protostar.
Whereas the spatial distributions of these components appear to trace the shell surrounding the northern redshifted outflow, these components are blueshifted, and the velocity features of these components appear distinct from those of the outflows. In addition, these components have $^{13}$CO and C$^{18}$O counterparts. We will discuss the natures of these components in the subsequent sections.

% A zoom-in view of the moment 1 map of the $^{12}$CO emission also reveals that
%the $^{12}$CO emission traces the velocity structure in the dusty disk.
%In the dusty disk, the $^{12}$CO emission is blueshifted to the northeast and
%redshifted to the southwest, and thus the $^{12}$CO emission exhibits a velocity
%gradient along the disk major axis. Such a velocity gradient can be interpreted
%as a rotational motion in the disk, which will also be discussed in more detail
%in the C$^{18}$O (2--1) emission.

% \subsection{Spatial Structure of Infalling Envelope and Disk Rotation Around CB 68}
% \subsubsection{$^{13}$CO (J=2--1) Emission}
%Spatial and Velocity Structures of the 
\subsection{$^{13}$CO and C$^{18}$O (J=2--1) Emission} \label{subsec:c18o}

%%%%%%%%%%%%
%final ver.%
%%%%%%%%%%%%
\begin{figure*}
\centering
\includegraphics[width=180mm, angle=0]{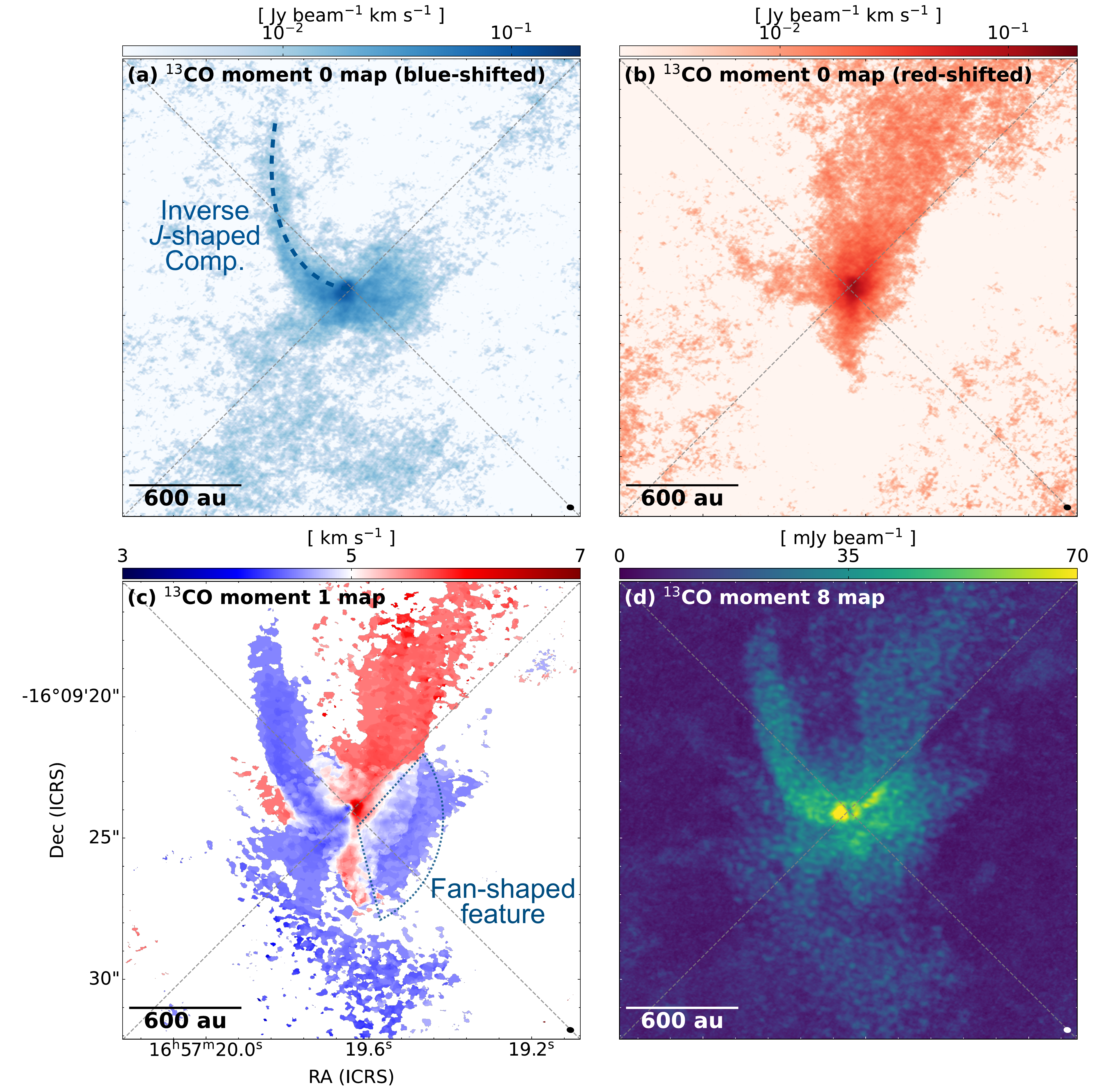}\\
\caption{Same as Figure \ref{fig_12CO} but for the $^{13}$CO (2--1) emission.
The integrated velocity ranges of the moment 0 maps of the blueshifted and redshifted emission are $V_{\rm LSR}=$ 0.18--4.85 km s$^{-1}$ and 5.19--9.36 km s$^{-1}$, respectively. The color ranges from 2.0$\times$$10^{-3}$ to 2.0$\times$$10^{-1}$ Jy beam$^{-1}$ km s$^{-1}$. 
5$\sigma$ clipping is adopted to make the moment 1 map (1$\sigma$ $=$ 2.5 mJy beam$^{-1}$).}
% \textcolor{teal}{The gray dashed lines are the the same as in Figure \ref{fig_cont}.}}
\label{fig_13CO}
\end{figure*}

%%%%%%%%%%%%
%final ver.%
%%%%%%%%%%%%
\begin{figure*}
\centering
\includegraphics[width=180mm, angle=0]{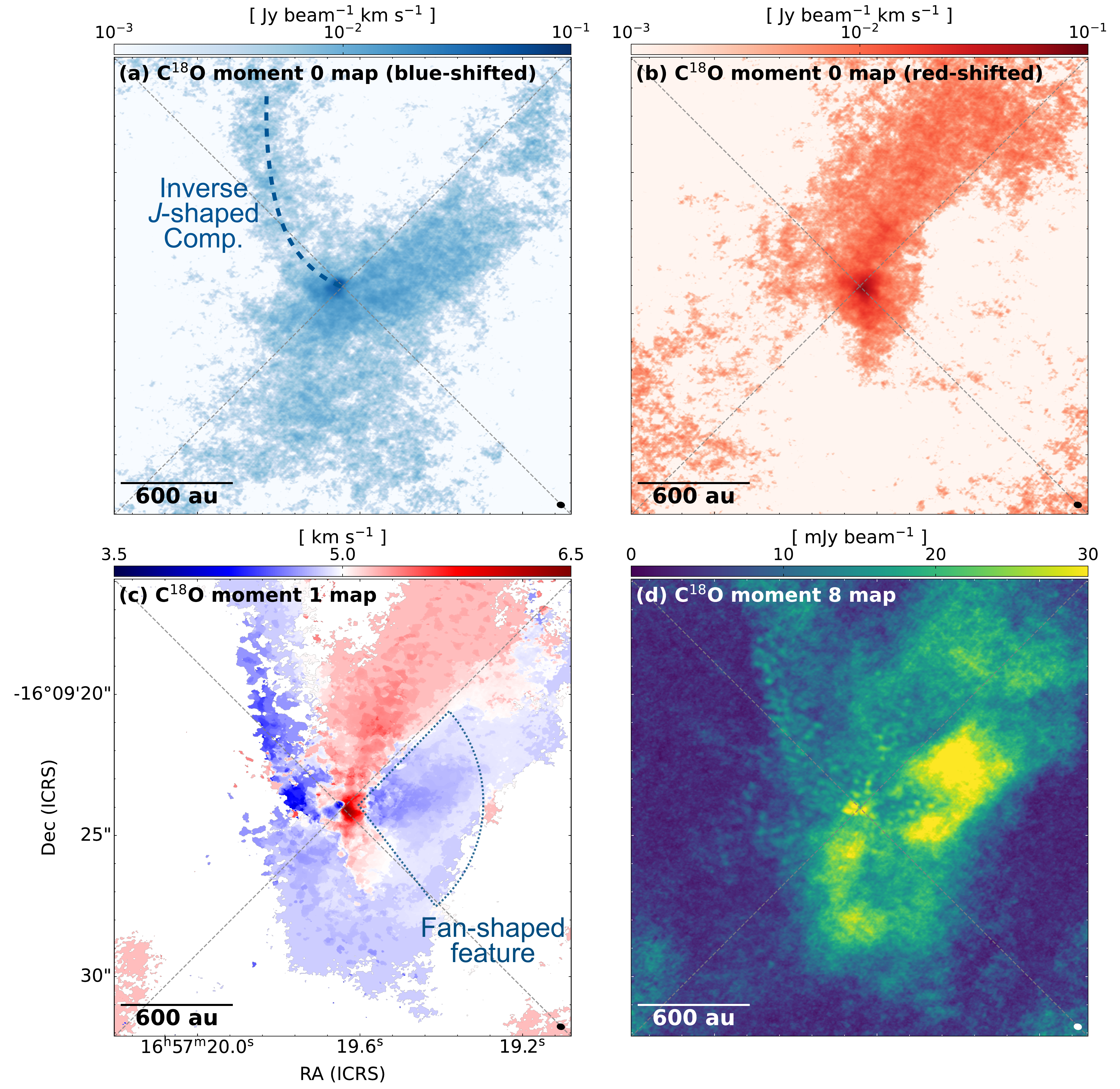}\\
\caption{Same as Figure \ref{fig_12CO} but for the C$^{18}$O (2--1) emission.
The integrated velocity ranges of the moment 0 maps of the blueshifted and redshifted emission are $V_{\rm LSR}=$ 1.18--4.85 km s$^{-1}$ and 5.19--8.86 km s$^{-1}$, respectively.
5$\sigma$ clipping is adopted to make the moment 1 map (1$\sigma =$ 1.9 mJy beam$^{-1}$).}
\label{fig_C18O}
\end{figure*}

%Figure \ref{fig_13CO} shows moment 0 (upper panels) and 1 maps (lower) of the $^{13}$CO (2--1) emission
%in the entire (left panels) and zoom-in views (right). The contours of the 1.3-mm dust-continuum emission is overlaid for comparison.
%\textcolor{red}{The velocity of the system doesn't seem to be given anywhere. This is important to specify.}
Figures \ref{fig_13CO} and \ref{fig_C18O} show the $^{13}$CO and C$^{18}$O counterparts of Figure \ref{fig_12CO}.
% In the vicinity ($\sim$300 au) of the central dusty disk,
%Both the blueshifted and redshifted \textcolor{red}{strong} $^{13}$CO  emission are present in the vicinity ($\sim$ 60 au) of the central dusty disk.
%The peaks of the blueshifted and redshifted emission are located to the southeast and northwest, respectively. 
While the extended blueshifted and redshifted $^{13}$CO and C$^{18}$O emission to the southeast and northwest, respectively, resemble the bipolar molecular outflow as traced by the $^{12}$CO emission, the velocity structures of these components are distinct from those of the $^{12}$CO outflows, as discussed below.
Note that the velocity range of the $^{13}$CO and C$^{18}$O emission is much narrower than that of the $^{12}$CO emission.
% which correspond to the bipolar molecular outflow as traced by the $^{12}$CO emission.}
Furthermore, there is %other, apparently distinct $^{13}$CO and C$^{18}$O components. 
a blueshifted, inverse $J$-shaped structure extending to the northeast (dashed curves in
Figures \ref{fig_13CO} and \ref{fig_C18O}).
The same component is also seen in the $^{12}$CO emission (Figure \ref{fig_12CO}).
There is another $^{13}$CO and C$^{18}$O emission component located to the west of the protostar, which exhibits a blueshifted, fan-shaped feature.
This component also has the $^{12}$CO counterpart.

\begin{figure*}
\centering
\includegraphics[width=170mm, angle=0]{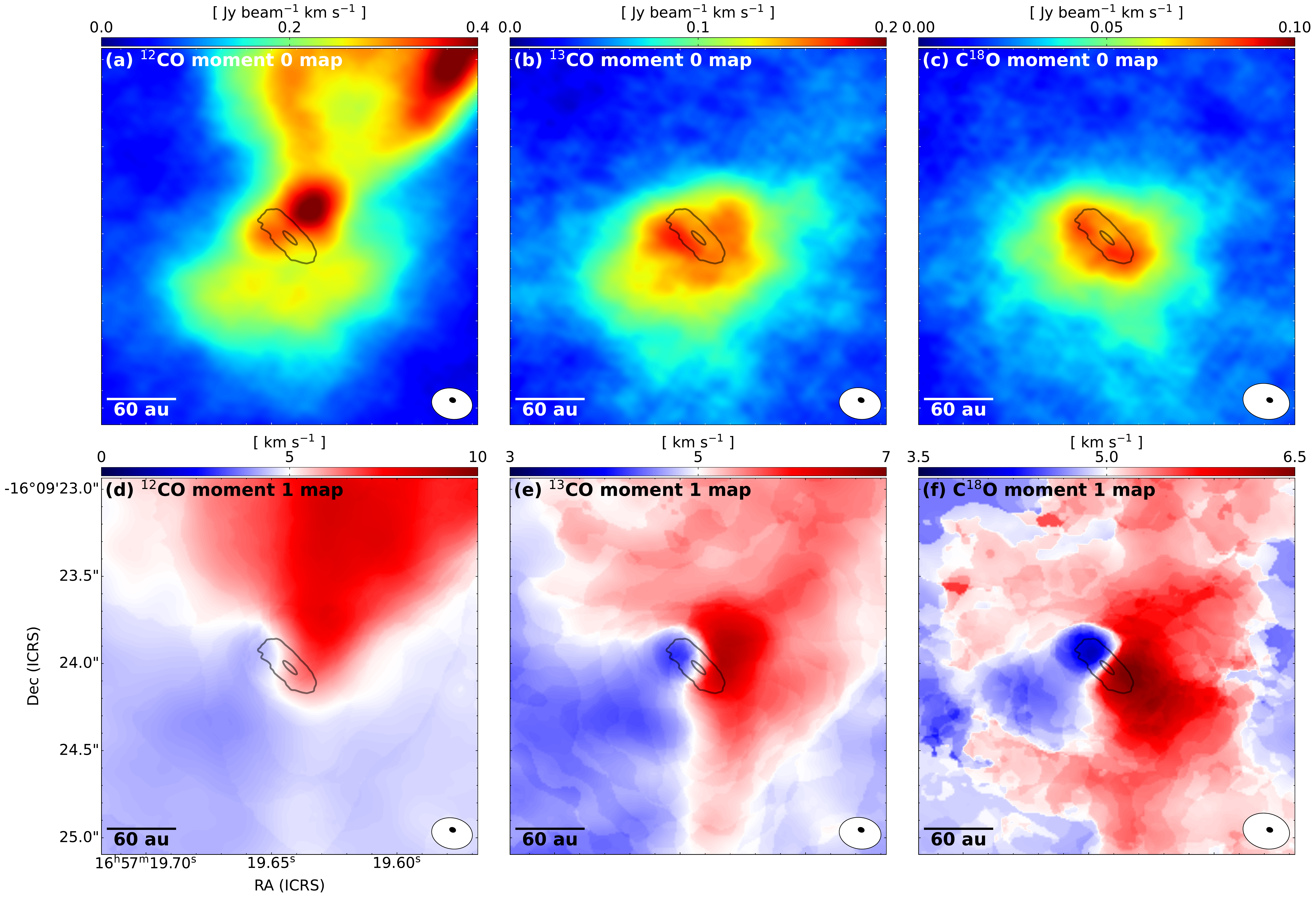}\\
\caption{Moment 0 and 1 maps of the $^{12}$CO, $^{13}$CO, and the C$^{18}$O (2--1) emission in the vicinity of the protostar as labeled.
The integrated velocity ranges of the moment 0 maps are $-$12.37--23.19 km s$^{-1}$, 0.01--10.03 km s$^{-1}$, and  1.01--9.36 km s$^{-1}$ for the $^{12}$CO, $^{13}$CO, and, the C$^{18}$O emission, respectively. 5$\sigma$ clipping is adopted to make moment 1 maps (1$\sigma =$ 1.2 mJy beam$^{-1}$, 2.5 mJy beam$^{-1}$, and 1.9 mJy beam$^{-1}$ for the $^{12}$CO, $^{13}$CO, and the C$^{18}$O maps, respectively).
Contours denote the 5$\sigma$ and 150$\sigma$ levels of the 1.3-mm continuum emission.
White and black filled ellipses at the bottom-right corners denote the synthesized beams of the line and 1.3-mm continuum images, respectively.
}
\label{fig_CO_iso_moom0}
\end{figure*}

The moment 0 and 1 maps of the CO isotopologue lines in the vicinity of the dust disk are shown in Figure \ref{fig_CO_iso_moom0}.
The $^{13}$CO and C$^{18}$O emission associated with the dust disk are visible, and in particular, the C$^{18}$O emission is elongated to the same direction as that of the dust emission (Figure \ref{fig_CO_iso_moom0}$b$ and $c$). 
This suggests that the C$^{18}$O emission traces the molecular gas in the protostellar disk.
On the other hand, the $^{13}$CO and particularly the $^{12}$CO emission show more extended structures along the outflow direction, which may suggest that the $^{13}$CO and $^{12}$CO emissions have more contamination from the outflows.
%The central disk component are clearly seen in the $^{13}CO$ and C$^{18}$O emission (Figure \ref{fig_CO_iso_moom0}(b) and (c)).
%In particular, C$^{18}$O emission shows a structure in the same direction as the major axis of the dust disk, while $^{12}$CO and $^{13}$CO components show more extended structures along the minor axis of the disk.}
Inside the dust disk these CO isotopologue emission are redshifted to the southwest and blueshifted to the northeast.
%As described in the previous subsection, 
This velocity gradient along the disk major axis can be interpreted as the rotational motion in the disk.
On the contrary, outside the dust disk the CO isotopologue emission is red- and blueshifted to the northwest and southeast, respectively.
In the $^{12}$CO outflow map, the northwestern side is redshifted and the southeastern side blueshifted.
Thus, the northwestern and southeastern sides of the disk correspond to the near- and far-sides of the disk plane, on the assumption that the disk plane is perpendicular to the outflow axis. If the $^{13}$CO and C$^{18}$O components originate from the disk plane, the redshifted and blueshifted emission arise from the near- and far-sides,
and such a velocity feature can be interpreted as an infalling motion.

%\textcolor{teal}{Moreover, the high-velocity $^{13}$CO and C$^{18}$O emission are located closer to the dusty disk and the low-velocity emission are further from the disk.}
%Thus, such a velocity feature can be interpreted as an infalling motion coplanar to the disk plane.}

%The zoom-in view of the C$^{18}$O emission reveals velocity structures
%consistent with those of the $^{13}$CO emission; Inside the dusty disk
%the C$^{18}$O emission also traces the disk rotation, and to the northwest
%and southeast of the dusty disk the C$^{18}$O emission are red- and blueshifted,
%respectively.

\begin{figure*}
\centering
\includegraphics[width=180mm, angle=0]{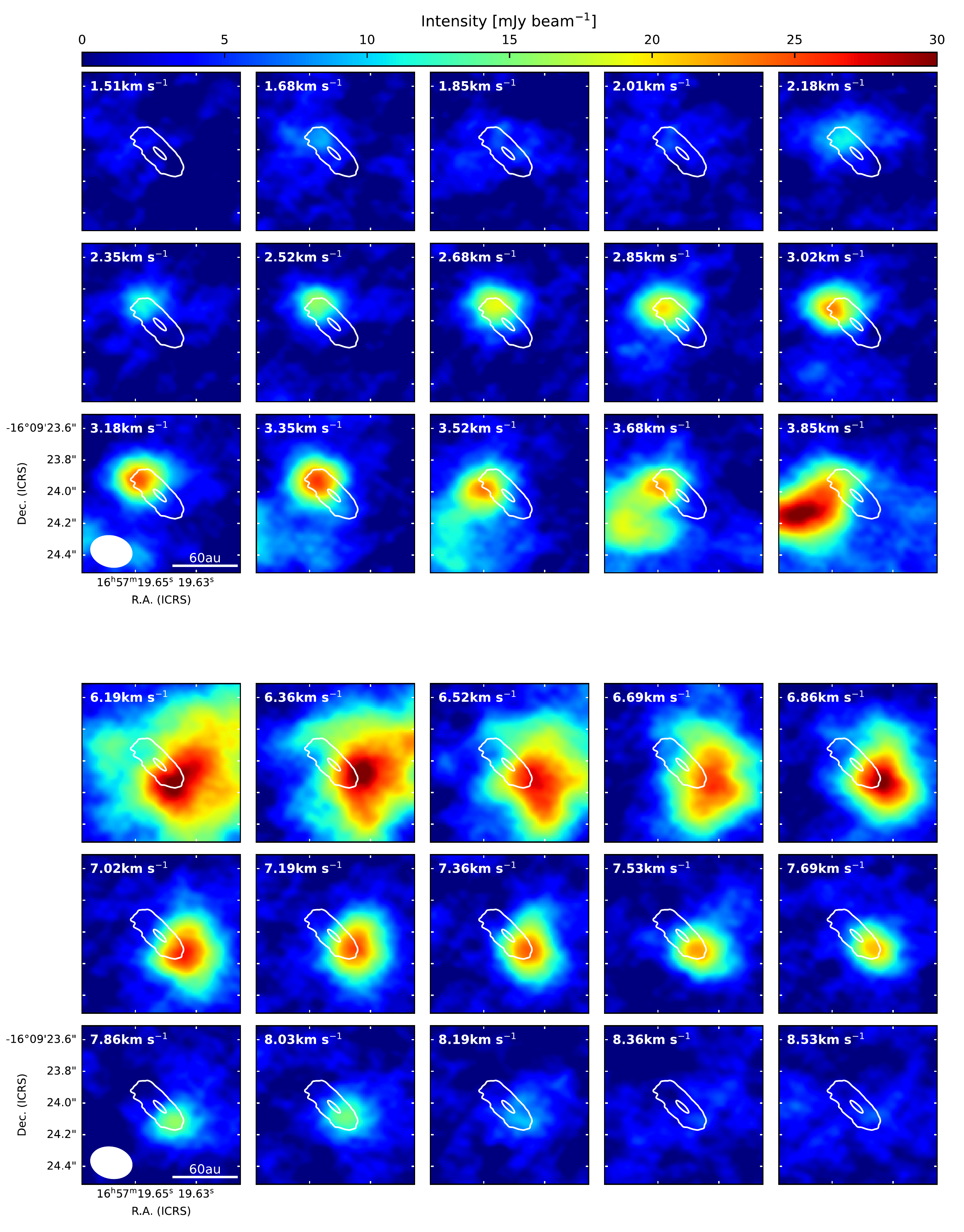}\\
\caption{Channel maps of the C$^{18}$O (2--1) emission in the high-velocity blueshifted (upper panels) and redshifted ranges (lower panels).
%Crosses show the position of the protostar, 
Numbers in the upper-left corners denote the LSR velocities.
The systemic velocity is 5.0 km s$^{-1}$.
Contours denote the distribution of the 1.3-mm continuum emission at 5$\sigma$ and 150$\sigma$ levels (1$\sigma =$ 21 $\mu$Jy beam$^{-1}$).}
\label{fig_C18O_high}
\end{figure*}

\begin{figure*}
\centering
\includegraphics[width=180mm, angle=0]{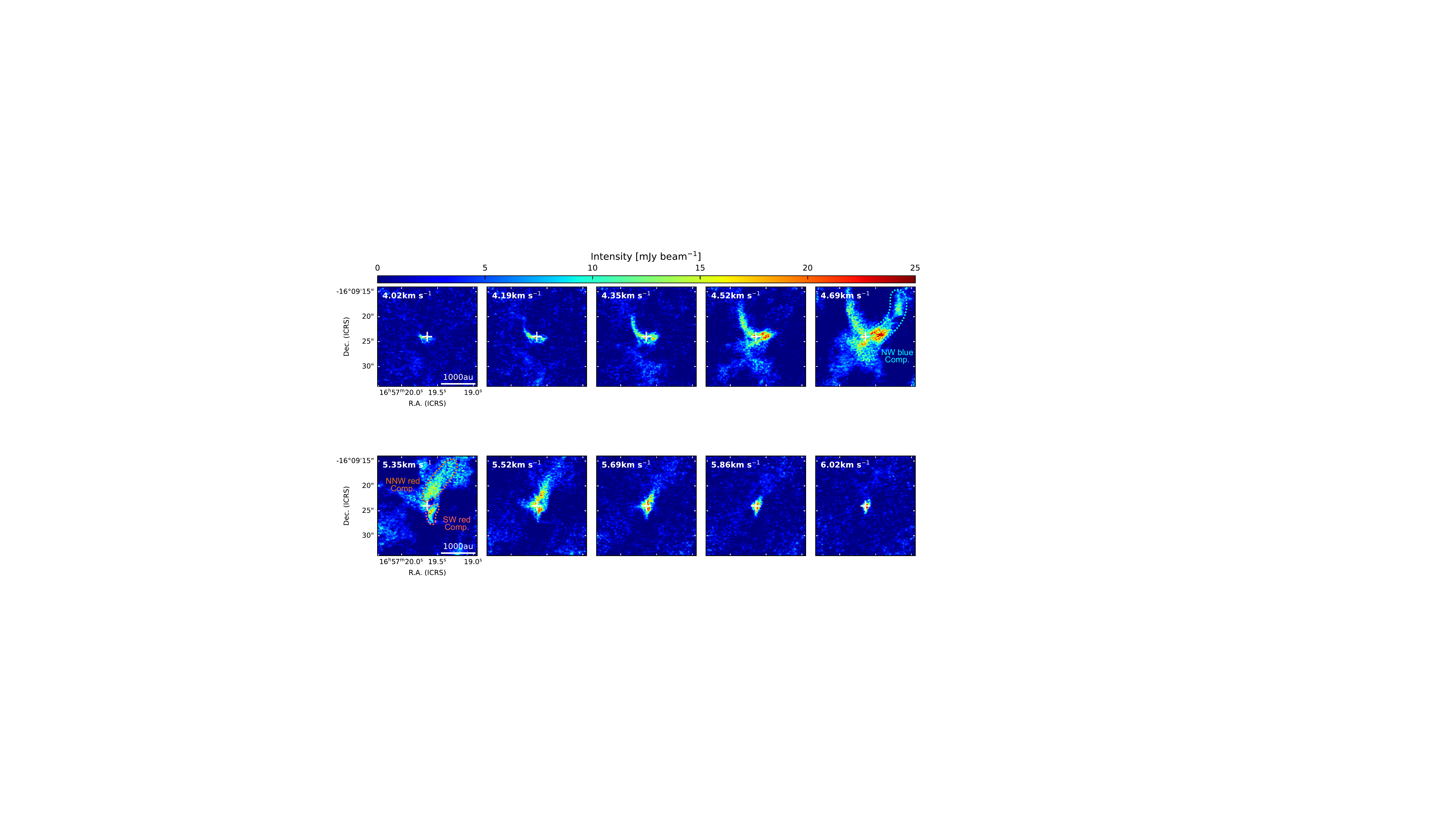}\\
\caption{Channel maps of the C$^{18}$O (2--1) emission in the low-velocity blueshifted (upper panels) and redshifted ranges (lower panels).
Crosses show the position of the protostar, and numbers in the upper-left corners denote the LSR velocities.}
\label{fig_C18O_low}
\end{figure*}

%To investigate the velocity features of the molecular gas surrounding the Class 0 protostar in more detail,
To investigate the velocity features of the molecular gas surrounding the protostar in more detail,
Figures \ref{fig_C18O_high} and \ref{fig_C18O_low} show the velocity channel maps of the C$^{18}$O emission in the high- and low-velocity ranges, respectively.
The full $^{13}$CO and C$^{18}$O velocity channel maps for the entire and zoom-in regions are Figure \ref{13co_ch}, \ref{c18o_ch}, \ref{13co_ch_high}, and \ref{c18o_ch_high}.
Note that the channel maps in the high-velocity range are presented in the close vicinity of the dust disk (white contours in Figure \ref{fig_C18O_high}). 
The high-velocity blueshifted C$^{18}$O emission is located to the northeastern part of the dust disk while the high-velocity redshifted emission to the southwest, suggesting the presence of the velocity gradient along the disk major axis and rotation in the disk.
%These spatial and velocity structures are consistent with Keplerian rotation of the disk. 
%\textcolor{red}{These spatial structures are consistent with rotation of the disk.}
In the blueshifted velocity range of 3.52--3.85 km s$^{-1}$, the location of the C$^{18}$O emission is shifted toward southeast, which connects to the emission component in the lower blueshifted velocities shown in Figure \ref{fig_C18O_low}.
Similarly, in the redshifted velocity range of 6.19--6.52 km s$^{-1}$, the C$^{18}$O emission protrusion
toward the west emerges, which connects to the lower redshifted emission in Figure \ref{fig_C18O_low}.

%The observed outermost radius of the C$^{18}$O emission is $\sim$0$\farcs$2 or $\sim 30$ au, which can be regarded as the radius of the disk rotation \textcolor{blue}{(see section  \ref{subsec:discdisc})}.
%Furthermore, the emission extents become larger toward the lower velocities. 

The C$^{18}$O velocity channel maps at lower velocity ranges over the entire emission area (Figure \ref{fig_C18O_low}) present gas components distinct from the central disk. 
%%%%%%%%%%%%%%%%%%%%%%%%%%%%%%%%%%%%%%%%%%%%%%%%%%%%%%%%%%%%
%At $V_{\rm LSR}$ = 4.85--4.92 km s$^{-1}$, 
At $V_{\rm LSR}$ $=$ 4.02 km s$^{-1}$, a blueshifted component to the southeast of the disk is seen.
From $V_{\rm LSR}$ $=$ 4.19 km s$^{-1}$, a blueshifted, inverse $J$-shaped structure emerges. 
This component curls from the southeast of the disk to the northeast, and extends to the northeast progressively until $V_{\rm LSR}$ $=$ 4.69 km s$^{-1}$.
This velocity feature, that is, higher velocity components located closer to the protostar and lower velocity components further, appears to be distinct from that of the outflows (see also Figure \ref{C18O_streamer}b).
%Hereafter this component with a systematic velocity structure is called as ``NE streamer". 
As we described above, this component is seen in all the CO isotopologue lines.
%and exhibits a higher velocity closer to the protostar (see also Figure \ref{C18O_streamer}b).}
%consistent with the picture of gas accretion.
%This component is also seen in the $^{13}$CO emission.
%but the spatial distribution
%of the 12CO emission in these velocity ranges is different from those of the
%The low-velocity of this blueshifted component is different from that of the blueshifted $^{12}$CO outflow.}
%This low-velocity blueshifted emission as seen in the C$^{18}$O and $^{13}$CO emission
%is unlikely to be the outflow origin.
In the velocity range of 4.19--4.69 km s$^{-1}$, there is another blueshifted component to the west of the protostellar disk.
This component is the origin of the fan-shaped blueshifted signature seen in Figure \ref{fig_C18O}.
This western component also appears to curl to northwest, and  presents a similar spatial-velocity feature to that of the northeastern blueshifted component, again opposite to that expected from the outflows.
In the velocity range of 4.35--4.69 km s$^{-1}$ the C$^{18}$O emission to the south is present. The extent of this blueshifted component is largest at the lowest velocity.

In the redshifted velocity of 6.02 km s$^{-1}$, the C$^{18}$O emission is located to the west of the protostar.
In the lower redshifted velocities (5.35--5.86 km s$^{-1}$), this component appears to be divided into two components; one to the southwest of the disk and the other extending to the north-northwest (NNW).
While the spatial location of the NNW component matches with that of the redshifted outflow lobe as seen in the $^{12}$CO emission, this component extends further to the NNW at lower velocities. %and vice versa.
%\textcolor{teal}{is compact at higher velocities.}
A similar velocity feature of the redshifted $^{13}$CO emission to the NNW is identified.
The sense of this velocity structure seen in the C$^{18}$O and $^{13}$CO emission is opposite to that of the Hubble flow of molecular outflows \citep{2007Arce}.
%It is possible that the northwestern component also traces another accretion streamer.

%The NW streamer, starting from $V_{LSR}$ = 4.1 km s$^{-1}$ also shows a similar
%velocity structure. It extends toward northwest further at lower velocities
%of $V_{LSR}$ = 4.4 km s$^{-1}$ and 4.8 km s$^{-1}$.
%On the redshifted side, the extended NW streamer is seen at $V_{LSR}$ = 5.4 km s$^{-1}$,
%and is shrinks toward the protostellar position progressively to higher velocities of
%$V_{LSR}$ = 5.8 km s$^{-1}$ and 6.1 km s$^{-1}$. While the spatial location
%of NW streamer is similar
%to that of the redshifted molecular outflow as seen in the $^{12}$CO emission,
%the NW streamer exhibits higher velocities closer to the protostar, opposite to the
%molecular outflow. All the streamers show this velocity feature,
%and the natures of these streamers are unlikely the outflow or the outflow cavities.

%%%%%%%%%%%%
%final ver.%
%%%%%%%%%%%%

%%%%%%%%%%%%
%final ver.%
%%%%%%%%%%%%

%%%%%%%%%%%%
%final ver.%
%%%%%%%%%%%%

\section{ANALYSIS}\label{sec:discussion}
\subsection{Keplerian Protostellar Disk} \label{subsec:discdisc}

\begin{figure*}
\centering
\includegraphics[width=180mm, angle=0]{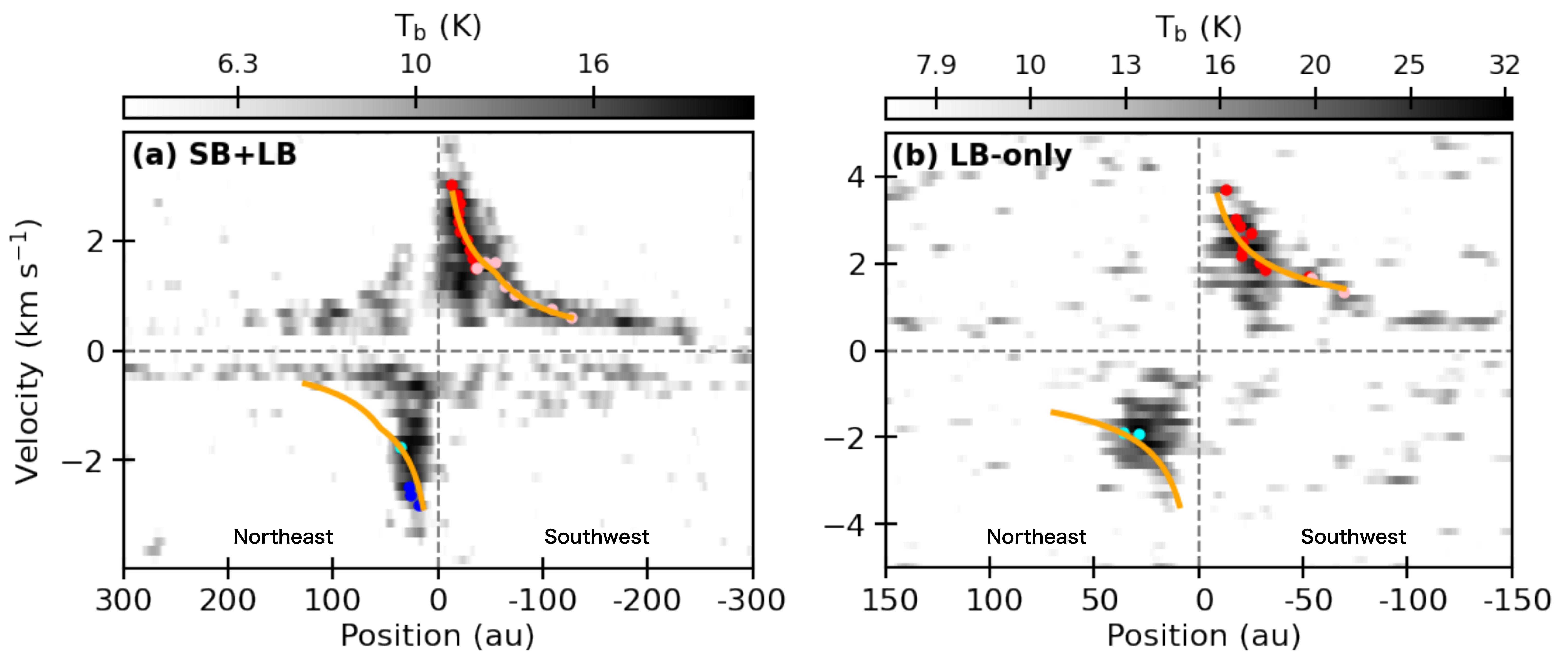}\\
\caption{
Position-Velocity diagrams of the C$^{18}$O (2--1) emission along the disk major axis (P.A. $=$ 45$\degr$) using the SB+LB data ($a$) and only the LB data ($b$).
%the northeast and the right 
%\textcolor{teal}{The left side of the figures correspond to side to the southwest.}
Blueish and reddish markers on the P-V diagrams show the representative blueshifted and redshifted data points as derived from SLAM, and light and thick color markers denote the points derived from the profiles along the velocity and positional axes, respectively.
% the positional and velocity axes, respectively. 
Orange solid curves in panel ($a$) denote the result from the double power-law fitting of the rotational profile, while those in panel ($b$) from the single power-law fitting (see text for details).}
\label{fig_C18O_pv}
\end{figure*}

%The CO isotopologue lines exhibit a velocity gradient along the northeast (blueshifted) to southwest (redshifted) direction in the dusty disk.
The CO isotopologue emission directly associated with the dust disk and most probably tracing the gas disk exhibits a velocity gradient along the northeast (blueshifted) to southwest (redshifted) direction.
The velocity channel maps of the C$^{18}$O (2--1) emission at high velocities  (Figure \ref{fig_C18O_high}) show archetypal signatures of the rotation. 
Figure \ref{fig_C18O_pv} shows Position-Velocity (P-V) diagrams of the C$^{18}$O (2--1) emission along the major axis of the dust disk (P.A.$=$ 45$\degr$) using the SB+LB data ($a$) and the LB data only ($b$).
The SB+LB and LB-only P-V diagrams are made by averaging the spectra over the transverse width of 0$\farcs$15 and 0$\farcs$05, respectively.
%The SB+LB P-V diagram exhibits velocity features both in the disk and the outer molecular gas, whereas the LB-only P-V diagram selectively traces the disk velocity structure.
The SB+LB P-V diagram shows the molecular emission not only in the first and third quadrants but also in the second and forth quadrants. 
This result implies that the SB+LB P-V diagram exhibits velocity structures of the protostellar disk as well as
those of the protostellar envelope. On the other hand, the LB only P-V diagram shows the emission predominantly in the first and third quadrants, suggesting that it traces the disk component only. 
Such a difference between the SB+LB and LB-only P-V diagrams can be attributed to the different degree of the missing flux.
%which shows emission, in which not only the first and third quadrants but also the second and forth quadrants, exhibits velocity features both in the disk and the gas surrounding the disk, whereas the LB-only P-V diagram, in which shows emission only the first and third quadrants, selectively traces the disk velocity structure.}
%\textcolor{teal}{This suggests that LB-only data shows emission only from the disk, while the gas surrounding the disk is resolved out.}
In the high-velocity blueshifted and redshifted components (1 km s$^{-1}$ $\leq$ $|$V$_{LSR}$$-$V$_{sys}$$|$ $\leq$ 3 km s$^{-1}$), the C$^{18}$O emission are well separated to the northeast and southwest, respectively.
On the other hand, at lower velocities ($|$V$_{LSR}$$-$V$_{sys}$$|$ $\lesssim$ 1 km s$^{-1}$) the SB+LB P-V diagram shows overlaps of the northeastern and southwestern emission components at the same
velocity.
%redshifted and blueshifted emission components at the same positions.
These results suggest that the high-velocity components trace the rotating disk, while the lower-velocity components correspond to the rotating and infalling protostellar envelope surrounding the disk as already demonstrated by the FAUST results \citep{2022Imai}.

Fitting of rotation curves to the C$^{18}$O P-V diagrams along the disk major axis was performed, using the Spectral Line Analysis/Modeling (SLAM; \cite{yusuke_aso_2023_7783868}) code \citep{2023Ohashi}.
SLAM identifies an emission peak along each velocity or positional axis with the Gaussian fitting, and derives the best parameters of the rotational profile, $i.e.,$ the central stellar mass
($\equiv M_{\star}$), the power-law index of the rotation velocity ($\equiv p$, where $v_{rot}\propto r^{-p}$), and the outermost radius of the disk ($\equiv R_{\rm disk}$),
%rotational velocity at a certain radius, power-law index, and the outermost radius of the disk, 
through the Markov Chain Monte Carlo (MCMC) algorithm.
%To avoid contamination from
%the low-velocity envelope components, the velocity range for the fitting is
%restricted to ? km s$^{-1}$ to ?km s$^{-1}$ in the blueshifted range and
%? km s$^{-1}$ to ?km s$^{-1}$ in the redshifted range.
The SLAM fitting to the emission ridge in the LB-only P-V diagram with an intensity threshold of 3$\sigma$ (1$\sigma =$ 2.0 mJy beam$^{-1}$) indicates $p =$ 0.451$\pm$0.015 and $R_{\rm disk} =$ 25.69$\pm$0.50 au (solid orange curves in Figure \ref{fig_C18O_pv}$b$).
%The SLAM fitting to the emission edges (``edge method") provides
%$p = ?\pm ?$ and $R_{disk} = ? +- ?$ au
%(dashed green curves in Figure \ref{fig_C18O_pv}b).
The rotational power-law index is close to 0.5 of Keplerian rotation, which implies that the high-velocity C$^{18}$O emission associated with the dust disk indeed traces the Keplerian rotation.
The central protostellar mass is then derived to be $M_{\star}$ = 0.158$\pm$0.003 $M_{\odot}$.
We note that these quoted errors are only statistical and arise from the MCMC process,
and that SLAM defines 68$\%$ of the posterior distribution as the range of the 1$\sigma$ error.
The actual error of the protostellar mass should be larger. For example,
\cite{2015Czekala} adopted a more sophisticated fitting to the visibilities in the 3-dimensional space
and quoted 4$\%$ error bars on dynamical mass measurements of the pre-main sequence spectroscopic binary AK Sco.
Measurement of the dynamical mass of IRAS 16544 through such a
full 3-dimensional fitting is deferred to our next eDisk papers.

The SB+LB P-V diagram, on the other hand, should include both the central Keplerian disk and the outer envelope.
We thus adopted two rotational profiles using a double power-law function, $i.e.,$ the outer rotational profile in the envelope and the inner disk rotational profile. To avoid contamination from extended low-velocity gas,
the fitting velocity range is restricted to 0.5 km s$^{-1}$$\leq$ $|$V$_{LSR}$$-$V$_{sys}$$|$$\leq$ 4 km s$^{-1}$, and the intensity threshold of 5$\sigma$ (1$\sigma =$ 1.3 mJy beam$^{-1}$) is adopted.
The double power-law fitting to the P-V diagram with SLAM yields the power-law indices in the inner and outer regions of $p =0.514 \pm 0.019$ and $p =1.014 \pm 0.039$, respectively.
% \textcolor{red}{$p =0.593 \pm 0.019$ and $p =0.819 \pm 0.044$} respectively. The inner power-law index is still consistent with that of the Keplerian rotation.
The central protostellar mass is derived to be $M_{\star}$ = 0.137$\pm$0.003 $M_{\odot}$, consistent with that from the LB-only fitting.
%These derived stellar masses from SB+LB and LB-only are also consistent with the FAUST result, which estimated the stellar mass to be 0.08-0.30 $M_{\odot}$.} 
The outer most disk radius, or the radius of the breaking point of the rotational profile, is derived to be $R_{\rm disk}$ = 54.55$\pm$0.50 au. 
The larger disk radius as derived from the SB+LB data than that derived from the LB data only is likely attributed to the effect of the missing flux.
%\textbf{ 
%The quoted error bar of the protostellar mass arises from MCMC fitting process, so only fitting errors are included, notsystematic errors.
%On the other hand, \cite{2015Czekala} infered 4\% on stellar mass obtained by a sophisticated fitting to the visibilities in the 3-dimensional space, so this should be more accrate.}
Previous FAUST observations of IRAS 16544 have also derived the central stellar mass in the range of $M_{\star}$ = 0.08-0.3 $M_{\odot}$, from the comparison of the P-V diagram of a model infalling-rotating envelope to that of the observed C$^{18}$O (2--1), CH$_3$OH (4$_{2}$--3$_{1}$ $E$), and OCS (19--18) emission. 
While this mass estimate is consistent with that of the present work, the FAUST estimate does not pinpoint the mass range and assumes that the protostellar envelope is free-falling, which is not necessarily the case \citep{2013Takakuwa,2014Ohashi,2015Aso,2017Aso}.
Our higher-resolution eDisk observations have succeeded to resolve the disk structure and to identify the Keplerian rotation in the protostellar disk, which enables us to refine the value of the central protostellar mass.

\begin{figure*}
\centering
\includegraphics[width=185mm, angle=0]{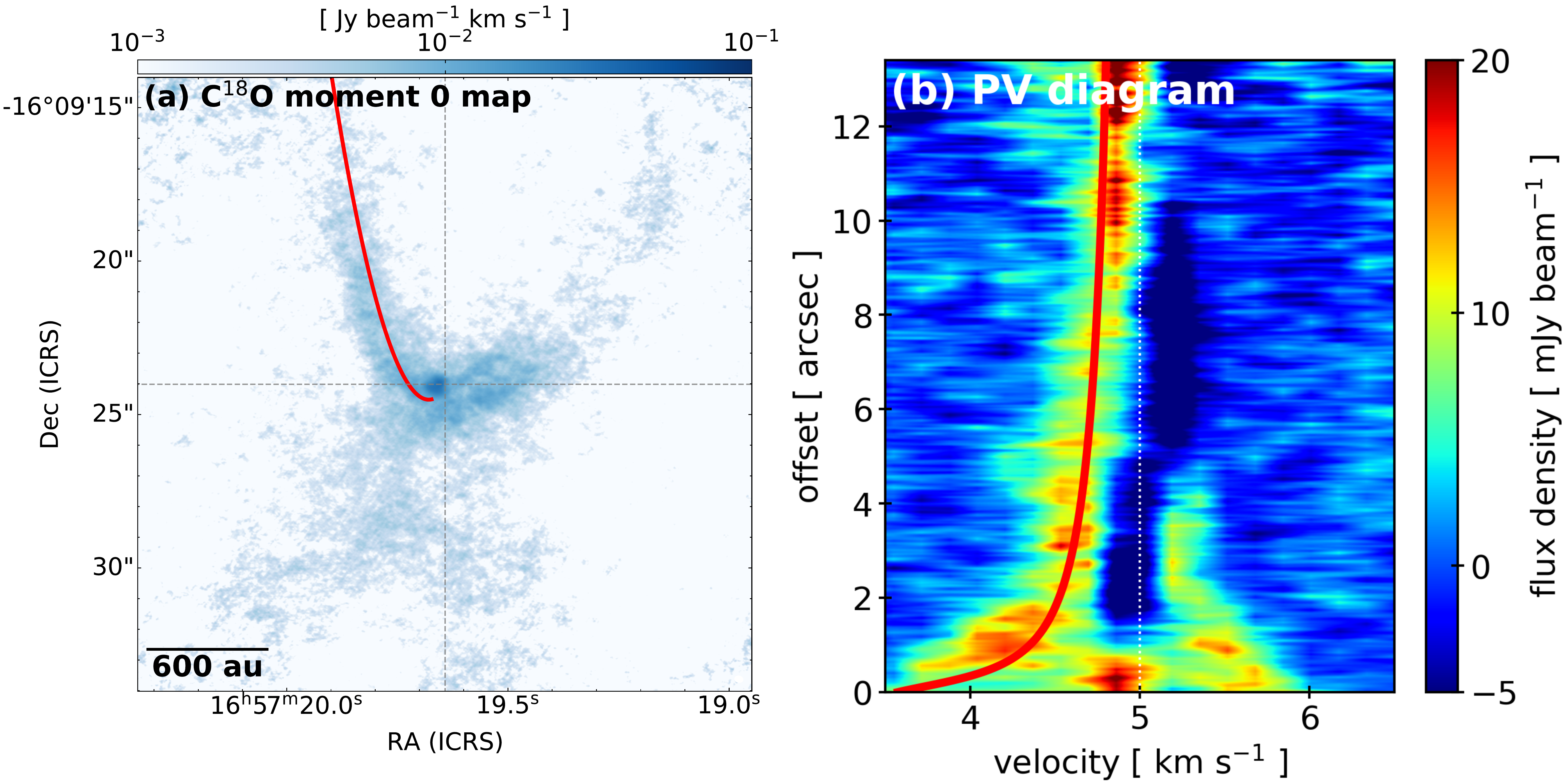}
\caption{a) Comparison of the solution of an accretion streamer to the spatial distribution of NE streamer as seen in the moment 0 map of the C$^{18}$O emission in the blueshifted velocity range (3.5--4.7 km s$^{-1}$). 
% A blue curve delineates the emission ridge by eye, which is adopted as the cut of the P-V diagram shown in panel b).
%From right to left, 
A red curve represents the solution at
$\phi_0 =$ 64$\degr$, $i_s =$ 73$\degr$, and $\theta_s =$ 60$\degr$,
with a central stellar mass of 0.14 $M_{\odot}$ and a centrifugal radius of 100 au.
b) P-V diagram of the C$^{18}$O emission along the red curve shown in panel a).
A red curve shows a spatial and velocity trail of the accretion streamer shown in panel ($a$).}
%$\theta_0$=\textcolor{teal}{17}$\degr$.}
\label{C18O_streamer}
\end{figure*}

\begin{figure*}[t]
    \centering
    \includegraphics[width=180mm, angle=0]{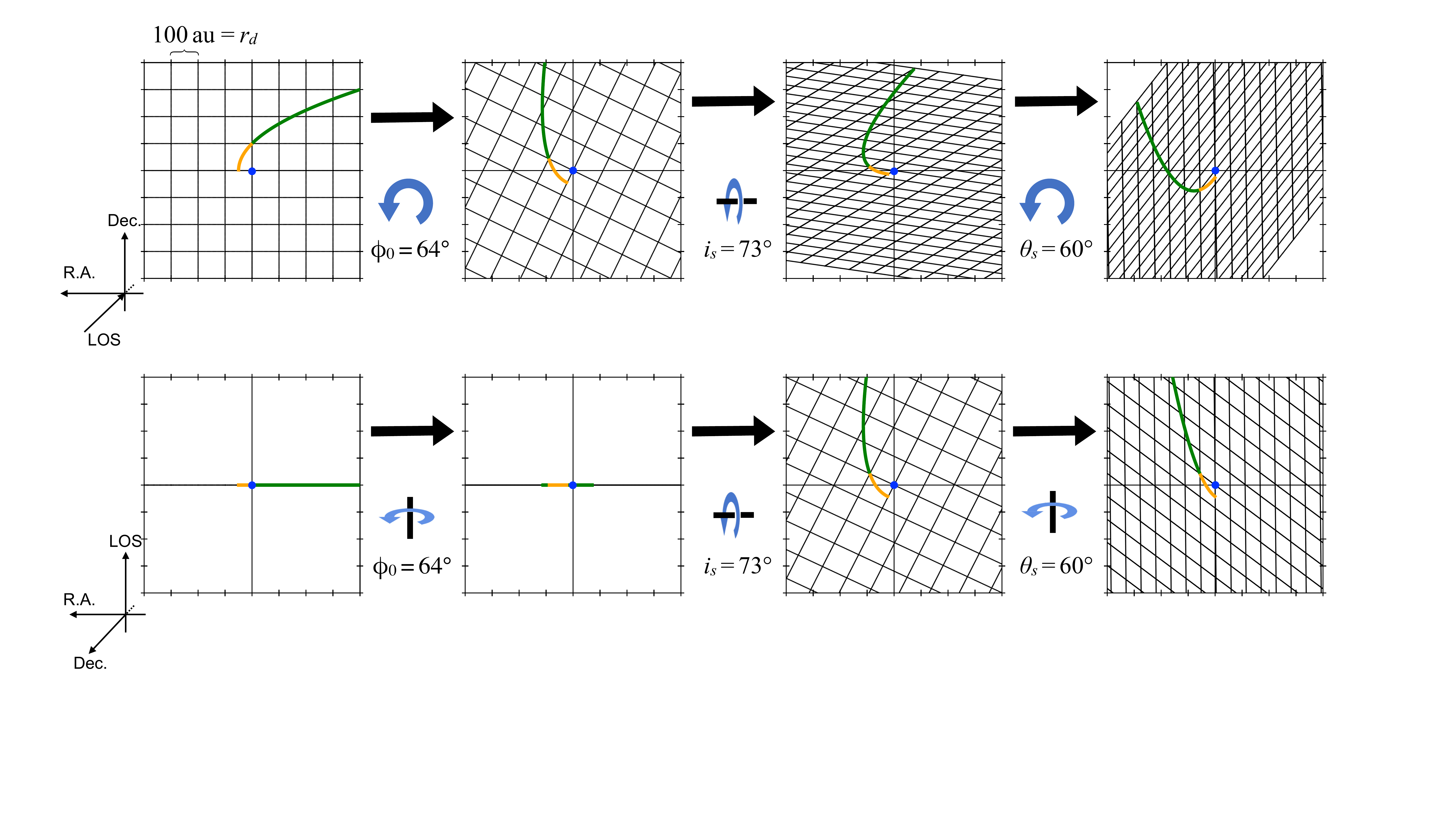}
    \caption{$Top:$ Illustrations of the streamer line projected on the plane of the sky. The x-, y-, and z-axes show Right Ascension, Declination, and the line-of-sight (LOS). In the images one cell size corresponds to the centrifugal radius ($\equiv r_d$ = 100 au). The blue and orange curves show the parabola
    outside/inside $r_d$, respectively. Filled blue circles show the position of the central protostar.
    From left to right rotation around the parabola axis, and rotation of the inclination and position angles are depicted for demonstration purpose. 
    $Bottom:$ Top view of the spatial structure of the streamer.
    % NE streamer accretes from the back side of the paper to the front side of it.
    The x-, y-, and z-axes show Right Ascension, line-of-sight, and Declination.}
    %b) Top view of the spatial structure of the streamer and disk. NE streamer accretes from the back side of the paper to the front side of it.
    %The right panel is the same as Figure \ref{fig_SO_SiO_CH3OH}b for direct comparison with the schematic picture to the left.
    \label{streamer_parameter}
\end{figure*}

% \subsection{The Bending Outflow}
%\subsection{Accretion Streamers toward the Disk} \label{subsec:streamer}
\subsection{Modeling of the Elongated Molecular Structure} \label{subsec:streamer}

%As well as the Keplerian protostellar disk,
In addition to the Keplerian protostellar disk, we have also identified intriguing elongated features at $\sim$1500 au scales seen in the lines of the CO isotopologues.
The velocity channel maps of the C$^{18}$O emission at low velocities (Figure \ref{fig_C18O_low}) exhibit at least three such features, inverse $J$-shaped strucrure, blueshifted component to the northwest of the disk, and redshifted one extending to the NNW.
% The most conspicuous one is \textcolor{red}{the inverse-$J$ shaped feature to the northeast.
The higher-velocity parts in these features are located closer to the protostar, which is opposite to the molecular outflow. Those high-velocity parts apparently deviate from the disk components as shown in Figures \ref{fig_C18O_high} and \ref{fig_C18O_low} at velocities of 3.52--3.85 km s$^{-1}$ and 6.19--6.52 km s$^{-1}$.
%These features are reminiscent of accretion streamers recent observations have identified \citep[$e.g.,$][]{2022Thieme}, and one of the intriguing interpretations for the nature of these components is the gas streamer accreting toward the central protostellar disk.
These features are reminiscent of accretion streamers recently observed in young, Class 0 protostars \citep[$e.g.,$][]{2022Thieme}.

%In this subsection, we discuss the nature of the most evident streamer,
%NE streamer.
We here discuss the spatial and velocity structure of the inverse $J$-shaped structure to the northeast (hereafter we call this component ``NE streamer").
Figure \ref{C18O_streamer}$a$ shows the moment 0 map of NE streamer as seen in the C$^{18}$O emission.
A prominent inverse $J$-shaped feature is evident in the moment 0 map.
The P-V diagram along the model curve that traces the emission ridge (see below) is shown
in Figure \ref{C18O_streamer}$b$.
There appears to be a long chain of the C$^{18}$O emission to the northeast of the source that is blueshifted, which we denote as NE streamer.
%from north to south in the blueshifted side, which corresponds to NE streamer.
NE streamer exhibits a gradual increase of the line-of-sight velocity toward the central disk.

To reproduce these observational results, we calculated spatial and velocity trails of ballistic accretion based on the formulae by \citet{1976Ulrich} and \citet{1981Cassen} (hereafter CMU model).
% For simplicity, the accretion trail is assumed to be on the same plane as that of the central disk.
%The inclination and position angles of the disk plane are assumed to be 73$\degr$ and 45$\degr$, respectively, as derived from the 2-dimensional Gaussian fitting to the 1.3-mm continuum image.
The radial ($\equiv v_r$) and azimuthal velocities ($\equiv v_\phi$) and the trail of the accretion are expressed as;
\begin{comment}
\begin{equation}
v_r=-\left(\frac{GM_{\star}}{r}\right)^{1/2}\left(1+\frac{\cos\theta}{\cos\theta_0}\right)^{1/2},
\label{eq3}
\end{equation}
\begin{equation}
v_\theta=\left(\frac{GM_{\star}}{r}\right)^{1/2}
\left(cos\theta_0-cos\theta\right)
\left(\frac{cos\theta_0+cos\theta}{cos\theta_0sin\theta}\right)^{1/2},
\label{eq4}
\end{equation}
\begin{equation}
v_\phi=\left(\frac{GM_{\star}}{r}\right)^{1/2}
\left(\frac{\sin\theta_0}{\sin\theta}\right)
\left(1-\frac{\cos\theta}{\cos\theta_0}\right)^{1/2},
\label{eq5}
\end{equation}
\begin{equation}
r=\frac{r_d sin^2\theta_0}{1-cos\theta/cos\theta_0},
\label{eq6}
\end{equation}
\end{comment}
\begin{equation}
v_r=-\left(\frac{GM_{\star}}{r}\right)^{1/2}\left(1-\cos\phi\right)^{1/2},
\label{eq3}
\end{equation}
\begin{equation}
v_\phi=\left(\frac{GM_{\star}}{r}\right)^{1/2}
\left(1+\cos\phi\right)^{1/2},
\label{eq5}
\end{equation}
\begin{equation}
r=\frac{r_d}{1+\cos\phi},
\label{eq6}
\end{equation}
where $G$ is the gravitational constant, $M_{\star}$ is the mass of the central protostar, $r$ is the radius, $\phi$ is the azimuth angle on the plane of the streamer, and $r_d$ is the centrifugal radius.
The mass of the central protostar is adopted as $M_{\star} =$ 0.14 $M_{\odot}$, which is derived from the SLAM fitting (see section \ref{subsec:discdisc}).
We then varied $r_d$ and $\phi_0$, the incident azimuthal angle, as well as the inclination $i_s$ and position angles $\theta_s$ of the rotational axis of the streamer, to match the observed spatial and velocity structures with the calculated trail by visual inspection.
$i_s$ is defined as the angle between the rotational axis and the line-of-sight.
%$\theta_s$ is defined as the rotation axis of the streamer with right-handed screw rotation,
%where 0$^\circ$ is north and increases counter-clockwise to the east.
$\phi_0$ is the angle of the streamer to which the parabolic trajectory opens,
and $\phi_0 =$ 0$\degr$ indicates west when
$\theta_s$ is 0 degree and increases counter-clockwise.
The visually determined parameter set is $\phi_0 =$ 64$\degr$, $r_d =$ 100 au, $i_s =$ 73$\degr$, and $\theta_s =$ 60$\degr$, with $M_{\star} =$ 0.14 $M_{\odot}$ (red curves in Figure \ref{C18O_streamer}).
Illustration of the derived streamer line in the 3-dimensional space is shown in Figure 
\ref{streamer_parameter}.
%In other words, the streamer direction can be written as follows:
%\begin{equation}
%\begin{split} 
%x_s &= -sin(\phi)cos(i_s)cos(P.A.)\\ &\quad- cos(\phi)sin(P.A.) = 0.008,
%\end{split}   
%\end{equation}
%\begin{equation}
%\begin{split} 
%y_s &= sin(\phi)cos(i_s)sin(P.A.)\\ &\quad- cos(\phi)cos(P.A.) = 0.511,
%\end{split}   
%\end{equation}
%\begin{equation}
%z_s = sin(\phi)sin(i_s) = 0.860,   
%\end{equation}
%where x, y, z are R.A, Dec., and the line of sight coordinate.
%}
%$\phi_0$ is defined as the angle on the streamer plane, where 0$^\circ$ is east and increases counter-clockwise to the south.
%Since $\theta_s$ is the position angle of the streamer plane, the position angle of the rotational axis of the streamer is perpendicular to $\theta_s$.}
Note that $r_d$ is the parameter of the rotational angular momentum of the infalling material, and thus this can be different from the present radius of the disk toward which the materials are accreting.
The calculated curve reproduces both the observed inverse $J$-shaped feature and the gradual increase of the velocity as the position gets closer to the disk.
This result implies that NE streamer could be interpreted as an accretion streamer toward the central disk, although the curve in the P-V diagram does not perfectly trace the observed spatial/velocity locations of the emission ridge.
\section{Discussion}
\subsection{Accretion Streamer to the Protostellar Disk}\label{subsec:discussion_streamer}
\begin{figure*}[t]
\centering
\includegraphics[width=180mm, angle=0]{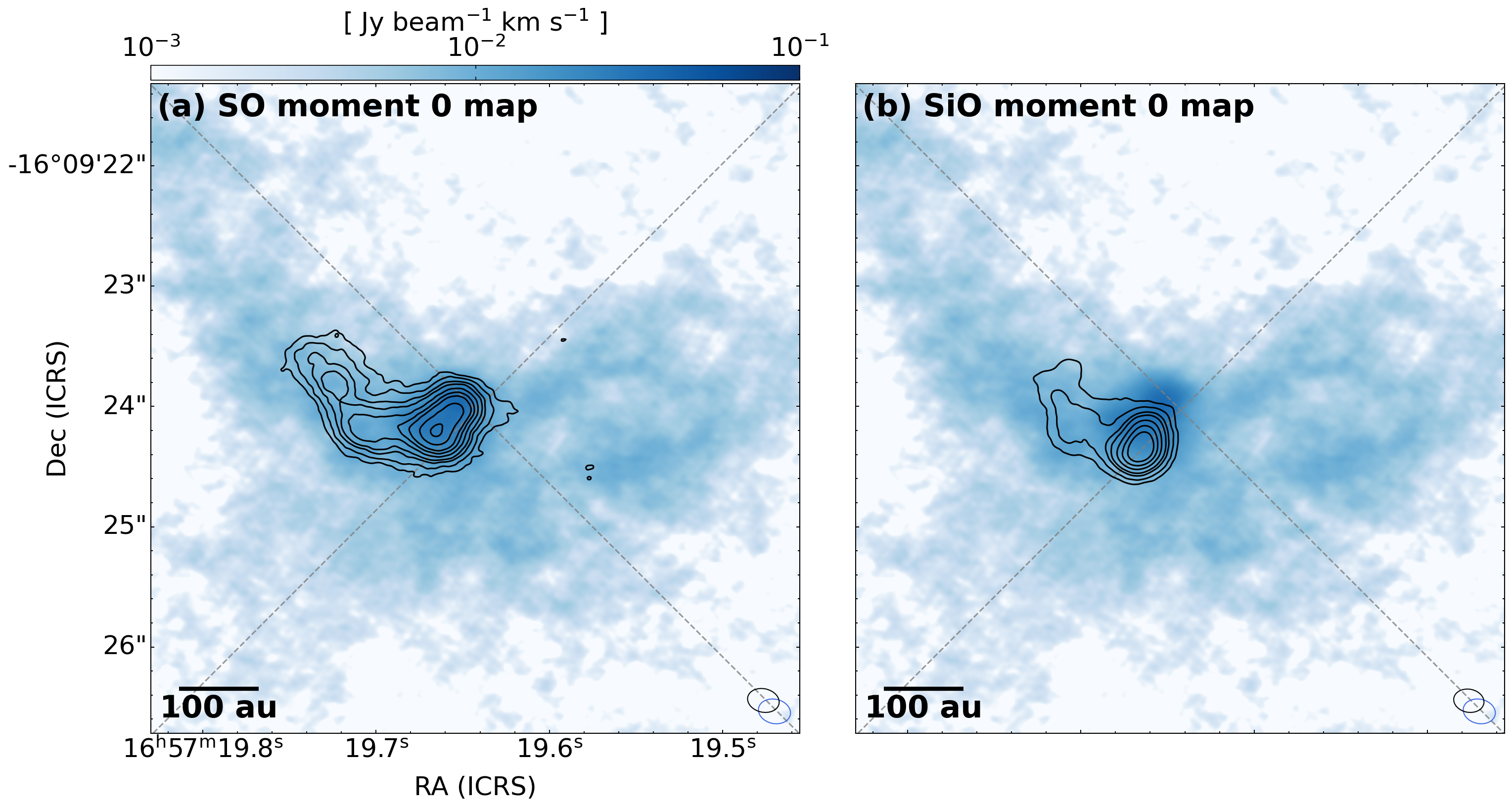}
\caption{
%Comparison of the spatial distributions of NE streamer as seen in the C$^{18}$O (2--1) emission (colors) and the shock tracers of $a$) SO (6$_5$--5$_4$), $b$) SiO (5--4), and $c$) CH$_3$OH (4$_2$--3$_1$ $E$) emission (contours).
%These shock tracers all peak on the disk and might be the streamer's landing point.
%The integrated velocity ranges of the C$^{18}$O, SO, SiO, and CH$_3$OH emission are 3.2--4.5 km s$^{-1}$, 1.5--9.2 km s$^{-1}$, $-$6.2--13.9 km s$^{-1}$, and $-$0.6--14.1 km s$^{-1}$ respectively.
%Contour levels are 5$\sigma$, 10$\sigma$, 15$\sigma$, 20$\sigma$, 30$\sigma$, 40$\sigma$, 50$\sigma$, and 60$\sigma$ (SO and SiO), and 5$\sigma$, 10$\sigma$, 20$\sigma$, and 40$\sigma$ (CH$_3$OH). 1$\sigma$ = 2.4, 0.7, and 0.6 mJy beam$^{-1}$ km s$^{-1}$ in panels ($a$), ($b$), and ($c$) respectively.
%Blue and white filled ellipses at the bottom-right corners denote the beam sizes of the C$^{18}$O and the relevant molecular lines in the panels.
%The gray dashed lines are the the same as in Figure \ref{fig_12CO}.
Comparison of the spatial distributions of NE streamer as seen in the C$^{18}$O (2--1) emission (colors) and the shock tracers of $a$) SO (6$_5$--5$_4$) and $b$) SiO (5--4) emission (contours).
These shock tracers all peak to the southeast of the disk and might be the streamer's landing point.
The integrated velocity ranges of the C$^{18}$O, SO, and SiO emission are 
%3.18386-4.51986:C18O, 3.184-4.520:SO, 3.18-4.52:SiO
3.184--4.520 km s$^{-1}$, 3.184--4.520 km s$^{-1}$, and 3.18--4.52 km s$^{-1}$, respectively.
%, 1.5--9.2 km s$^{-1}$, and $-$6.2--13.9 km s$^{-1}$, respectively.
Note that the SiO map is made from the integration over the two velocity channels at $V_{LSR}$ $=$ 3.18 and 4.52 km s$^{-1}$, because of the coarser velocity resolution of the SiO data ($=$ 1.34 km s$^{-1}$).
Contour levels are 5$\sigma$, 8$\sigma$, 12$\sigma$, 15$\sigma$, 20$\sigma$, 30$\sigma$, 40$\sigma$, 50$\sigma$, and 60$\sigma$. 1$\sigma =$ 1.2 and 1.3 mJy beam$^{-1}$ km s$^{-1}$ in panels ($a$) and ($b$), respectively.
Blue and black open ellipses at the bottom-right corners denote the beam sizes of the C$^{18}$O and the relevant molecular lines in the panels.
The gray dashed lines are the the same as in Figure \ref{fig_12CO}.}
\label{fig_SO_SiO_CH3OH}
\end{figure*}

\begin{figure*}[t]
    \centering
    \includegraphics[width=150mm, angle=0]{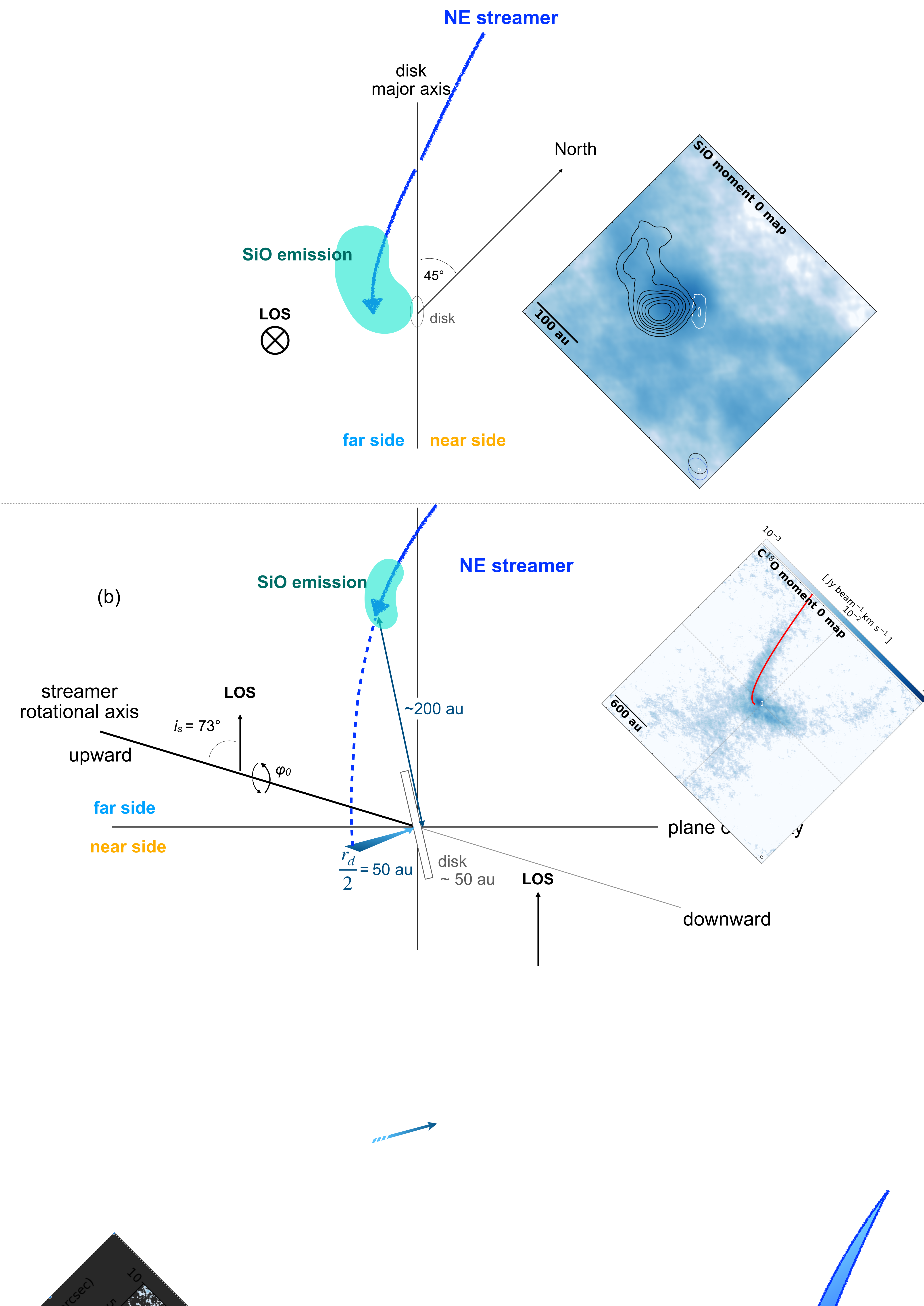}
    \caption{Schematic picture of the protostellar disk and NE streamer in IRAS 16544 projected onto the plane of the sky     obtained with the eDisk observations.
    % Observer view of the spatial structure of the streamer and the disk.
    For comparison, a zoom-in view of Figure \ref{fig_SO_SiO_CH3OH}b, with the distribution of the dust-continuum emission in white contours overlaid, is also shown aside. 
    The contour levels of the continuum emission
    are 5$\sigma$ and 150$\sigma$ (1$\sigma$ $=$ 21 $\mu$Jy beam$^{-1}$).
    The white ellipse at the lower right corner shows the beam size of the dust continuum emission.
    %b) Top view of the spatial structure of the streamer and disk. NE streamer accretes from the back side of the paper to the front side of it.
    %The right panel is the same as Figure \ref{fig_SO_SiO_CH3OH}b for direct comparison with the schematic picture to the left.
    }
    \label{scheme}
\end{figure*}

%一文入れる
%section4.2の解析の結果、NE elongated structureはdiskに降着するガスの弾道を捉えていることが明らかとなった。
From our model fitting in section \ref{subsec:streamer},
it is likely that the NE elongated structure traces the trajectory of
the material accreting to the disk.
Figure \ref{fig_SO_SiO_CH3OH} compares the spatial distribution of NE streamer and the moment 0 maps of a) the SO (6$_5$--5$_4$) and b) SiO (5--4) emission (contours) at the common velocity range.
%Note that the coarser velocity resolution
%of the SiO data prevents the exact match of the integrated velocity range
%(see Figure \ref{SiO_ch} in Appendix).
%Note that the velocity range of the SO and SiO emission (3.2--4.5 km s$^{-1}$) overlaps with that of NE streamer seen in the C$^{18}$O emission (3.2--4.5 km s$^{-1}$) (see Figure \ref{SiO_ch} in Appendix).
%While the SO emission exhibits multiple components, 
%The velocity range of the SiO emission component which coincides with the SO emission components (3.2--4.5 km s$^{-1}$) also overlaps with that of NE streamer.
Figure \ref{fig_SO_SiO_CH3OH} shows that the SO and SiO emission appear to trace the tip of NE streamer. %, which connects to the protostellar disk.
Furthermore, the SO and SiO emission appears to curl toward the northeast, following the trail of NE streamer.
Since the SO and SiO lines are known to be shocked gas tracers \citep{1998Bachiller,2001Bachiller,2006Hirano,2014Sakai,2018Oya,2021Okoda}, one of the possible interpretations for these emission distributions is that these emission trace the accretion shock at the landing point of the streamer. The centrifugal
radius of NE streamer is estimated to be $\sim$100 au (section \ref{subsec:streamer}).
This radius is larger than the radius of the Keplerian rotating disk as inferred from the SLAM fitting (section \ref{subsec:discdisc}).
Thus NE streamer is likely to accrete onto the envelope
outside the Keplerian rotating disk.
A schematic picture of the streamer plus disk system is shown in Figure \ref{scheme}.

Previous ALMA observations of protostellar envelopes have also found similar accretion streamers. 
\citet{2014Yen} have revealed blueshifted ($\sim$2000 au in length) and redshifted ($\sim$5000 au) gas streamers in the C$^{18}$O (2--1) emission toward the Class I protostar L1489 IRS. The blueshifted and redshifted streamers are found to accrete onto the plane of the large ($r \sim$300 au) Keplerian protostellar disk above and below the disk plane, respectively.
The velocity structures in these two steamers are consistent with free-falling gas flows with the centrifugal radius of 300 au and the mass accretion rate of 4--7$\times$10$^{-7}$ $M_{\odot}~yr^{-1}$.
In HL Tau, \citet{2019Yen} have found an intriguing one-arm spiral with a length of 520 au in the HCO$^{+}$ (3--2) emission, which extends from southwest to northwest of the planet-forming disk, and curls toward northwestern vicinity of the disk center. 
Kinematical analyses of this HCO$^{+}$ component reveals that the spiral is a rotating and infalling flow above the disk surface.
In the Class 0 protostar Lupus 3-MMS, \citet{2022Thieme} found multiple extended accretion flows along the outflow cavities in the C$^{18}$O (2--1) emission. 
The flows matched well with the CMU model with the mass accretion rates of 0.5--1.1$\times$10$^{-6}$ $M_{\odot}~yr^{-1}$. 
While those flows follow the edges of the outflows, they have revealed that those structures are not outflow cavities but accretion flows, with the help of their kinematical model.
Note that %the streamers 
the streamer found in IRAS 16544 also resembles the outflow cavity at a first glance.
However, the channel map shows that the high-velocity component is detected near the central star and the low-velocity component is away from the IRAS 16544, suggesting that this
component is not outflow-related.
%Similar accretion flows are also identified in the low-mass protostar IRAS 03292+3039 in HC$_3$N (10--9) \citep{2020Pineda}.

These growing pieces of evidence imply that accretion streamers in protostellar envelopes are not rare, but could be a common astrophysical phenomenon. Recent high-resolution, high dynamic-range observations of protostellar envelopes with ALMA have been finding these accretion streamers, which are in contrast with the classical picture that protostellar envelopes are continuous gas structures with rotation and infalling motions. 
If non-uniform, filamentary or fiber-like structures surround the natal dense cores \citep{2011Hacar,2013Hacar,2017Hacar}, it is natural that these gas structures exhibit accretion streamers in the course of protostellar formation. Numerical simulations of magnetized turbulent cloud cores indeed show such kinds of filamentary structures \citep{2017Kuffmeier}. 
Furthermore, a flattened envelope formed in the magnetized turbulent core is warped and exhibits spiral structures around a centrifugally-supported rotating disk \citep{2014Li}. 
Non-ideal MHD simulations with ambipolar diffusion also predict infalling spirals connecting to the central disk \citep{2016Zhao,2018Zhao}.
\subsection{Physical Properties of the Disk Associated with the Class 0 Protostar IRAS 16544} \label{subsec:planet}
Our high-resolution eDisk observations have succeeded to identify the Keplerian disk around the Class 0 protostar IRAS 16544.
Previous interferometric observations have been finding a number of Keplerian rotating disks around Class I protostars \citep[$e.g.,$][]{2012Takakuwa,2014Yen,2015Aso,2017Yen}, but only a handful of Class 0 protostars associated with the Keplerian disks are identified \citep[$e.g.,$][]{2012Tobin,2013Murillo,2014Ohashi,2017Aso}. 
%eDisk has also found Keplerian rotating disks around other Class 0 protostars such as Ced 110 IRS 4A \citep{2022Sai}, IRAS 16253-2429 \citep{2022Aso}, and L1527 IRS \citep{2022vant}.
A statistical study of Class 0 disks by CALYPSO argues that disk sizes around Class 0 sources could be smaller than those around Class I sources \citep[see $e.g.,$][]{2019Maury}.
ALMA observations of a Class 0 protostar B335 identified the upper limit of the Keplerian disk of $\lesssim$5 au \citep{2015Yen,2019Yen_B335,2019Bjerkeli,2019Imai}.
Systematic studies of disk sizes as a function of the protostellar evolutionary sequence will be the subject to the forthcoming eDisk papers.

%Our high-sensitivity, high-resolution eDisk observations of the Class 0 protostar IRAS 16544 have revealed a dust disk with Keplerian rotation.
%molecular outflows with a number of flow components, and conspicuous accretion streamers.
%In this final subsection, we discuss these observational results in the context of the physical picture of IRAS 16544.

\begin{figure*}[t]
\centering
\includegraphics[width=180mm, angle=0]{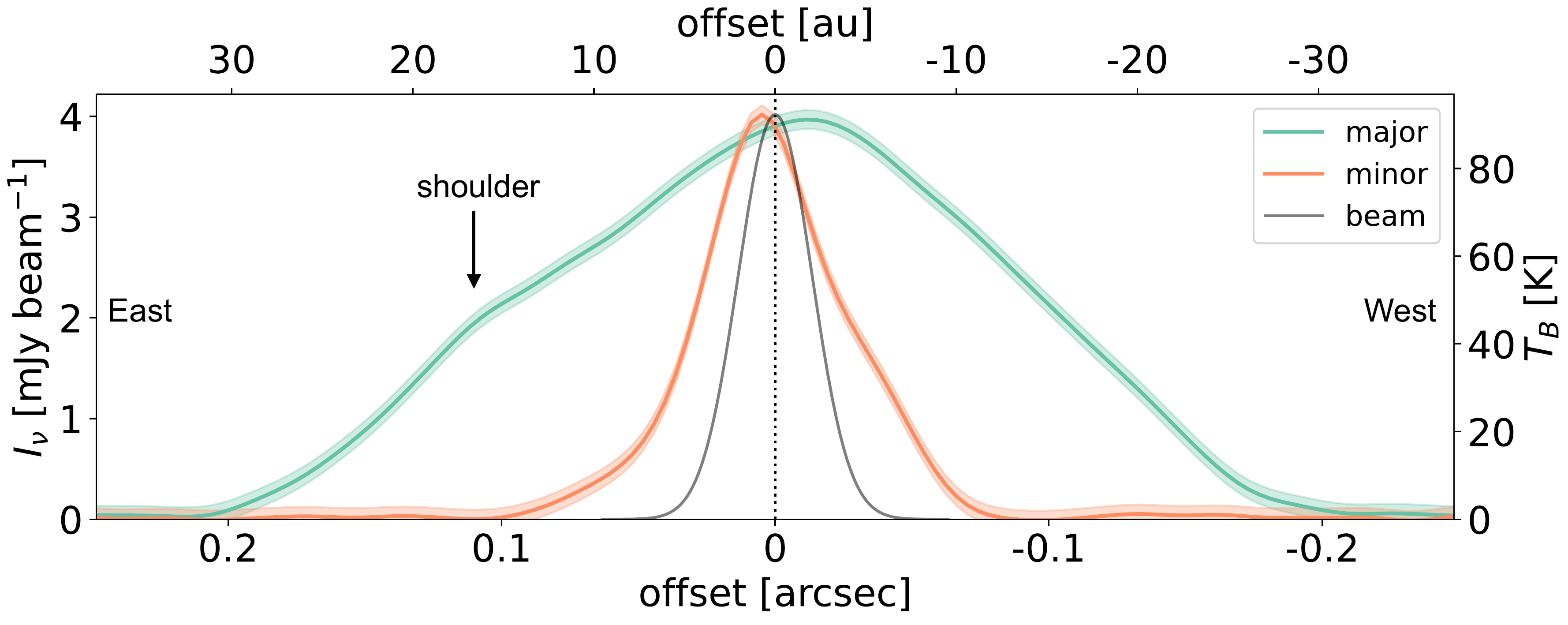}\\
\caption{Intensity profiles of the 1.3-mm dust-continuum emission along the major (light green curve) and minor (orange) axes in IRAS 16544. 
The origin of the coordinates, denoted by a vertical dotted line, is set to be the centroid position of the continuum emission as derived from the two-dimensional Gaussian fitting.
The positive direction of the major axis corresponds to the northeast, and the positive direction of the minor axis corresponds to southeast.
The right vertical axis indicates the brightness temperature.
Shaded areas in the profiles denote the $\pm$3$\sigma$ ranges.
For comparison, the profile of the geometrically-averaged beam is shown in a gray curve.}
\label{fig_radial_profile}
\end{figure*}

%In the 1.3-mm dust-continuum emission, there is an asymmetric emission distribution along the minor axis of the disk, and the peak of the 1.3-mm dust-continuum emission is skewed toward southeast (see Figures \ref{fig_cont} and \ref{fig_radial_profile}).
The dust disk exhibits non-Gaussian, asymmetric structures, as shown in the residual image after the subtraction of the fitted Gaussian (Figure \ref{fig_cont}$b$).
%Figure \ref{fig_cont}$b$ compares the residual image after the subtraction of the fitted Gaussian (color) to the original 1.3-mm continuum image (white contours).
%The residual image reflects the peak offset and the asymmetric emission distribution along the minor axis.
%In addition, the northeastern residual emission is more intense and extended than the southwestern one, suggesting that the emission is distributed asymmetrically along the major axis too.
To investigate the distribution of the dust emission more closely, the intensity profiles of the 1.3-mm dust-continuum emission
along the major and minor axes are shown in Figure \ref{fig_radial_profile}.
The origin of the profiles is set to be the centroid position as derived from the Gaussian fitting.
It is obvious that the Gaussian centroid does not match with the position of the emission peaks both along the major and minor axes.
Along the major axis, the northeastern and southwestern parts are not mirror-symmetric, and there is a possible ``shoulder'' around $\sim$0$\farcs$11 in the northeastern side.
This may imply that the disk is not azimuthally symmetric.
Along the minor axis, the peak location is $\sim$0$\farcs$01 offset toward southeast from the Gaussian centroid, and the northwestern profile is shallower than the southeastern profile within $\sim$0$\farcs$05.
%Such an asymmetry along the disk minor axes has also been detected in the other eDisk targets; $i.e.,$ Class 0 protostar L1527 IRS \citep{2022vant} and Class I protostar IRAS 04302+2247 \citep{2022Lin}, IRAS 04169+2702 \citep{2022Ilseung}, GSS30 IRS3 \citep{2023Santamaria-Miranda et} and R CrA IRS 7B \citep{2023Ohashi}.
%Such an asymmetry along the disk minor axes has also been detected in the other eDisk targets.
%As discussed in detail in our eDisk modeling paper by \citet{2022Takakuwa}, the asymmetric emission distribution along the disk minor axes implies that the 1.3-mm dust-continuum emission is optically thick and traces the disk flaring.

The observed peak brightness temperature of the 1.3-mm dust-continuum emission exceeds $\gtrsim$90 K.
While detailed radiative transfer modeling is required, such a high brightness temperature of the 1.3-mm dust-continuum emission likely indicates that the 1.3-mm emission is optically thick.
The observed outflows are redshifted to the northwest and blueshifted to the southeast, and thus the northwestern side of the dust disk is on the near side while the southeastern side is the far side.
If the 1.3-mm dust-continuum emission is optically thick and the dust distribution is flared, toward the southeastern, far side, the flared, warm disk surface is directly seen without the cold dust component intervening along the line of sight.
On the other hand, toward the northwestern, near side, we should see the colder portion of the disk close to the midplane along the line of sight. 
Thus the observed shift of the peak position of the 1.3-mm dust-continuum emission along the minor axis can be interpreted as the disk flaring.
%and beyond the peak the emission profile becomes steeper.
%Besides, toward the northwest the emission profile becomes shallower.
%These interpretations are in good agreement with the observations as shown
%in Figures \ref{fig_cont} and \ref{fig_radial_profile}.
%In such a case, the observed asymmetry along the minor axis suggests the presence of the dust flaring in this young Keplerian protostellar disk.
Such a dust flaring has been directly imaged by the VLA observations of the optically-thin 7-mm dust-continuum emission toward the Class 0 protostar L1527 IRS by \citet{2022Sheehan}.

%\textcolor{red}{There is also an asymmetric or non mirror-symmetric emission distribution along the disk major axis (Figure \ref{fig_radial_profile}).}
%A possible shoulder-like emission component is also seen to the $\sim$17 au northeast.
%In contrast to the asymmetric distribution along the minor axis, such a distribution along the major axis cannot be reproduced with any axisymmetric feature around the disk rotational axis. 

%Thus, the asymmetry along the major axis implies presence of the real non-axisymmetric structure, such as a spiral or crescent, in the protostellar disk.

Stability of a protostellar disk can be estimated with the Toomre $Q$ parameter as,
    \begin{equation}
    Q \sim \frac{2M_{\star}}{M_{\rm disk}}\frac{H}{R},
    \label{eq7}
    \end{equation}
where $M_{\rm disk}$ is the mass of the disk, $M_{\star}$ is the mass of the star, $R$ is the radius, and $H$ is the scale height at a radius $R$ \citep[$e.g.,$][]{2016Kratter, 2020Tobin}. 
From the results of the dust continuum emission and the SLAM fitting of IRAS 16544,
we derived that $M_{\rm disk}$ is 1.63$\times$10$^{-3}$--1.02$\times$10$^{-2}$ $M_{\odot}$ and $M_{\star}$ is $\sim$ 0.14 $M_{\odot}$.
% Assuming the sound speed $c_s$ = 0.45 km s$^{-1}$ ($T_K$ = 50 K),
Assuming the typical value of $\frac{H}{R}$ $=$ 0.1, the range of Toomre $Q$ is 17--2.7.
These nominal range of the $Q$ value suggests that the protostellar disk is gravitationally stable.
On the other hand, the 1.3-mm dust emission is probably optically thick as discussed above and the derived disk mass should be regarded as a lower limit. 
If the true disk mass is several times higher, the protostellar disk should be gravitationally unstable.
Observations at longer wavelengths, such ALMA Band 1--3, are required to properly estimate the disk mass and the stability of the disk.
%It is typically considered
%that if the $Q$ value is less than $\sim$1 the disk is gravitationally unstable and if Q is greater than ~1, then the disk is gravitationally stable (There’s another special state where if Q is ~ 1, then it’s marginally gravitationally stable).
If the true $Q$ value is less than $\sim$1, the major axis asymmetry and the detected shoulder-feature could be due to spirals induced by the gravitational instability.
%A similar asymmetric feature along the disk major axis has been identified by the VLA 7-mm observations of the Class 0 protostar L1527 IRS by \citet{2022Sheehan}.
%The detected non-axisymmetric structures in these Class 0 disks is probably the excess component above the axisymmetric, continuous disk feature.
%Spirals and crescent features have been identified in protoplanetary disks around Class II sources \citep[$e.g.,$][]{2013Fukagawa,2015vandermarel,2018Andrews,2018HuangIII}.
%In these protoplanetary disks the regions outside
%the spirals or crescent are almost devoid of the dust emission.
%The asymmetric features found in the Class 0 disks are likely at the initial stage of the development of those features.
%\textcolor{red}{Thus, it is not definitive to judge the stability of
%this protostellar disk.}

Whereas the gravitational instability in a massive disk is a possible cause of the observed non-axisymmetric dust feature, there are other physical mechanisms to produce such a disk asymmetry.
In the case of IRAS 16544, the accretion streamers are observed, and the shock could affect the protostellar disk. The accretion shock can generate strong spiral density waves, which could exhibit the non-axisymmetric feature in the disks \citep{2015Lesur}.
\cite{2022Kuznetsova} have incorporated heterogeneous infall based on the CMU model and an embedded disk, and performed hydrodynamic simulations to investigate the effect of the anisotropic accretion steamers on the disk structure. 
Their results demonstrate that the anisotropic infall induces the Rossby Wave Instability (RWI) in the disk, and forms a vortex and azimuthally asymmetric feature in the disk. 
Recent growing number of observational evidence for accretion streamers, combined with this hydrodynamic simulation, implies that the streamers could be one of the main physical mechanisms to form the asymmetric structure in the disks. 
This in turn could affect future evolution of disks and planet formation therein. 
Within their parameter space, \cite{2022Kuznetsova} also argued that disk self-gravity does not play an important role, with the Toomre $Q$ parameter above the marginal stability criterion.
Another physical mechanism of formation of azimuthal asymmetric structure in the disk incorporates presence of planetary and substellar companions.
The companions in the disk can also induce RWI, vortex and gas horseshoes, and spiral density waves, which show asymmetric millimeter dust-continuum emission \citep{2021vanderMarel}.

IRAS 16544 is associated with an active, prominent molecular outflow.
%with the axes of the redshifted and blueshifted outflow lobes bent.
%The redshifted outflow lobe
%in particular has a number of flows toward different directions.
A rotating and infalling protostellar envelope has also been identified in IRAS 16544 \citep{2022Imai}, and we have found a possible accretion streamers.
These results suggest that IRAS 16544 is in the active mass accretion phase, typical of Class 0 sources. 
Our high-resolution eDisk observations have unveiled that such an active Class 0 protostar also has a well-developed Keplerian rotating disk with hints of flaring and non-axissymmetric substructure, which could be related to future planet formation in the disk.

%\textcolor{red}{adding mechanisms to make the substructure other than planet formation}
%Formation of a Keplerian disk and subsequent disk substructure can be regarded as the initiation of planet formation. Our observational results of the early Class 0 protostar CB 68 imply that planet formation should be initiated from the main accretion phase of the protostar.

%%%%%%%%%%%%%%%%%%%%%%%%%%%%%%%%%%%%%%%%%%%%%%%%%%%%%%%%%%%%%%%%%%%%%%%%%%%%%%%%%%%%%%%%%%%%%%%%%%%%%%%%%%%%%%%%%%%%%%%%%%%%%%%%%%%%%%%%%%%%%%%%%%%%%%%%%%%%%%%%%%%%%%%%%%%%%%%%%%%%%%%%%%%%%%%%%%%

\section{Summary} \label{sec:summary}

We have carried out high-resolution (0$\farcs$036 $\times$ 0$\farcs$027  $\sim$5 au) and high-sensitivity ALMA observations of the young Class 0 protostar IRAS 16544-1604 embedded in Bok Globule CB 68 with the 1.3-mm dust-continuum emission, $^{12}$CO, $^{13}$CO, C$^{18}$O ($J=$ 2--1), SO ($J_N=$ 6$_5$--5$_4$), and other Band 6 lines as a part of the ALMA Large Program, eDisk.
%Our eDisk observations of CB 68 have provided us with a number of intriguing features in this protostellar source, as summarized below.
The main results are as follows:

\begin{itemize}
    \item [1.]
    The 1.3-mm dust-continuum emission reveals a $r \sim$30 au protostellar disk along the northeast to southwest direction at a position angle of $\sim$45$\degr$. 
    The aspect ratio indicates that the disk is near to edge-on, with an inclination angle of $\sim$73$\degr$.
    Along the minor axis, the emission peak is skewed toward southeast, and beyond the peak the emission profile is steeper in the southeastern side than that in the northwestern side.
    The skewed intensity profile implies that the 1.3-mm dust-continuum emission is optically thick and the dust distribution is flared, $i.e.,$ dusts are yet to be settled onto the midplane.
    The intensity profile along the major axis is not mirror-symmetric but asymmetric, with a possible shoulder to the $\sim$17 au northeast.
    This suggests the presence of non-axissymetric structure in this Class 0 protostellar disk.
%    The dust mass in the protostellar disk is estimated to be ? - ? at $T_{dust}$ = 20 K
%    and ? K, which is derived from the canonical relation of
%    $T_{\rm d}$ = 43 K $\times$ $(\frac {L_{bol}}{L_{sun}})^{0.25}$.
    \item [2.]
    The $^{12}$CO (2--1) images show conspicuous molecular outflows, with the redshifted outflow lobe to the northwest and the blueshifted lobe to the southeast. The $^{12}$CO emission in the redshifted lobe exhibits a number of flow-like features pointing to various directions.
%    \textcolor{red}{The $^{12}$CO emission in the redshifted lobe largely aligns with the overall outflow, and it exhibits a number of flow-like features pointing to various directions. }
    Furthermore, the axis of the redshifted outflow lobe appears not
    perpendicular to the disk, and the redshifted and blueshifted
    outflow lobe show a bending morphology. This could reflect misalignment between the magnetic fields and rotational axes in the natal core.
    \item [3.]
    In the dust disk, the $^{12}$CO, $^{13}$CO, and C$^{18}$O (2--1) emission are blueshifted to the northeast and redshifted to the southwest, which is interpreted as the rotation in the disk. %The outermost radius of the molecular emission is $r \sim$30 au.
    Our analysis of the P-V diagram of the C$^{18}$O emission along the disk major axis proves that the rotational profile is consistent with the Keplerian rotation, 
    %where the central protostellar mass is estimated to be $\sim$0.1$M_{\odot}$.
    around a central protostar with a mass of $\sim$0.14 $M_{\odot}$.
    \item [4.]
    %\textcolor{red}{The P-V diagram along the NE streamer 
    In the outer region, the C$^{18}$O emission shows multiple streamer-like features. The most prominent one is an inverse $J$-shaped, blueshifted component elongated toward the northeast (NE streamer), which has $^{12}$CO and $^{13}$CO counterparts. 
    There are other such features; redshifted component to the north-northwest and blueshifted one to the northwest.
%    and NNE and NW streamers have the $^{13}$CO counterparts.
    In these features, the higher-velocity components are located closer to the protostellar disk, suggesting accretion motion.
%    This velocity structure is opposite to that of the molecular outflow and outflow cavities, and suggests gas motion of accretion flows.
%    On the assumption that NE streamer is co-planar to the protostellar disk,
    We searched for the trajectory of the ballistic infalling flow which reproduces the moment 0 map and the P-V diagram of NE streamer.
%    The best parameters are a central star mass of \textcolor{teal}{0.14} $M_{\odot}$ and a centrifugal barrier of \textcolor{teal}{140} au.}
    We found that a centrifugal barrier of 100 au reasonably reproduces the spatial and velocity structure of NE streamer, with a central protostellar mass of 0.14 $M_{\odot}$ derived from the P-V analysis of the Keplerian disk.
    %In the outer region, the C$^{18}$O emission shows multiple streamer-like features, one blueshifted to the northeast (NE streamer), redshifted to the north-northwest, and another one blueshifted to northwest. The NE streamer has $^{12}$CO and $^{13}$CO counterparts.
    Furthermore, the SO (6$_5$--5$_4$) and SiO (5--4) emission, shock tracers, are seen at the tip of NE streamer. Since the centrifugal radius of NE streamer
    %, $\sim$100 au,
    is larger than the radius of the Keplerian disk ($\sim$50 au), NE streamer is landing
    onto the envelope outside the Keplerian rotating disk, where the accretion shock takes place.
    %    which is an apparent landing point of the streamer onto the midplane. 
%    These molecular emission curl to northeast from the south of the protostar.
%    A possible interpretation for these results is that the streamer is accreting,
%    and that the accretion shock takes place on the landing point of NE streamer onto the disk.
%    The spatial and velocity structure of the NE streamer can indeed be reproduced with a simple model of a ballistic infalling flow.
 
%    Our eDisk observations of the young Class 0 protostar CB 68 have found that a Keplerian disk is well-developed in the early Class 0 stage, and the disk also starts forming non-axissymmetric structure often seen in Class II sources.
%    At the same time, active mass outflow and mass accretion are ongoing through the powerful molecular outflow and accretion streamers, respectively.
%    Our results imply that initiation of planet formation has already started in the main accretion phase of the protostellar evolution if the non-axisymmetric dust substructure is a signpost of planet formation.
\end{itemize}
    Our high-resolution eDisk observations of the Class 0 protostar IRAS 16544 have identified a compact Keplerian-rotating disk associated with flared and non-axisymmetric dust distribution. At the same time, active mass outflow and mass accretion are ongoing through the powerful molecular outflows and accretion streamers, respectively.
    The non-axisymmetric signature in the dust disk could
    reflect the spiral features induced by the gravitational instability in the disk,
    or interaction with the accretion streamers, although
    the optical thickness of the 1.3-mm dust-continuum emission prevents us from
    directly investigating the disk mass and its gravitational stability.
    These results present an updated physical picture of the Class 0 stage, when formation of substructures often seen in Class II disks has started.

\section*{Acknowledgments}
%We are grateful to the anonymous referee for insightful
%comments and detailed reading of the manuscript.
We are grateful to N. Harada and M. Omura for technical assistance with the Python codes.
We would like to thank all the ALMA staff supporting this work. 
S.T. is supported by JSPS KAKENHI grant Nos. 21H00048 and 21H04495, and by NAOJ ALMA Scientific Research grant No. 2022-20A.
K.S. is supported by JSPS KAKENHI grant Nos. 21H04495.
N.O. acknowledges support from National Science and Technology Council (NSTC) in Taiwan through the grants NSTC 109-2112-M-001-051 and 110-2112-M-001-031.
J.J.T. acknowledges support from NASA RP 80NSSC22K1159.
J.K.J. acknowledges support from the Independent Research Fund Denmark (grant No. 0135-00123B).
Y.A. acknowledges support by NAOJ ALMA Scientific Research Grant code 2019-13B, Grant-in-Aid for Scientific Research (S) 18H05222, and Grant-in-Aid for Transformative Research Areas (A) 20H05844 and 20H05847.
F.J.E. acknowledges support from NSF AST-2108794.
S.G. acknowledges support from the Independent Research Fund Denmark (grant No. 0135-00123B).
I.D.G.-M. acknowledges support from grant PID2020-114461GB-I00, funded by MCIN/AEI/10.13039/501100011033.
P.M.K. acknowledges support from NSTC 108-2112- M-001-012, NSTC 109-2112-M-001-022 and NSTC 110-2112-M-001-057.
W.K. was supported by the National Research Foundation of Korea (NRF) grant funded by the Korea government (MSIT) (NRF-2021R1F1A1061794).
S.-P.L. and T.J.T. acknowledge grants from the NSTC of Taiwan 106-2119-M-007-021-MY3 and 109-2112-M-007-010-MY3.
C.W.L. is supported by the Basic Science Research Program through the NRF funded by the Ministry of Education, Science and Technology (NRF- 2019R1A2C1010851), and by the Korea Astronomy and Space Science Institute grant funded by the Korea government (MSIT; Project No. 2022-1-840-05). 
J.-E.L. is supported by the NRF grant funded by the Korean government (MSIT) (grant number 2021R1A2C1011718).
Z.-Y.L. is supported in part by NASA NSSC20K0533 and NSF AST-1910106.
Z.-Y.D.L. acknowledges support from the Jefferson Scholars Foundation, the NRAO ALMA Student Observing Support (SOS) SOSPA8-003, the Achievements Rewards for College Scientists (ARCS) Foundation Washington Chapter, the Virginia Space Grant Consortium (VSGC), and UVA research computing (RIVANNA).
L.W.L. acknowledges support from NSF AST-2108794.
S.M. is supported by JSPS KAKENHI grant Nos. JP21J00086 and 22K14081.
S.N. acknowledges support from the National Science Foundation through the Graduate Research Fellowship Program under Grant No. 2236415. 
R.S. acknowledges support from the Independent Research Fund Denmark (grant No. 0135-00123B).
P.D.S. acknowledges support from NSF AST-2001830 and NSF AST-2107784.
M.L.R.H. acknowledges support from the Michigan Society of Fellows.
J.P.W. acknowledges support from NSF AST-2107841.
Y.Y. is supported by the International Graduate Program for Excellence in Earth-Space Science (IGPEES), World-leading Innovative Graduate Study (WINGS) Program of the University of Tokyo.
H.-W.Y. acknowledges support from the NSTC in Taiwan through the grant NSTC 110-2628-M-001-003-MY3 and from the Academia Sinica Career Development Award (AS-CDA-111-M03).
This paper makes use of the following ALMA data: ADS/JAO.ALMA \#2019.1.00261.L and 2019.A.00034.S.
ALMA is a partnership of ESO (representing its member states), NSF (USA), and NINS (Japan), together with NRC (Canada), MOST and ASIAA (Taiwan), and KASI (Republic of Korea), in cooperation with the Republic of Chile. 
The Joint ALMA Observatory is operated by ESO, AUI/NRAO, and NAOJ. 
The National Radio Astronomy Observatory is a facility of the National Science Foundation operated under cooperative agreement by Associated Universities, Inc.

%%%%%%%%%%%%
%final ver.%
%%%%%%%%%%%%
\begin{figure*}[t]
\centering
\includegraphics[width=180mm, angle=0]{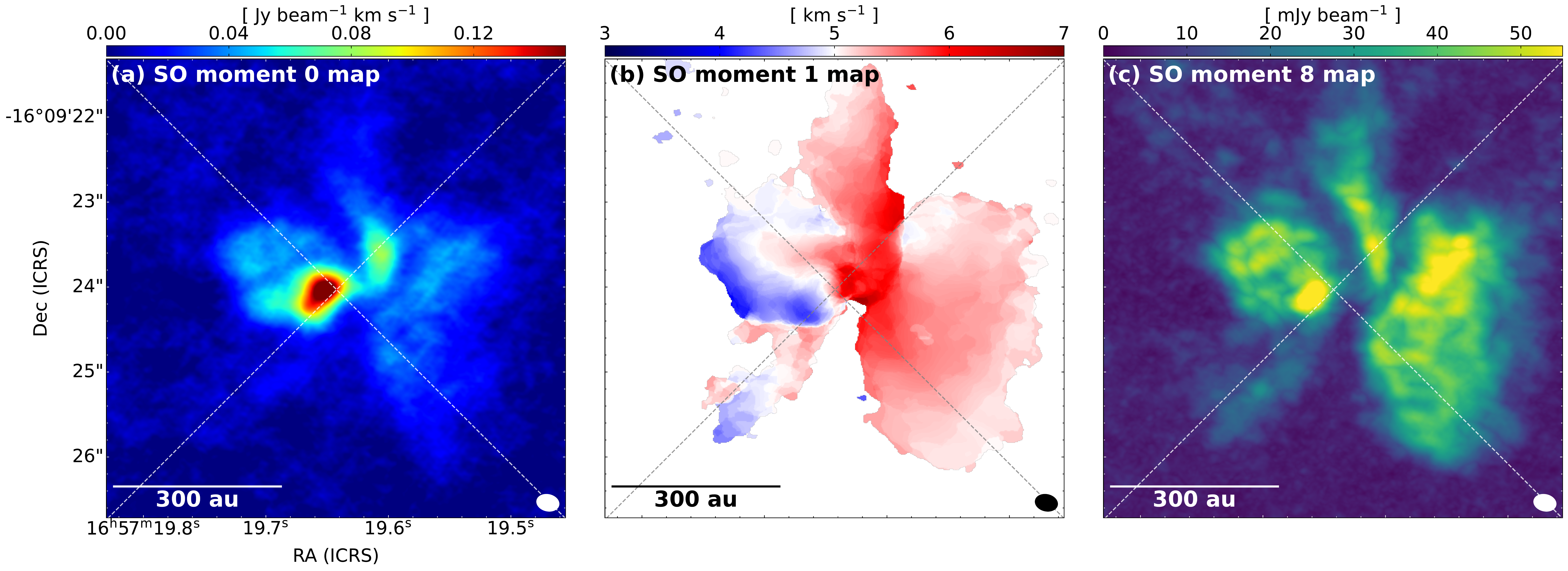}\\
\caption{
Moment 0, 1, and 8 maps of the SO (6$_5$--5$_4$) emission in IRAS 16544.
The integrated velocity range of the moment 0 map is from 1.514 km s$^{-1}$ to 9.196 km s$^{-1}$. 5$\sigma$ clipping is adopted to make the moment 1 map (1$\sigma =$ 2.4 mJy beam$^{-1}$). 
%Contours denote the distribution of the 1.3-mm dust-continuum emission, and the contour levels are 5$\sigma$, and 150$\sigma$ (1$\sigma$ = 21 $\mu$Jy beam$^{-1}$).
The white and gray dashed lines are the the same as in Figure \ref{fig_12CO}.}
\label{fig_SO_mom}
\end{figure*}

% \section*{Appendix} \label{sec:appendix}
\appendix
%\input{Appendix_A.tex}
%\section*{Appendix} \label{sec:appendix}
%\section{$^{13}$CO (2--1) Velocity Channel Maps} \label{sec:appendix13co}

%Figures and show velocity channel maps of the $^{13}$CO (2--1) emission
%at the high and low-velocity ranges, respectively, in CB 68.
%These figures are $^{13}$CO counterparts to Figures \ref{fig_C18O_high} and %\ref{fig_C18O_low}.
%Velocity structures traced by the $^{13}$CO emission are overall consistent
%with those traced by the C$^{18}$O emission.
%The high-velocity blueshifted $^{13}$CO emission is located to the northeast
%and redshifted $^{13}$CO emission to the southwest inside the dusty disk.
%The extents of the $^{13}$CO emission becomes larger at the velocities
%closer to the systemic velocity. These spatial and velocity structures
%are consistent with the Keplerian rotation in the protostellar disk.

%In the $^{13}$CO velocity channel maps at the lower velocities,
%the velocity structure of the NE streamer, as also seen in the C$^{18}$O emission,
%can be identified. In the NE streamer higher velocity $^{13}$CO emission is
%located closer to the central disk progressively.
%The velocity components of the NNW and NW streamers are also seen,
%and in the higher velocities the emission components are located closer to
%the central disk.
%These results imply that the accretion streamers are seen in the
%$^{13}$CO emission as well as the C$^{18}$O emission.
\section{Spatial and velocity Structures of the Detected Molecular Lines} \label{sec:appendixlines}

The spectral setting of eDisk enables us to observe a number of ancillary molecular lines simultaneously (see Table \ref{table:CB68_molecular_line}).
In this appendix, we present spatial and velocity structures of these molecular lines.

Figure \ref{fig_SO_mom} shows the moment 0, 1, and moment 8 maps of the SO ($J_N=$ 6$_5$--5$_4$) emission in IRAS 16544.
%The SO emission distribution appears to
%be somewhat different from that of the CO isotopologue lines.
The primary SO emission peak is located to southeast of the dust disk, and the secondary peak is located to the northwest. Additional emission peaks are seen to the east and northeast of the primary peak, and these emission components likely comprise the blueshifted, NE streamer seen in the CO isotopologue lines (see Figure \ref{fig_SO_mom}$b$).
On the other hand, the secondary SO peak to the northwest is redshifted.
%which curls toward the north. While this SO emission component spatially matches with
%the NNW streamer, the curling morphology in the SO emission is not identified in the
%CO isotopologue lines.
To the west of the protostar, an extended, fan-shaped redshifted
SO emission is present. 
%In the CO lines there is also a fan-shaped component to the west of the protostar.
The blueshifted component to the west of the protostar as seen in the CO isotopologue lines (see Figure \ref{fig_C18O_low}) is not seen in the SO emission, and difference of the distributions between the SO and CO isotopologue emission in the 3-dimensional space is present. 
This is the reason why the moment 1 map of the SO emission shows redshifted to the west while that of the CO isotopologue emission blueshifted.
%It is puzzling, however, that the western CO emission component is blueshifted while the SO emission component is redshifted.

\begin{figure*}
\centering
\includegraphics[width=160mm, angle=0]{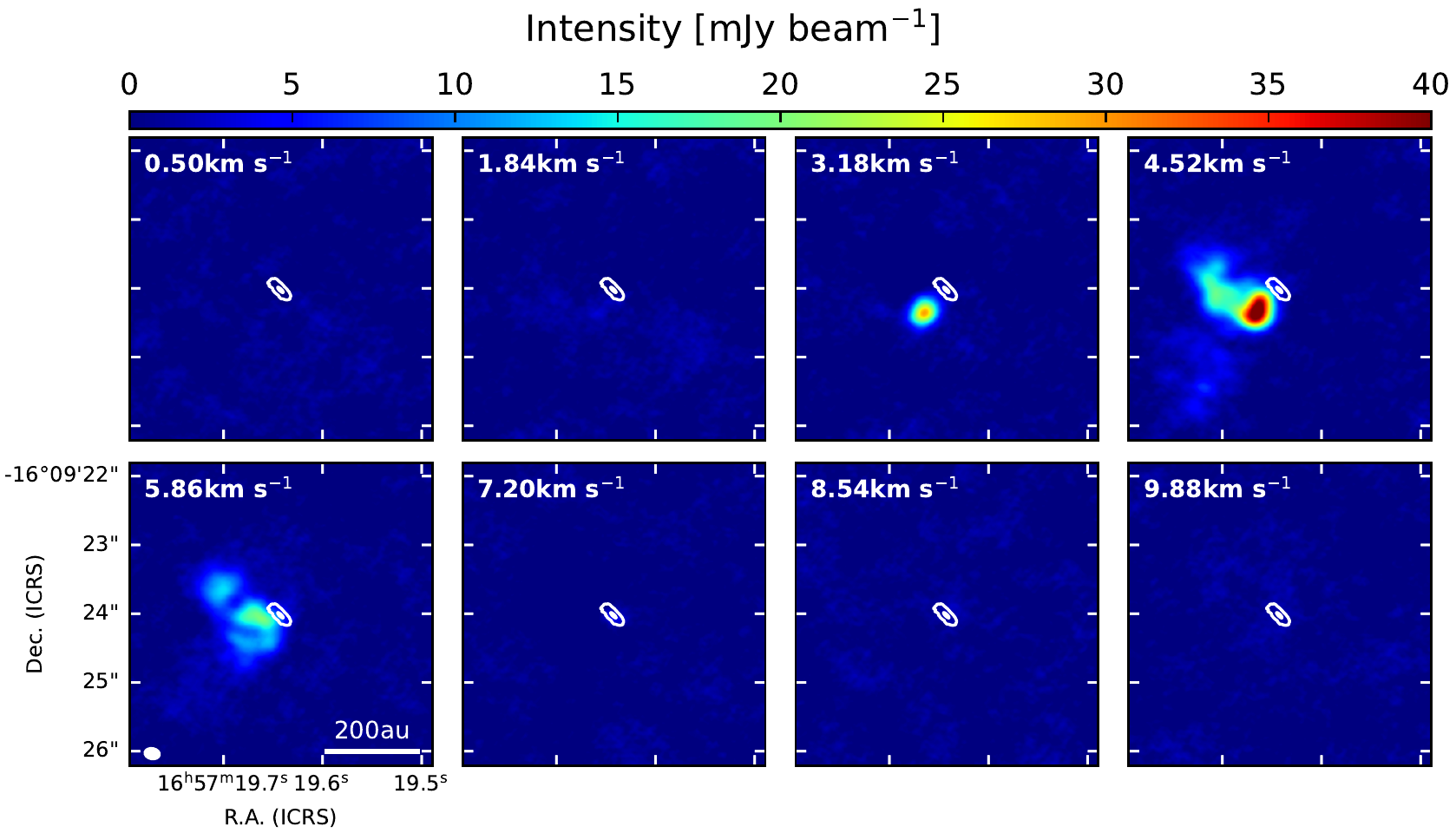}\\
\caption{Velocity channel maps of the SiO (5--4) emission in IRAS 16544.
Contours denote the distribution of the 1.3-mm dust-continuum emission, and the contour levels are 5$\sigma$ and 150$\sigma$ (1$\sigma =$ 21 $\mu$Jy beam$^{-1}$).
}
\label{SiO_ch}
\end{figure*}

\begin{figure*}
\centering
\includegraphics[width=160mm, angle=0]{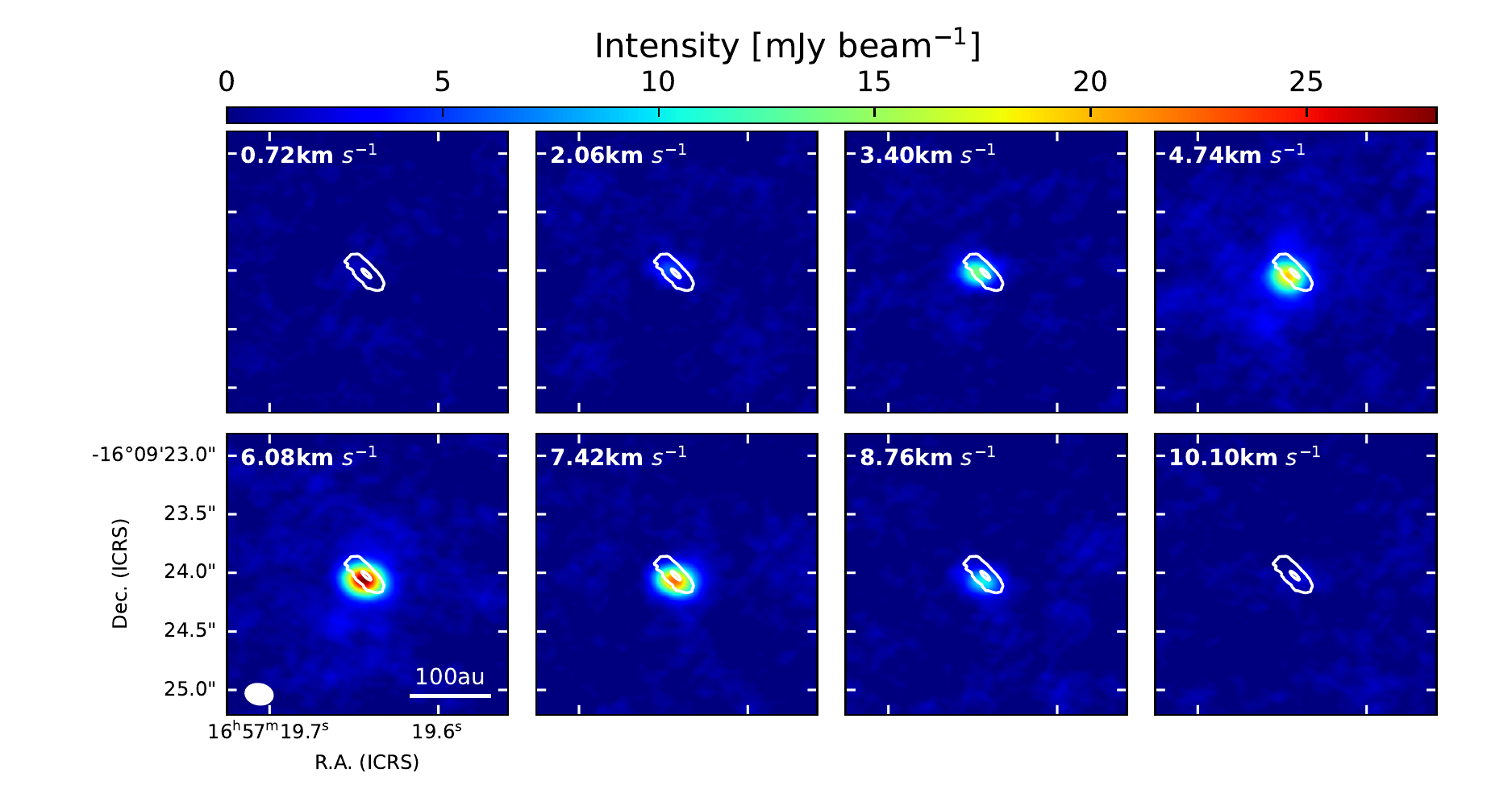}\\
\caption{Velocity channel maps of the CH$_3$OH (4$_2$--3$_1$ $E$) emission in IRAS 16544.
Contours denote the distribution of the 1.3-mm dust-continuum emission, and the contour levels are 5$\sigma$ and 150$\sigma$ (1$\sigma =$ 21 $\mu$Jy beam$^{-1}$).
}
\label{CH3OH_ch}
\end{figure*}

\begin{figure*}
\centering
\includegraphics[width=160mm, angle=0]{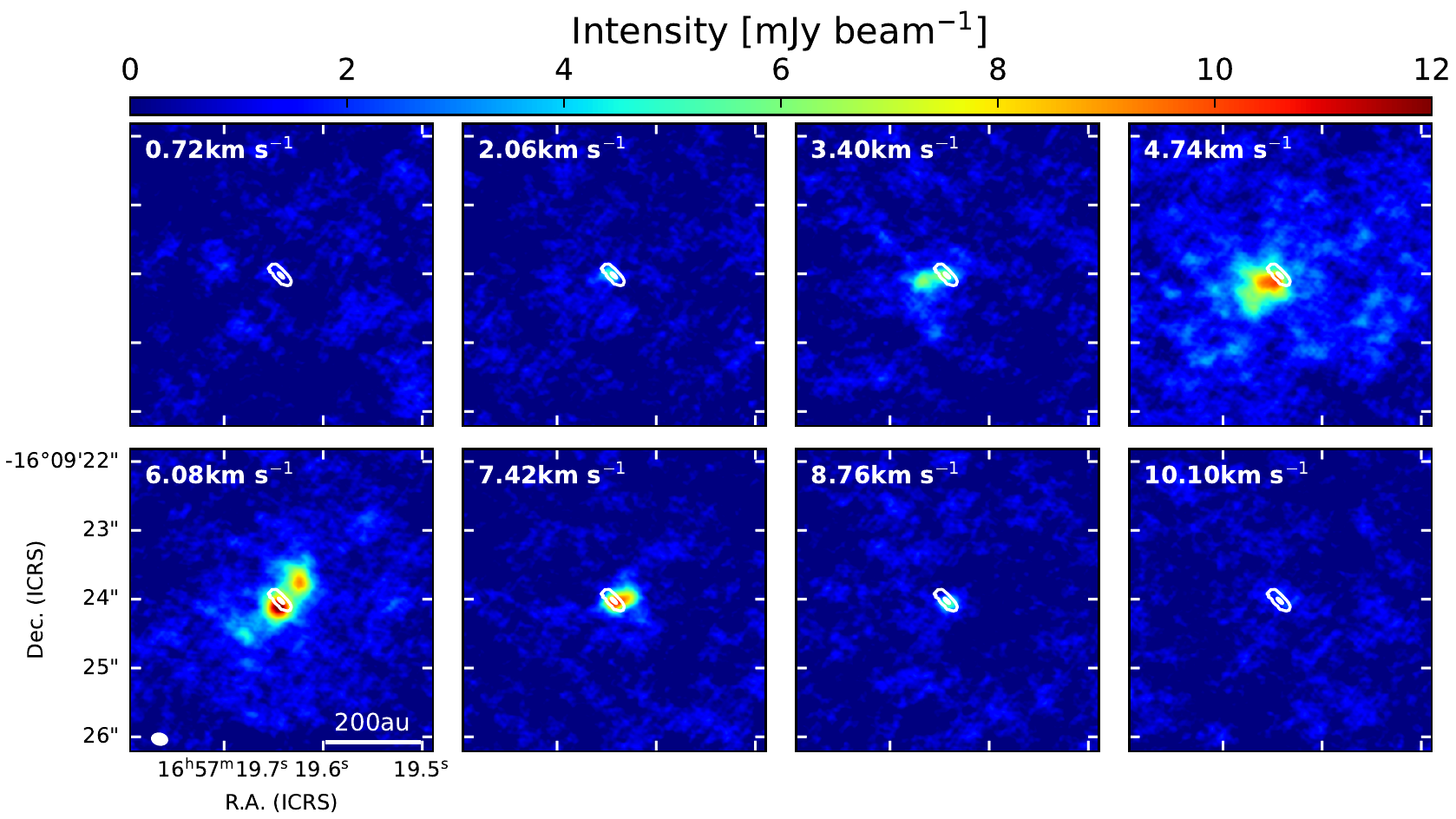}\\
\caption{
Velocity channel maps of the DCN (3--2) emission in IRAS 16544.
Contours denote the distribution of the 1.3-mm dust-continuum emission, and the contour levels are 5$\sigma$ and 150$\sigma$ (1$\sigma$ = 21 $\mu$Jy beam$^{-1}$).
}
\label{DCN_ch}
\end{figure*}

Figures \ref{SiO_ch}, \ref{CH3OH_ch}, and \ref{DCN_ch} show velocity channel maps of the SiO (5--4), CH$_{3}$OH (4$_2$--3$_1$ $E$), and DCN (3--2) emission, respectively.
In the blueshifted velocity of 3.18--4.52 km s$^{-1}$, the SiO emission peaks toward the southeast of the dust disk.
At $V_{LSR} =$ 4.52 km s$^{-1}$, the SiO emission distribution curls toward the northeast, consistent with our interpretation that the SiO emission traces the tip of NE streamer.
The SiO emission is also present in the southeastern side at the redshifted velocity $V_{LSR} =$ 5.86 km s$^{-1}$.
The CH$_{3}$OH emission is located toward the southeast of the disk, but the peak location is closer to the disk than that of the SiO emission.
In addition,
%The presence of the CH$_{3}$OH emission only toward the southeastern part can be interpreted that the CH$_{3}$OH emission traces the landing point of the accretion flow onto the disk.
the CH$_{3}$OH emission shows a clear velocity gradient along the disk major axis, consistent with the disk rotation. 
Thus the CH$_{3}$OH emission could be originated from the protostellar disk.
%the warm surface of the disk on the far side, while the CH$_{3}$OH emission on the near side is obscured by the optically-thick dust emission.
%Due to the limited spatial and velocity resolutions it is not straightforward to pinpoint the origin of the CH$_{3}$OH emission.
The DCN emission shows a similar emission component to the southeast with a velocity gradient along the disk major axis,
while another redshifted component to the northwest is also seen.

\begin{figure*}
\centering
\includegraphics[width=150mm, angle=0]{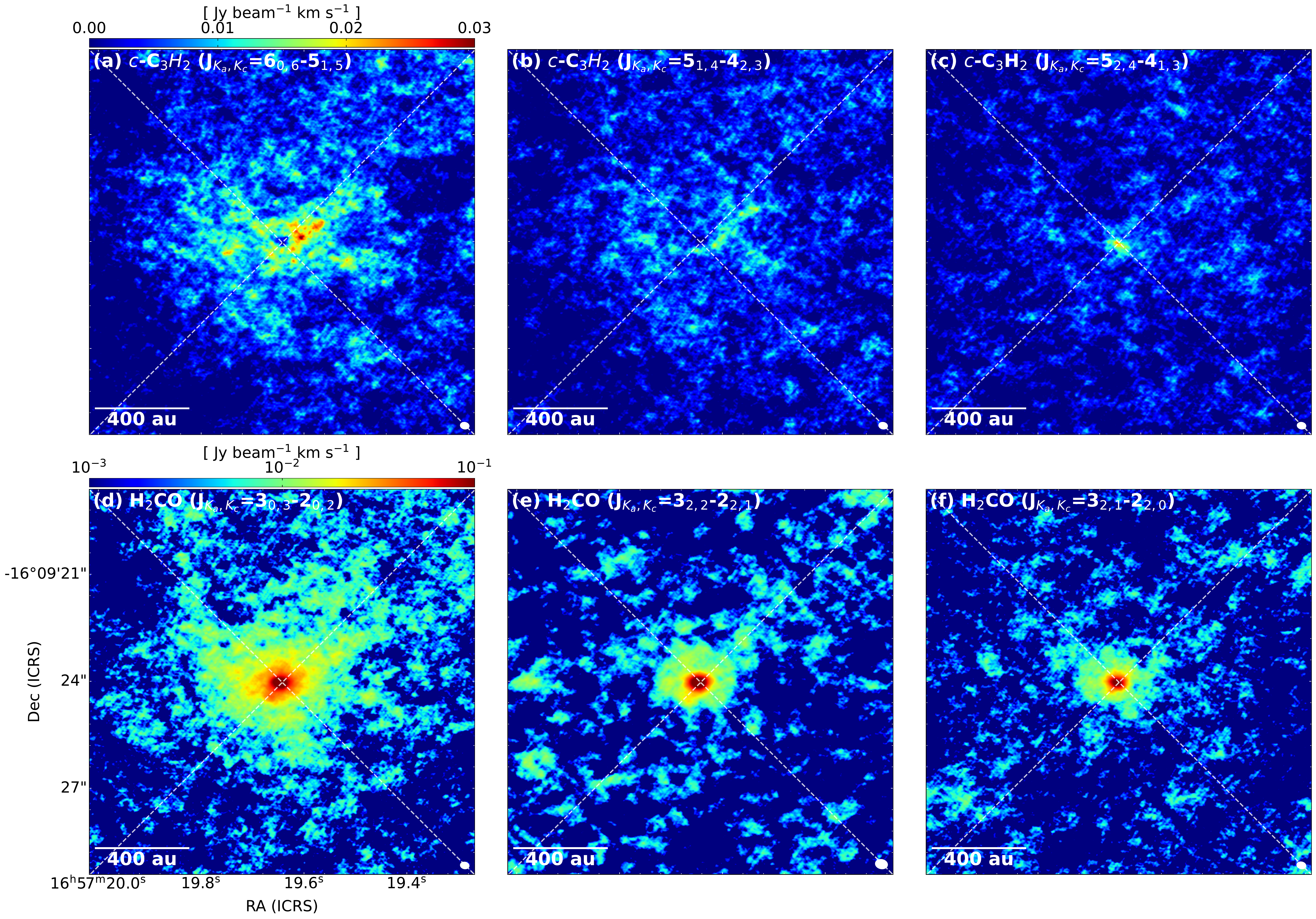}\\
\caption{
Moment 0 maps of the $c$-C$_3$H$_2$
($J_{K_a,K_c}$ = 6$_{0,6}$--5$_{1,5}$; 5$_{1,4}$--4$_{2,3}$;
5$_{2,4}$--4$_{1,3}$) and H$_{2}$CO
($J_{K_a,K_c}$ = 3$_{0,3}$--2$_{0,2}$; 3$_{2,2}$--2$_{2,1}$;
3$_{2,1}$--2$_{2,0}$) emission as labelled.
The integrated velocity ranges of the
$c$-C$_3$H$_2$
(6$_{0,6}$--5$_{1,5}$; 5$_{1,4}$--4$_{2,3}$;
5$_{2,4}$--4$_{1,3}$) and H$_{2}$CO
(3$_{0,3}$--2$_{0,2}$; 3$_{2,2}$--2$_{2,1}$;
3$_{2,1}$--2$_{2,0}$) emission are $-$0.62--18.14 km s$^{-1}$, $-$0.62--18.14 km s$^{-1}$, $-$0.62--18.14 km s$^{-1}$, $-$1.96--14.12 km s$^{-1}$, $-$1.96-- 15.46 km s$^{-1}$, and 0.01--10.03 km s$^{-1}$, respectively. 
%Contours denote the distribution of the 1.3-mm dust-continuum emission, and the contour levels are 5$\sigma$ and 150$\sigma$ (1$\sigma$ = 21 $\mu$Jy beam$^{-1}$).
The white dashed lines are the the same as in Figure \ref{fig_12CO}.}
\label{otherlines}
\end{figure*}

Figure \ref{otherlines} compares the moment 0 maps of the three $c$-C$_3$H$_2$ and H$_2$CO lines.
The $c$-C$_3$H$_2$ emission are extended ($\sim$2000 au), and a number of patchy $c$-C$_3$H$_2$ emission components are present.
The 6$_{0,6}$--5$_{1,5}$ and 5$_{1,4}$--4$_{2,3}$ transitions are weak toward the protostellar disk, while the 5$_{2,4}$--4$_{1,3}$ transition peaks at the disk location.
On the other hand, all the three H$_2$CO transitions exhibit strong peaks toward the protostar.
%as in the case of the SiO, CH$_3$OH, SO, and DCN lines.
The H$_2$CO lines also trace the extended component surrounding the disk, and the 3$_{0,3}$--2$_{0,2}$ transition also appears to trace NE streamer.
Multi-transitional analysis using these three transitions of $c$-C$_3$H$_2$ and H$_2$CO should provide us with important insights on the physical conditions of the molecular gas, which should be the subject to the subsequent papers.

%%%%%%%%%%%%
%final ver.%
%%%%%%%%%%%%

%%%%%%%%%%%%
%final ver.%
%%%%%%%%%%%%

%%%%%%%%%%%%
%final ver.%
%%%%%%%%%%%%

%%%%%%%%%%%%
%final ver.%
%%%%%%%%%%%%

\section{Velocity Channel Maps of the CO isotopologue Lines} \label{sec:appendixchannelmap}
\begin{figure*}
\centering
\includegraphics[width=180mm, angle=0]{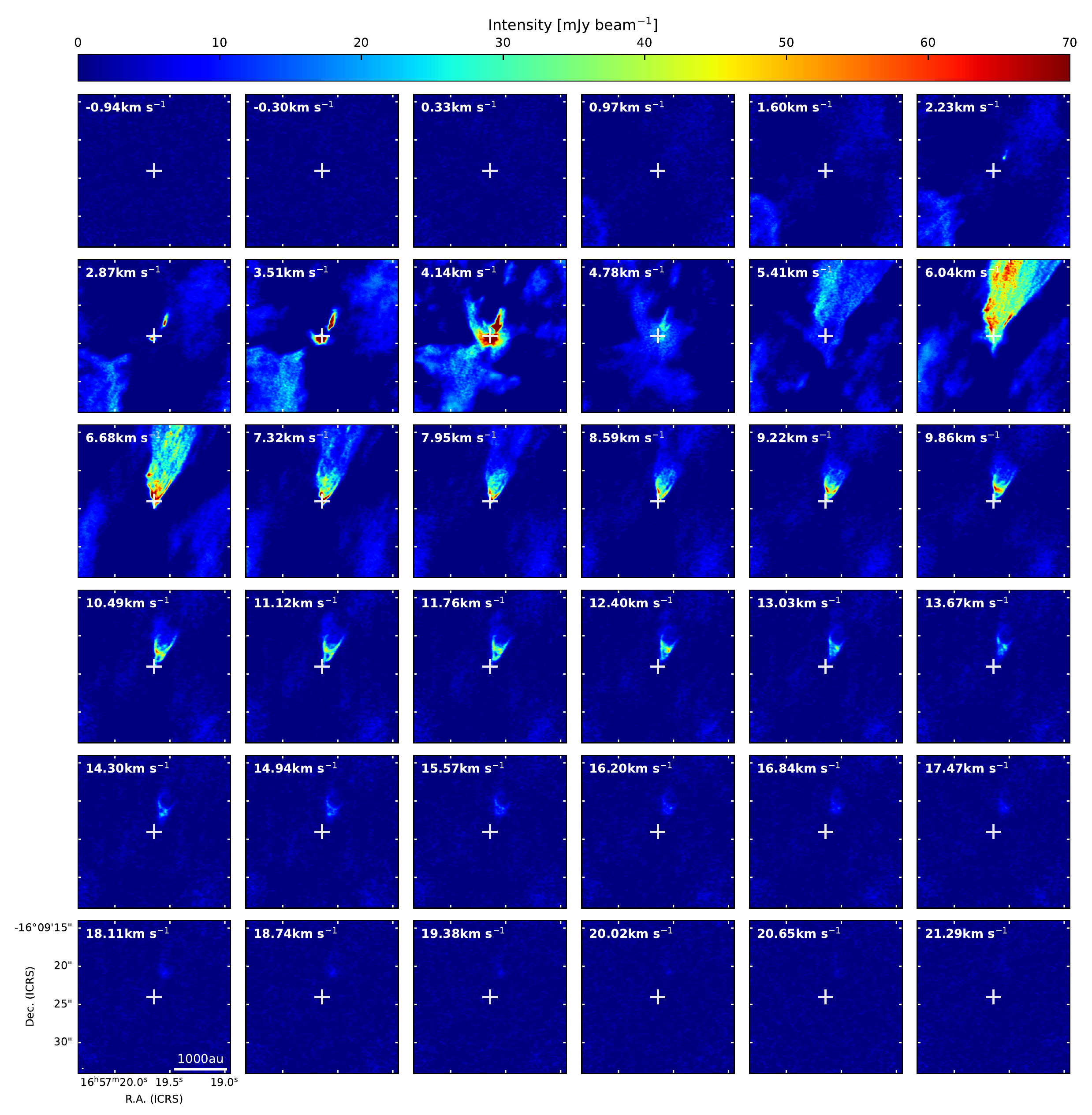}\\
\caption{ 
Velocity channel maps of the $^{12}$CO (2--1) emission toward IRAS 16544 over the entire region. Crosses indicate the position of the protostar. A white ellipse at the lower-left corner in the lowest-left panel shows the synthesized beam.
}
\label{co_ch}
\end{figure*}

\begin{figure*}
\centering
\includegraphics[width=150mm, angle=0]{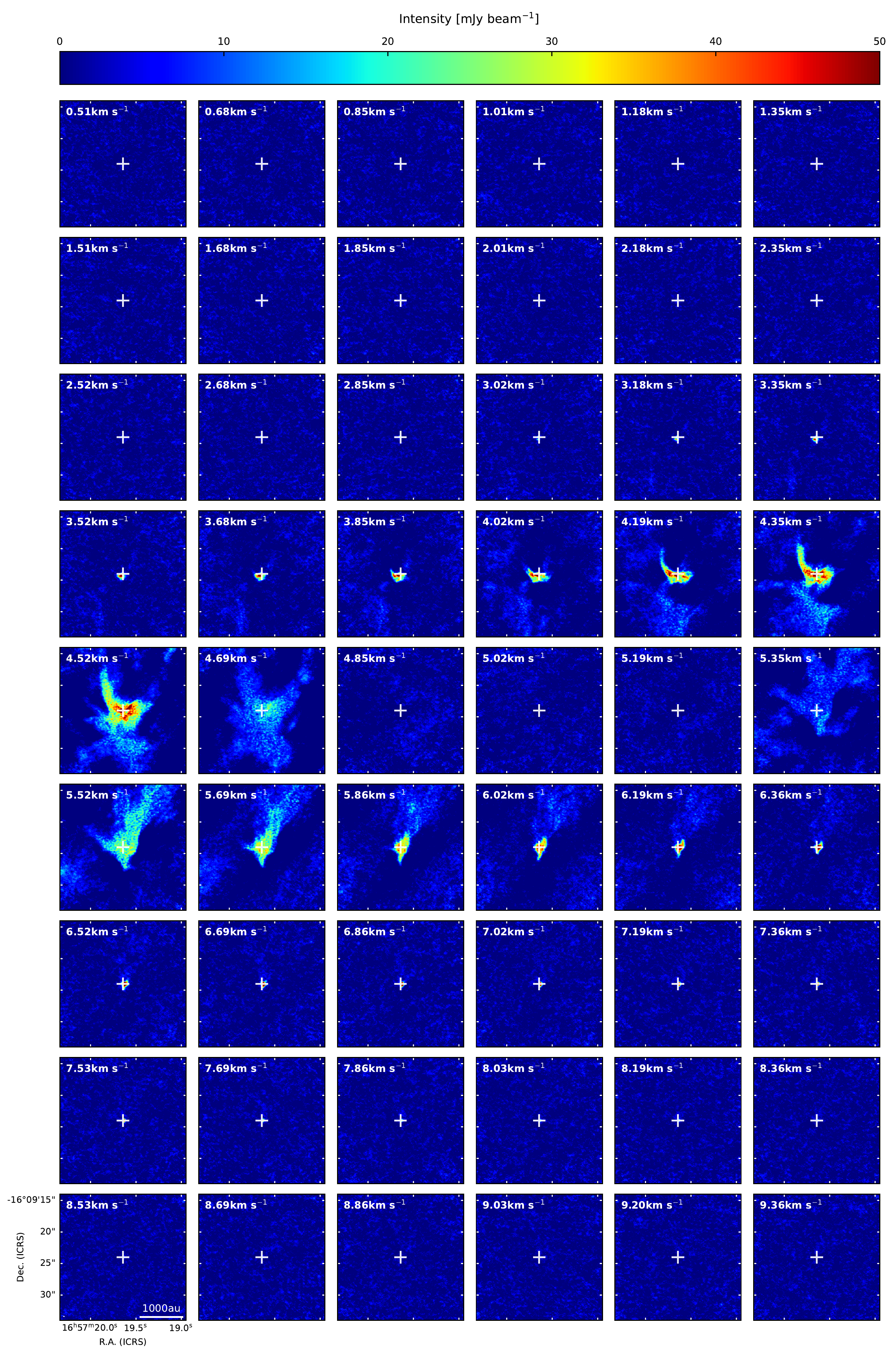}\\
\caption{ 
Velocity channel maps of the $^{13}$CO (2--1) emission toward IRAS 16544 over the entire region. Crosses indicate the position of the protostar. A white ellipse at the lower-left corner in the lowest-left panel shows the synthesized beam.
}
\label{13co_ch}
\end{figure*}

\begin{figure*}
\centering
\includegraphics[width=180mm, angle=0]{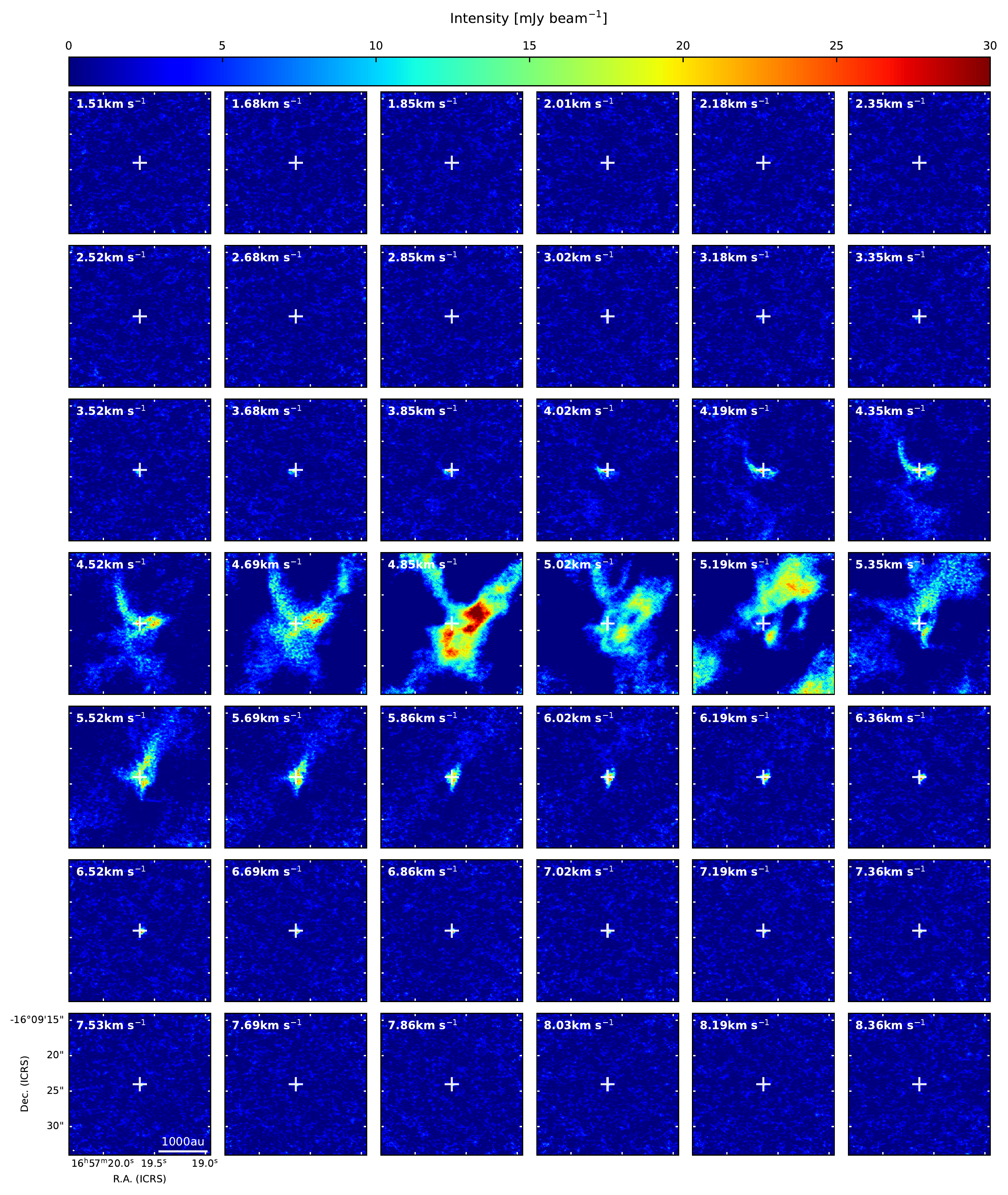}\\
\caption{ 
Velocity channel maps of the C$^{18}$O (2--1) emission toward IRAS 16544 over the entire region. Crosses indicate the position of the protostar. A white ellipse at the lower-left corner in the lowest-left panel shows the synthesized beam.
}
\label{c18o_ch}
\end{figure*}

\begin{figure*}
\centering
\includegraphics[width=180mm, angle=0]{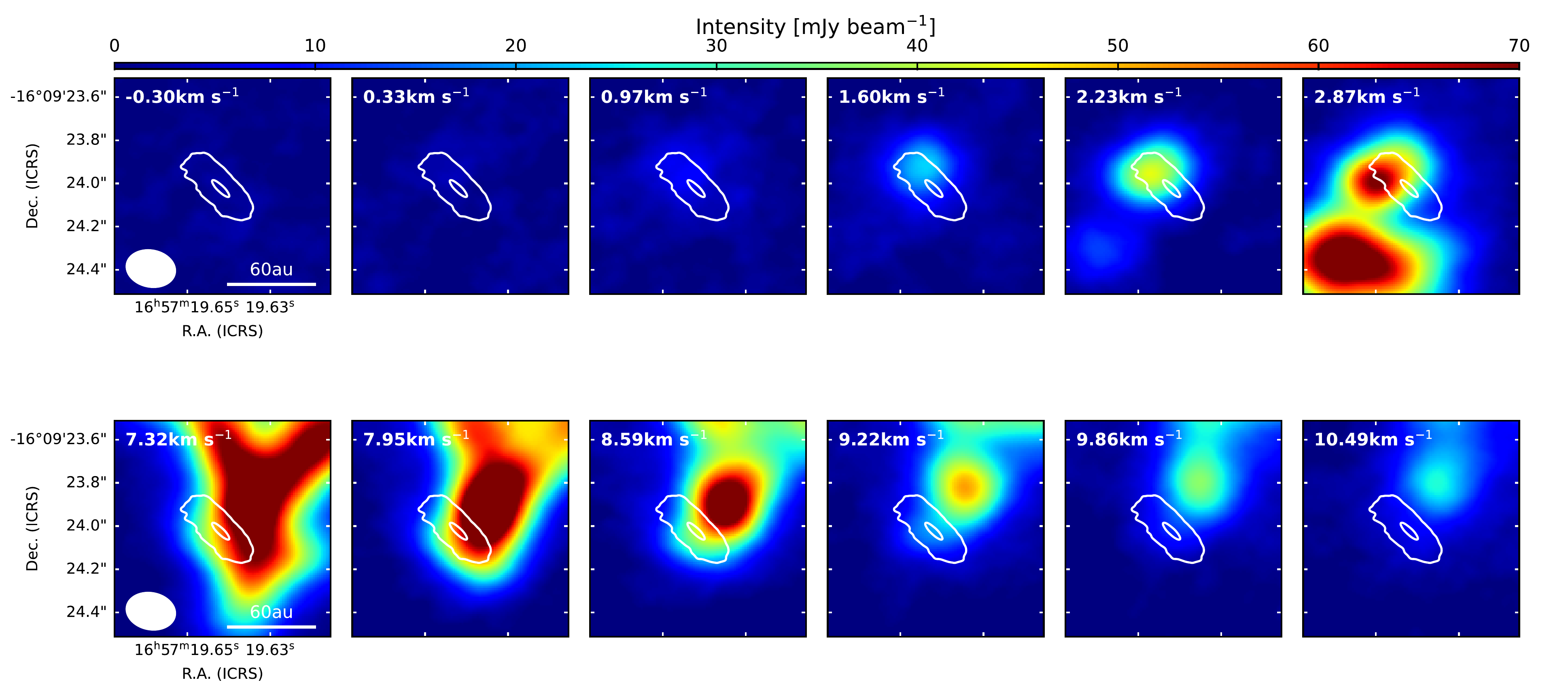}\\
\caption{ 
Velocity channel maps of the $^{12}$CO (2--1) emission in the high-velocity blue- (upper panels) and red-shifted ranges (lower panels) toward IRAS 16544 in the zoom-in region. Contours denote the distribution of the 1.3-mm dust-continuum emission, and the contour levels are 5$\sigma$ and 150$\sigma$ (1$\sigma$ = 21 $\mu$Jy beam$^{-1}$).
}
\label{12co_ch_high}
\end{figure*}

\begin{figure*}
\centering
\includegraphics[width=180mm, angle=0]{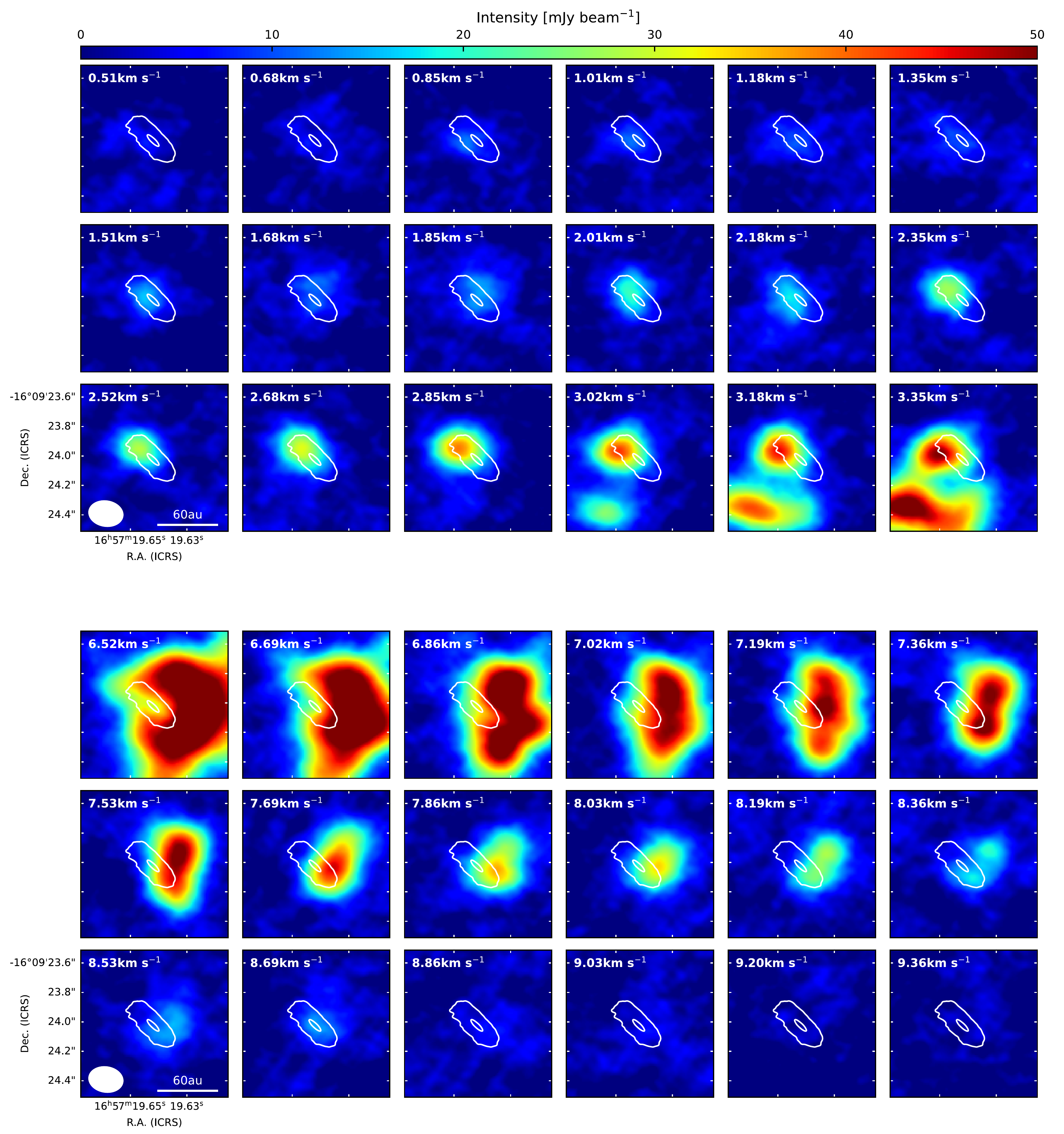}\\
\caption{ 
Velocity channel maps of the $^{13}$CO (2--1) emission in the high-velocity blue- (upper panels) and red-shifted ranges (lower panels) toward IRAS 16544 in the zoom-in region. Contours denote the distribution of the 1.3-mm dust-continuum emission, and the contour levels are 5$\sigma$ and 150$\sigma$ (1$\sigma$ = 21 $\mu$Jy beam$^{-1}$).
}
\label{13co_ch_high}
\end{figure*}

\begin{figure*}
\centering
\includegraphics[width=180mm, angle=0]{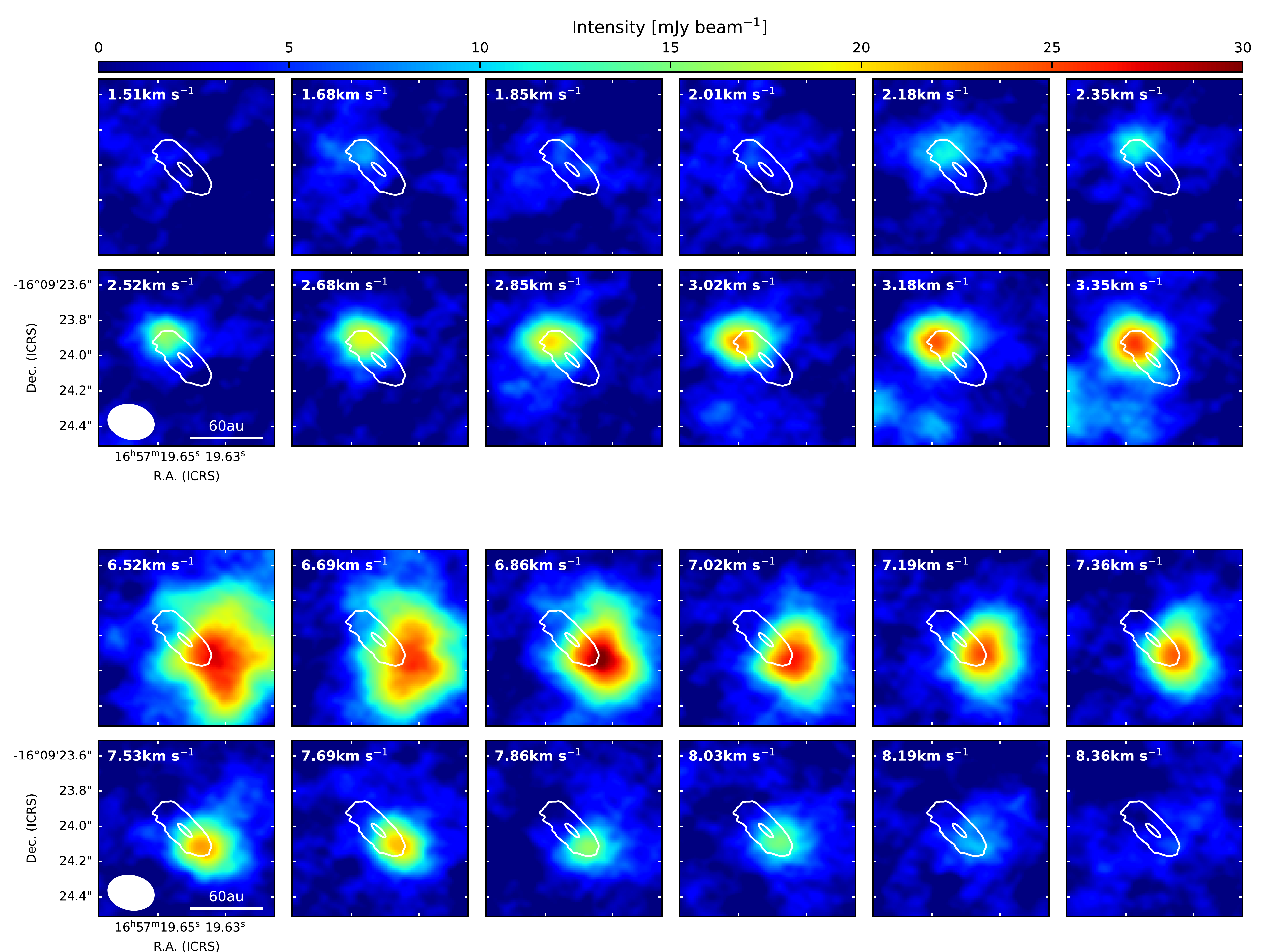}\\
\caption{ 
Velocity channel maps of the C$^{18}$O (2--1) emission in the high-velocity blue- (upper panels) and red-shifted ranges (lower panels) toward IRAS 16544 in the zoom-in region. Contours denote the distribution of the 1.3-mm dust-continuum emission, and the contour levels are 5$\sigma$ and 150$\sigma$ (1$\sigma$ = 21 $\mu$Jy beam$^{-1}$).
}
\label{c18o_ch_high}
\end{figure*}

The velocity channel maps of the CO isotopologue lines in IRAS 16544, both in the entire and zoom-in regions (six images), are shown in Figures \ref{co_ch}, \ref{13co_ch}, \ref{c18o_ch}, \ref{12co_ch_high}, \ref{13co_ch_high}, and \ref{c18o_ch_high}.% is accessible through the online journal.

%\figsetstart
%\figsetnum{6}
%\figsetnum{6}
% \figsettitle{CO channel maps.zip}
%\figsettitle{Velocity channel maps of the CO isotopologue lines}

\software{CASA \citep{Mcmullin2007}, matplotlib \citep{Hunter2007},
bettermoments \citep{2018Teague,2019Teague}, PVextractor \citep{2016Ginsburg}, APLpy \citep{aplpy2012,aplpy2019}, SLAM \citep{yusuke_aso_2023_7783868}, astropy \citep{2022Astropy}}
\facility{ALMA}

\bibliographystyle{aasjournal}
\bibliography{CB68_paper}



\end{document}